\documentclass[aps,prl,twocolumn,superscriptaddress,floatfix]{revtex4-1}
\usepackage{graphicx}
\usepackage{latexsym}
\usepackage{amssymb}
\usepackage{amsmath}
\usepackage{amsfonts}
\usepackage{bm}
\usepackage{multirow}
\usepackage{color}
\usepackage{hyperref}
\usepackage{manfnt} 

\newcommand{\ket}[1]{|#1 \rangle}
\newcommand{\dsZ}{\mathbb{Z}}
\newcommand{\scP}{\mathcal{P}}
\newcommand{\scD}{\mathcal{D}}

\newcommand{\refcite}[1]{Ref.\,\cite{#1}}
\newcommand{\eqnref}[1]{Eq.\,(\ref{#1})}
\newcommand{\figref}[1]{Fig.\,\ref{#1}}
\newcommand{\tabref}[1]{Tab.\,\ref{#1}}
\newcommand{\secref}[1]{Sec.\,\ref{#1}}
\newcommand{\appref}[1]{Appendix~\ref{#1}}

\newcommand{\cube}{\text{\mancube}}
\newcommand{\ZTwo}{\texorpdfstring{$Z_2$}{Z2} }

\newcommand{\Et}{}
\newcommand{\Eh}{}

\newcommand{\h}{{h'}}

\newcommand{\is}{=}

\newcommand{\bhat}{\hat}
\newcommand{\bbar}{\hat}
\newcommand{\bdot}{\hat}

\newcommand{\EQyz}{E^\text{YZ}_\square}
\newcommand{\EXyz}{E^\text{YZ}_+}
\newcommand{\EQ}{E_\square}
\newcommand{\EX}{E_+}
\newcommand{\ES}{E_*}
\newcommand{\EC}{E_\cube}
\newcommand{\XYZ}[3]{^{#2_\text{#1}}_{#3}}
\newcommand{\sQ}[2]{[#1\sigma]\XYZ{XY}{\square}{#2}} 
\newcommand{\sX}[2]{[#1\sigma]\XYZ{XY}{+}{#2}}
\newcommand{\tQ}[2]{[#1\tau]\XYZ{XZ}{\square}{#2}}
\newcommand{\tX}[2]{[#1\tau]\XYZ{XZ}{+}{#2}}
\newcommand{\rQ}[2]{[#1\sigma,#1\tau]\XYZ{YZ}{\square}{#2}}
\newcommand{\rX}[2]{[#1\sigma,#1\tau]\XYZ{YZ}{+}{#2}}
\newcommand{\rC}[2]{[#1\sigma,#1\tau]^\cube_{#2}}
\newcommand{\rS}[2]{[#1\sigma,#1\tau]^*_{#2}}
\newcommand{\HsQ}[2]{[#1\rho]\XYZ{XY}{\square}{#2}}
\newcommand{\HsX}[2]{[#1\rho]\XYZ{XY}{+}{#2}}
\newcommand{\HtQ}[2]{[#1\rho]\XYZ{XZ}{\square}{#2}}
\newcommand{\HtX}[2]{[#1\rho]\XYZ{XZ}{+}{#2}}
\newcommand{\HrQ}[2]{[#1\rho]\XYZ{YZ}{\square}{#2}}
\newcommand{\HrX}[2]{[#1\rho]\XYZ{YZ}{+}{#2}}
\newcommand{\HrC}[2]{[#1\rho]^\cube_{#2}}
\newcommand{\HrS}[2]{[#1\rho]^*_{#2}}

\hypersetup{
  colorlinks = true,
  urlcolor = blue,
  pdfauthor = {Kevin Slagle, Yong Baek Kim},
  pdftitle = {Fractons from Two-Spin Interactions and Dualities}
}

\begin{document}

\title{Fracton Topological Order from Nearest-Neighbor Two-Spin Interactions \\ and Dualities}

\author{Kevin Slagle}
\affiliation{Department of Physics, University of Toronto, Toronto, Ontario M5S 1A7, Canada}

\author{Yong Baek Kim}
\affiliation{Department of Physics, University of Toronto, Toronto, Ontario M5S 1A7, Canada}
\affiliation{Canadian Institute for Advanced Research, Toronto, Ontario, M5G 1Z8, Canada}

\begin{abstract}

Fracton topological order describes a remarkable phase of matter which can be characterized by fracton excitations with constrained dynamics and a ground state degeneracy that increases exponentially with the length of the system on a three-dimensional torus.
However, previous models exhibiting this order require many-spin interactions which may be very difficult to realize in a real material or cold atom system.
In this work, we present a more physically realistic model which has the so-called X-cube fracton topological order \cite{VijayXCube} but only requires nearest-neighbor two-spin interactions.
The model lives on a three-dimensional honeycomb-based lattice with one to two spin-1/2 degrees of freedom on each site and a unit cell of 6 sites.
The model is constructed from two orthogonal stacks of $Z_2$ topologically ordered Kitaev honeycomb layers \cite{KitaevHoneycomb},
  which are coupled together by a two-spin interaction.
It is also shown that a four-spin interaction can be included to instead stabilize 3+1D $Z_2$ topological order.
We also find dual descriptions of four quantum phase transitions in our model, all of which appear to be discontinuous first order transitions.

\end{abstract}

\pacs{}

\maketitle


\section{Introduction}

Quantum phases with fracton topological order \cite{ChamonModel,Rasmussen2016,Xu2006,Bravyi2011,HaahCode,Yoshida2013,VijayFracton} have recently received much attention. \cite{HsiehPartons,PremHaahNandkishore,HaahBifurcation,WilliamsonUngauging}
These phases can be characterized as having topological excitations, called fractons \cite{VijayFracton}, which are immobile (when isolated from other fractons \cite{PretkoGravity}),
  and for having large degeneracies (on a torus) which can be sub-extensive: e.g. $\text{degen} \sim 2^L$ for a 3D torus of length $L$.
Fracton order has applications including self-correcting quantum memory for quantum computers \cite{HaahCode,HaahSelfCorrection,HaahEnergyLandscape},
  and even emergent gravity given a finite density of (possibly gapless) fractons \cite{PretkoGravity}.

However, most models of fracton order require many-spin interactions,
  which seems difficult to realize in a real material or cold atom system.
For example, the so-called X-cube order requires a 12-spin interaction term \cite{VijayXCube}.
Recently, it has been shown \cite{VijayLayer,MaLayers} that X-cube order can be constructed by coupling together three orthogonal stacks of intersecting two-dimensional toric codes \cite{KitaevToric}, requiring only 4-spin interactions.
Unfortunately, 4-spin interactions can also be difficult to realize in a physical system.
On the other hand, in \refcite{KitaevHoneycomb} Kitaev presents a honeycomb lattice model with only 2-spin interactions which, in a certain large coupling limit, has an effective Hamiltonian described by toric code.
In \secref{sec:main}, we combine these ideas to produce a more realistic model of fracton order which only requires nearest-neighbor 2-spin interactions.
Our model consists of two orthogonal stacks of Kitaev honeycomb models, which intersect on certain shared links.
To induce fracton order, we add a large two-spin coupling on the shared links to couple together the orthogonal layers.

\begin{figure*}
\begin{minipage}{.58\textwidth}
\includegraphics[width=\textwidth]{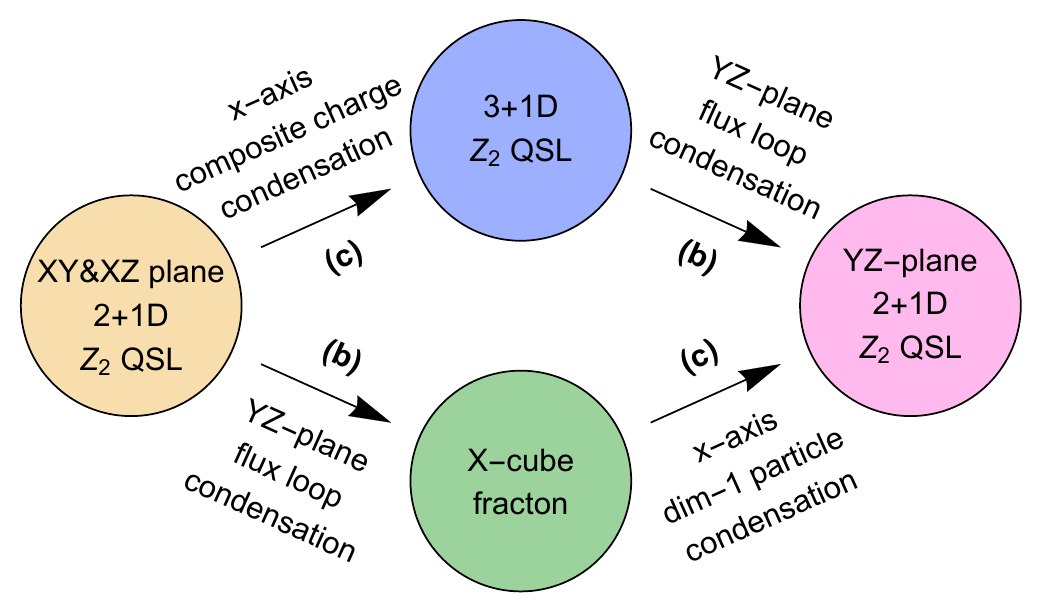} \\
\textbf{(a)} quantum phase transitions
\end{minipage}
\begin{minipage}{.38\textwidth}
\includegraphics[width=.85\textwidth]{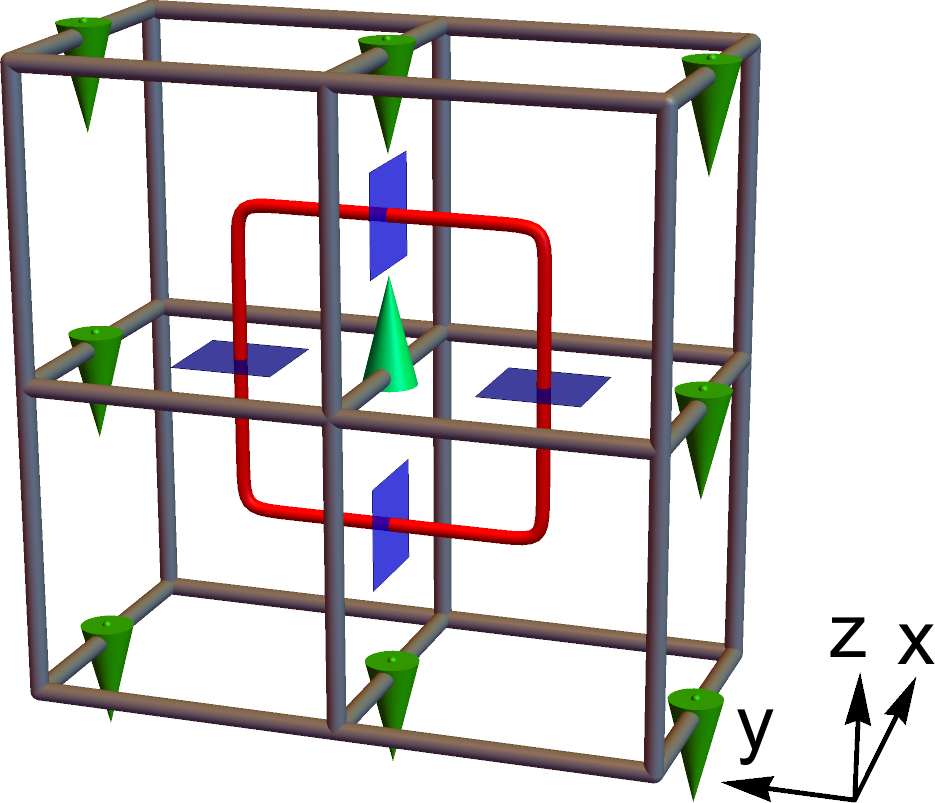} \\
\textbf{(b)} dual YZ-plane 2+1D domain wall proliferation
\includegraphics[width=.9\textwidth]{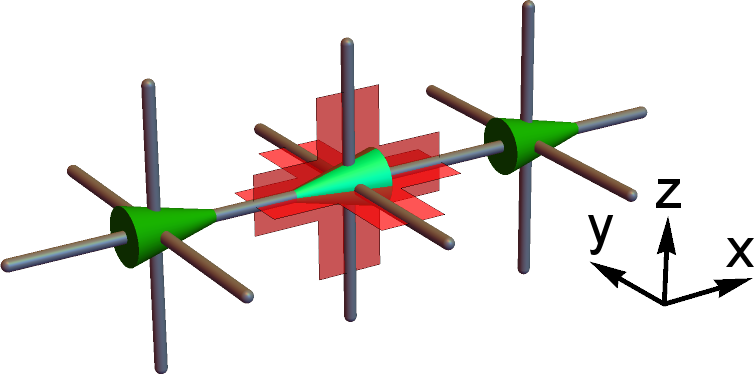} \\
\textbf{(c)} dual x-axis 1+1D Ising symmetry breaking
\end{minipage}
\caption{
Two series of condensation driven phase transitions with dual descriptions ((b) and (c)), which are described in \secref{sec:dualities} and \secref{sec:global duality}.
All four phase transitions appear to be first order when generic perturbations are added (\appref{app:MC}).
The transition from XY \& XZ plane 2+1D $Z_2$ quantum spin liquid (QSL) to X-cube fracton topological order is driven by a condensation of YZ-plane composite flux loops (red loop in (b)), which are excitations of a loop of XY and XZ--plane 2+1D toric code flux operators ($\sQ{\tilde}{p}$ and $\tQ{\tilde}{p}$ in \secref{sec:duality}).
In the dual theory, the YZ-plane flux loop is mapped to a YZ-plane domain wall excitation of Ising spins (green cones in (b), $\eta^z_{\hat\iota}$ in \secref{sec:duality}).
The transition from X-cube order to a YZ-plane 2+1D $Z_2$ QSL is driven by a condensation of
  dimension-1 X-cube particle excitations (red cross in (c), $\HsX{\bhat}{\iota}$ and $\HtX{\bhat}{\iota}$ in \secref{sec:duality'}) which are bound to 1D lines parallel to the x-axis.
In the dual theory, these x-axis particle excitations are mapped to excitations (light green cone $\ket{\leftarrow}$ in (c), $\mu^x_\iota$ in \secref{sec:duality'}) of 1+1D Ising paramagnets,
  which spontaneously break a $Z_2$ Ising symmetry when they condense since they are conserved modulo two.
The transition from XY \& XZ plane QSL to 3+1D QSL is driven by condensing a composite of XY and XZ--plane 2+1D toric code charge excitations (red cross in (c)),
  which is also dual to a 1+1D Ising paramagnet excitation.
The transition from 3+1D QSL to YZ-plane QSL is is driven by condensing YZ-plane flux loops (red in (b)) in 3+1D toric code,
  which are dual to 2+1D Ising domain walls.
These dualities also describe phase transitions shown in \figref{fig:phaseDiagram} and \figref{fig:extendedPhaseDiagram}.
}\label{fig:duality}
\end{figure*}

\refcite{VijayLayer} also showed that their isotropic coupled layer construction admits a duality describing a first order phase transition out of X-cube fracton topological order.
Additionally, \refcite{VijayXCube} gave dualities describing phase transitions out of an entire class of fracton phases;
  however, the nature of these phase transitions is currently unknown.
In this work, we find two new dualities which describe two different quantum phase transitions out of the X-cube phase which both appear to be discontinuous first order transitions.
The dual descriptions of the two phase transitions are described by a grid of weakly coupled 1+1D transverse-field Ising models,
  or a stack of decoupled 2+1D transverse-field Ising models.
These dualities are summarized in \figref{fig:duality} and are described in detail in \secref{sec:dualities}.

Although a single layer of a 1+1D or 2+1D transverse-field Ising model hosts a continuous quantum phase transition,
  this isn't necessarily the case when they are weakly coupled together.
The dual theories in our model allow the 1+1D or 2+1D Ising layers to be coupled together
  as long as a subdimensional Ising symmetry which flips all spins on any given layer is preserved.
Therefore, a $\phi_L^2 \phi_{L'}^2$ term (in a stack of $\phi^4$ field theories) or $\sigma_{LI}^z \sigma_{LJ}^z \sigma_{L'I}^z \sigma_{L'J}^z$ term
  (in a stack of transverse-field Ising models) is symmetry allowed,
  where $L$ and $L'$ denote neighboring layers, and $I$ and $J$ are nearest-neighbor sites on a layer.
($\phi_L \phi_{L'}$ or  $\sigma_{LI}^z \sigma_{L'I}^z$ are odd under subdimensional symmetry and thus not allowed.)
In \appref{app:MC} we give classical Monte Carlo evidence that these symmetry allowed perturbations are relevant under RG (renormalization group) 
  and that the transitions are likely to be discontinuous first order transitions.
That is, for small but finite layer coupling, we find evidence of first order transitions;
  but we can not rule out the possibility that smaller layer couplings could result in a continuous phase transition.
We also find intermediate phases with topological (possibly with fractons) order and spontaneous translation symmetry breaking when the sign of the layer coupling is negative.

In \secref{sec:extended} we also consider an extended model where we introduce a different interaction term to couple together the intersecting honeycomb layers (and an infinitesimal perturbation to break accidental symmetry).
Similar to the previous coupled layer constructions \cite{VijayLayer,MaLayers}, when this coupling is large, 3+1D $Z_2$ topological order is stabilized.
However, when both this coupling and the coupling that resulted in X-cube order are large, a stack of 2+1D $Z_2$ topological order is stabilized on planes orthogonal to the original two stacks of honeycomb lattices.
This result differs from the previous coupled layer constructions \cite{VijayLayer,MaLayers} which obtain a trivial state when both couplings are large.
This difference occurs because our construction is built from only two orthogonal stacks of 2+1D $Z_2$ topological order, while the previous constructions used three stacks.
In \secref{sec:global duality} we give a global duality description for four (most likely first order) phase transitions in the extended model by mapping to decoupled 2+1D and 1+1D transverse-field Ising models,
  which is summarized in \figref{fig:duality}.
In \secref{sec:previousCoupledLayer} we discuss connections to previous coupled layer constructions \cite{VijayLayer,MaLayers}.
Finally, in \secref{sec:conclusion} we conclude with possible future directions.

\section{Nearest-Neighbor Fracton Model}
\label{sec:main}

\begin{figure}
\includegraphics[width=\columnwidth]{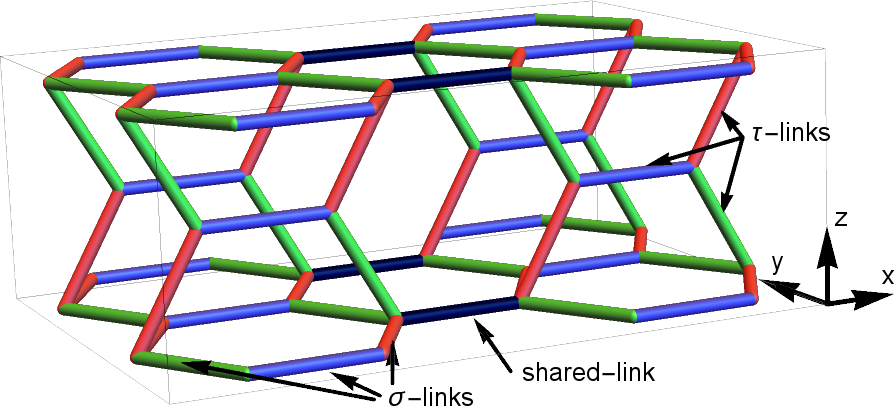}
\caption{
The lattice that we consider: orthogonal stacks of honeycomb lattices along the XY and XZ planes which intersect on the dark-blue links, which we refer to as shared links in the text.
$\sigma^\mu$ and $\tau^\mu$ spin-1/2 degrees of freedom reside on sites of the XY and XZ honeycomb lattices, respectively.
Thus, the shared sites on either end of a dark-blue link host both a $\sigma^\mu$ and $\tau^\mu$ spin.
The unit cell consists of 8 spins: the spins on both sides of a $\sigma$-link, a $\tau$-link, and a shared-link which has two spins on each side.
On each honeycomb lattice we impose a Kitaev honeycomb model (\eqnref{eq:H_K})
  where the links are colored red ($x$-link), green ($y$-link), or (dark) blue ($z$-link) to indicate the kind of coupling in \eqnref{eq:H_K}:
  $\sigma_i^x \sigma_j^x$, $\sigma_i^y \sigma_j^y$, $\sigma_i^z \sigma_j^z$ respectively (and similar for $\tau^\mu$).
To induce X-cube fracton topological order, we place a strong $\sigma_i^z \tau_i^z$ coupling on every shared site.
}\label{fig:lattice}
\end{figure}

The model lives on a 3 dimensional lattice, shown in \figref{fig:lattice}, of two orthogonal stacks of honeycomb lattices:
  one stack consisting of a honeycomb lattice on an XY plane for each $z \in \dsZ$,
  and the other stack on XZ planes for each $y \in \dsZ$.
Some of the links (dark-blue in \figref{fig:lattice}) of each honeycomb lattice will overlap with a link on an orthogonal honeycomb lattice.
We will refer to these links as ``shared links'', and the sites on each end of a shared link will be referred to as ``shared sites''.
There will be a spin-1/2 degree of freedom on every site of the honeycomb lattices,
  which means that there will be two spin-1/2 on each shared site.
The spins on the XY and XZ planes will be labeled $\sigma^\mu$ and $\tau^\mu$, respectively.

\begin{figure}
\includegraphics[width=\columnwidth]{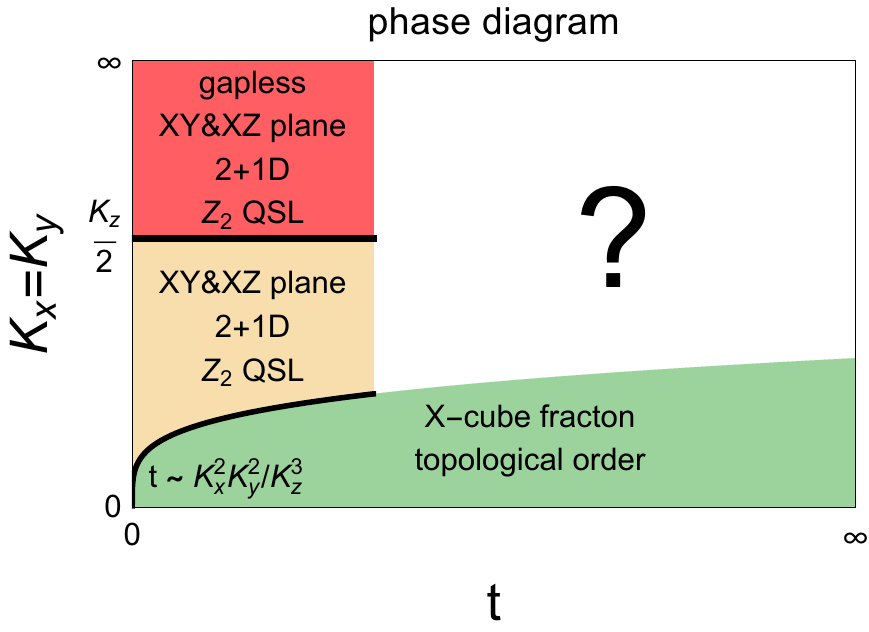}
\caption{
Rough phase diagram of the Hamiltonian in \eqnref{eq:H} as a function of $t$ (horizontal axis) and $K_x = K_y$ (vertical axis);
  $K_z$ will be used to set the energy scale.
When $t=0$, the model consists of two orthogonal stacks (along the XY and XZ planes) of decoupled honeycomb lattices (\figref{fig:lattice}),
  where each honeycomb lattice will host a quantum spin liquid (QSL) with $Z_2$ topological order,
  which will be gapped when $K_x = K_y < K_z/2$ and gapless when $K_x = K_y \ge K_z/2$ \cite{KitaevHoneycomb}.
These QSL phases are stable to a small coupling $t$ which couples the orthogonal honeycomb lattices together.
(The gapless QSL is stable since $t$ preserves time reversal symmetry; see e.g. Sec. 6.1 of \refcite{KitaevHoneycomb}.)
In \secref{sec:X-cube} and \appref{app:X-cube} we argue that X-cube fracton topological order \cite{VijayXCube} results when $\max(K_x, K_y) \ll \min(K_z, t)$.
We do not know what phase(s) occur in the white region.
In the bottom left corner, a phase transition between the QSL and fracton orders occurs near $t \sim K_x^2 K_y^2 / K_z^3$ (in the $\max(K_x,K_y) \ll K_z$ limit),
  which can be inferred from \figref{fig:largeKzDiagram} in the appendix.
In \secref{sec:duality} we propose that this transition is dual to a stack 2+1D Ising transitions;
  in \appref{app:MC} we find that the transition is likely first order and that (depending on the sign of certain perturbations) a topologically ordered intermediate phase is also possible.
}\label{fig:phaseDiagram}
\end{figure}

The Hamiltonian is given by
\begin{align}
  H   & = H_K + H_t \label{eq:H}\\
  H_K &= - \sum_{\mu = x,y,z} \sum_{\left< i j \right>}^{\mu\text{-links}}
           \begin{cases}
             K_\mu\, \sigma^\mu_i \sigma^\mu_j & \sigma\text{-link} \\
             K_\mu\, \tau  ^\mu_i \tau  ^\mu_j & \tau  \text{-link} \\
             K_z  \, (\sigma^z_i \sigma^z_j + \tau^z_i \tau^z_j) & \text{shared-link}
           \end{cases} \label{eq:H_K}\\
  H_t &= - \sum_{\left< i j \right>}^\text{shared-links} \frac{t}{2}\, (\sigma^z_i \tau^z_i + \sigma^z_j \tau^z_j) \label{eq:H_t}
\end{align}
In \secref{sec:extended}, we will discuss an alternate coupling that can produce 3+1D $Z_2$ topological order.
A rough phase diagram for this model is given in \figref{fig:phaseDiagram}.
The signs of all coefficients in $H$ (i.e. $K_\mu$ and $t$, and also $h$ and $\h$ in $H'$, which is defined later in \eqnref{eq:H'}) will be taken to be positive.

$H_K$ is a sum of Kitaev honeycomb models \cite{KitaevHoneycomb} on each honeycomb lattice.
$\sum_{\mu = x,y,z}$ sums over the different kinds of links ($x$=red, $y$=green, $z$=blue in \figref{fig:lattice}),
  and $\sum_{\left< i j \right>}^{\mu\text{-links}}$ sums over all links of type $\mu$.
A link is also referred to as a $\sigma$-link (or $\tau$-link) if it is a link on either a XY (or XZ) honeycomb plane.
If a link is shared by both an XY and XZ honeycomb plane, then it is instead referred to as a shared-link.
$K_x$, $K_y$, $K_z$ are couplings across the different bond types.
If we only consider $H_K$, then the ground state will have decoupled 2+1D $Z_2$ topological order on each honeycomb lattice, and will be gapped when $K_z > K_x + K_y$ \cite{KitaevHoneycomb}.

$H_t$ couples the 2D honeycomb lattices together.
$H_t$ consists of a $\sigma^z_i \tau^z_i$ coupling on every shared site.
In \secref{sec:X-cube} we argue that this term will cause loops of fluxes to condense, which will result in X-cube fracton topological order when $K_z$ and $t$ are large.

\subsection{Decoupled Phase: XY\&XZ plane 2+1D \ZTwo Topological Order}
\label{sec:decoupled}

\begin{figure}
\begin{center}\includegraphics[width=\columnwidth]{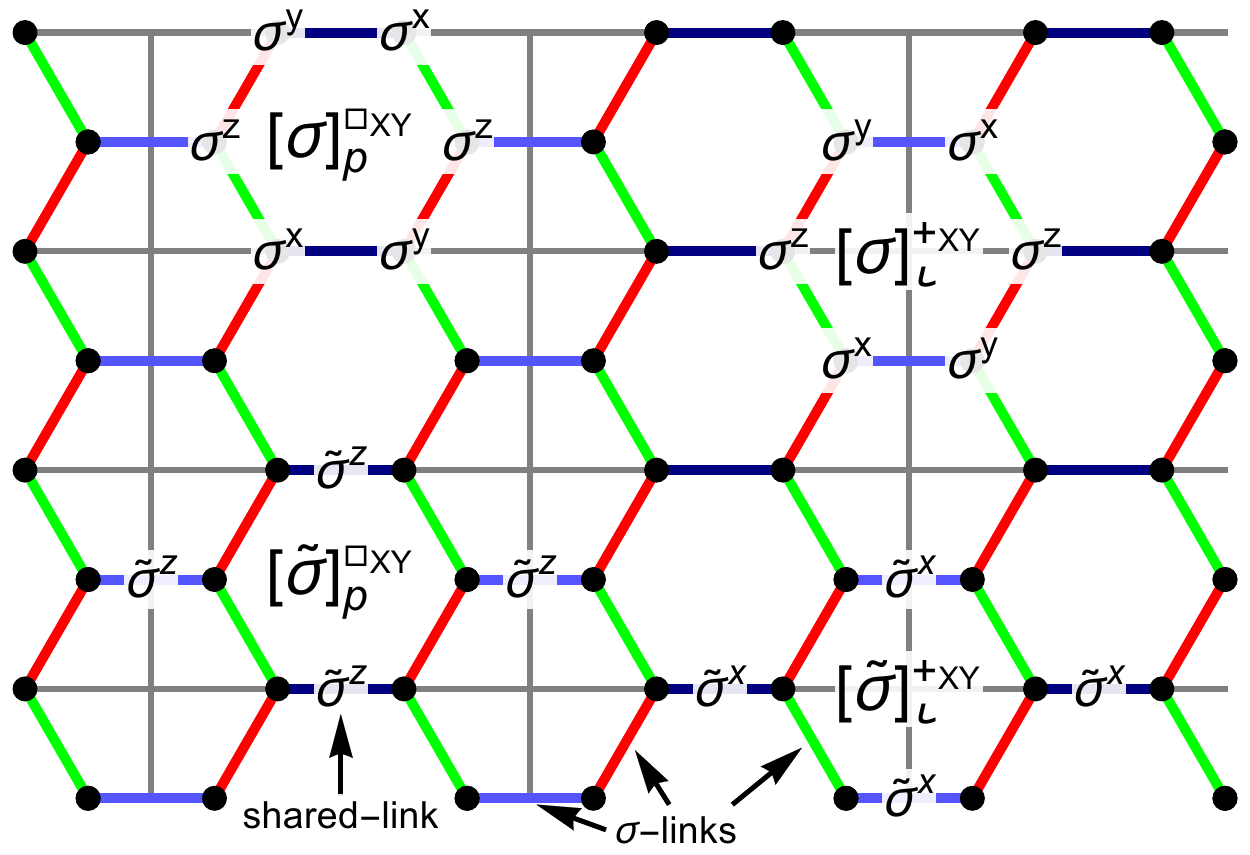}\end{center}
\caption{
An XY plane honeycomb lattice: a constant $z$ slice of \figref{fig:lattice}.
There is a $\sigma^\mu$ degree of freedom on every site,
  and also a $\tau^\mu$ on every shared-site: i.e. every site next to a dark-blue shared-link, which overlap gray lines.
Red, green, and blue links host a $K_x \sigma_i^x \sigma_j^x$, $K_y \sigma_i^y \sigma_j^y$, or $K_z \sigma_i^z \sigma_j^z$ coupling, respectively.
When $\max(K_x,K_y) \ll K_z$, degenerate perturbation theory will produce the operators $\sQ{}{p}$ (top-left) and $\sX{}{\iota}$ (top-right),
  which both take the form $\sigma_{i_1}^z \sigma_{i_2}^y \sigma_{i_3}^x \sigma_{i_4}^z \sigma_{i_5}^y \sigma_{i_6}^x$ for sites around a hexagon, as shown above.
The low energy Hilbert space will have $\sigma^z_i \sigma^z_j = 1$ across all blue links;
  and so we will perform a local change of basis (and projection) $\sigma \rightarrow \tilde\sigma$ given in \eqnref{eq:toric basis} to describe the effective low energy Hilbert space;
  the new $\tilde\sigma^\mu$ operators are centered on the blue links.
In the $\tilde\sigma$ basis, $\sQ{}{p} \rightarrow \sQ{\tilde}{p}$ (bottom-left) and $\sX{}{\iota} \rightarrow \sX{\tilde}{\iota}$ (bottom-right) take the form of toric code flux and charge operators, as shown above.
The gray lines indicate the rectangular lattice of the toric code.
The XZ planes are similar, but with the replacement $\sigma \leftrightarrow \tau$.
See \figref{fig:cubicLattice} for a 3 dimensional version.
}\label{fig:toricCode}
\end{figure}

\begin{figure}
\includegraphics[width=\columnwidth]{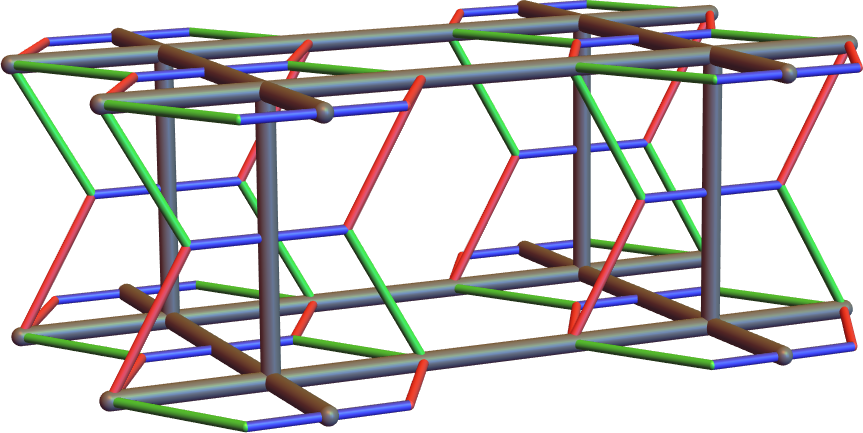}
\caption{
The lattice that we consider (\figref{fig:lattice}), except we also draw gray lines (as in \figref{fig:toricCode}) to indicate the intersecting rectangular lattices of the toric
code (\eqnref{eq:H toric}) on XY and XZ planes.
}\label{fig:cubicLattice}
\end{figure}

Before explaining the fracton phase, we will first review the $t = 0$ case with $\max(K_x,K_y) \ll K_z$.
In this limit, the XY and XZ--plane hexagon lattices are completely decoupled from each other, and each form a Kitaev honeycomb model \cite{KitaevHoneycomb}.
In \figref{fig:toricCode}, we show a honeycomb lattice on the XY plane.
When $\max(K_x,K_y) \ll K_z$, degenerate perturbation theory results in the following effective Hamiltonian (ignoring subleading terms) \cite{KitaevHoneycomb}:
\begin{align}
  H_\text{toric}           &= H^\text{XY}_\text{toric} + H^\text{XZ}_\text{toric} \label{eq:H toric}\\
  H^\text{XY}_\text{toric} &= - \sum_{p \is \square}^\text{XY planes} \EQ\, \sQ{}{p} - \sum_{\iota \is +}^\text{XY plane} \EX\, \sX{}{\iota} \nonumber\\
  H^\text{XZ}_\text{toric} &= - \sum_{p \is \square}^\text{XZ planes} \EQ\, \tQ{}{p} - \sum_{\iota \is +}^\text{XZ plane} \EX\, \tX{}{\iota} \nonumber\\
  \EQ \sim \EX &\sim \frac{K_x^2 K_y^2}{K_z^3} \quad (\text{when } t = 0) \nonumber
\end{align}
\[ \sQ{}{p}, \tQ{}{p}, \sX{}{\iota}, \tX{}{\iota} \quad \text{are defined in \figref{fig:toricCode}} \]
$\sum_{p \is \square}^\text{XY planes}$ sums over plaquettes $p$ that lie on an XY plane and
$\sum_{\iota \is +}^\text{XY plane}$ sums over vertices $\iota$ that lie on an XY plane (see \figref{fig:toricCode}).
$\sum_{p \is \square}^\text{XZ planes}$ and $\sum_{\iota \is +}^\text{XZ plane}$ are similar, but instead sum over XZ plalnes.
$\sQ{}{p}$ and $\sX{}{\iota}$ are defined in \figref{fig:toricCode} and are closely related to the plaquette/flux and vertex/charge operators in Kitaev's toric code \cite{KitaevToric}.
$\sQ{}{p}$ and $\sX{}{\iota}$ are centered on the plaquettes $p$ and vertices $\iota$, respectively, of the XY plane rectangular lattice (gray in \figref{fig:toricCode}).
$\tau$ operators are similar, but instead live on XZ planes.
In \figref{fig:cubicLattice}, we show the 3 dimensional lattice again, but with the intersecting rectangular toric code lattices also drawn.
A more detailed calculation is done in \appref{app:XY-XZ}, with excitation energies $\EQ$ and $\EX$ given for nonzero $t$ (\eqnref{eq:energies}).

It is useful to do a change of basis and projection into the low energy Hilbert space where the energy of the large $K_z$ term in $H$ (\eqnref{eq:H}) is minimized with $\sigma^z_i \sigma^z_j = 1$ across every $z$-link:
\begin{align}
  \sigma^z_i \sigma^z_j &\rightarrow 
    \begin{cases}
       1 & \sigma\text{-link} \nonumber\\
       1 & \text{shared-link}
    \end{cases} \\
  \sigma^z_i \sigma^0_j \;\;\&\;\; \sigma^0_i \sigma^z_j &\rightarrow
    \begin{cases}
       \tilde\sigma^z_\ell & \sigma\text{-link} \label{eq:toric basis}\\
       \tilde\sigma^x_\ell & \text{shared-link}
    \end{cases} \\
  \sigma^x_i \sigma^y_j    &\rightarrow
    \begin{cases}
       \tilde\sigma^x_\ell & \sigma\text{-link} \nonumber\\
       \tilde\sigma^z_\ell & \text{shared-link}
    \end{cases} \\
  \tau &\rightarrow \tilde\tau \text{ is obtained by } \sigma \leftrightarrow \tau \nonumber\\
  H_\text{toric} &\rightarrow \tilde{H}_\text{toric} = \tilde{H}^\text{XY}_\text{toric} + \tilde{H}^\text{XZ}_\text{toric} \label{eq:H toric'}
\end{align}
\begin{align}
  \tilde{H}^\text{XY}_\text{toric} &= - \sum_{p \is \square}^\text{XY planes} \EQ\, \sQ{\tilde}{p} - \sum_{\iota \is +}^\text{XY plane} \EX\, \sX{\tilde}{\iota} \nonumber\\
  \tilde{H}^\text{XZ}_\text{toric} &= - \sum_{p \is \square}^\text{XZ planes} \EQ\, \tQ{\tilde}{p} - \sum_{\iota \is +}^\text{XZ plane} \EX\, \tX{\tilde}{\iota} \nonumber
\end{align}
\begin{align}
  \sQ{}{p}     \rightarrow \sQ{\tilde}{p}     &= \prod_{\ell \in \square}^p \tilde\sigma^z_\ell \!&,\;\;
  \sX{}{\iota} \rightarrow \sX{\tilde}{\iota} &= \prod_{\ell \in +}^\iota   \tilde\sigma^x_\ell \nonumber\\
  \tQ{}{p}     \rightarrow \tQ{\tilde}{p}     &= \prod_{\ell \in \square}^p \tilde\tau  ^z_\ell \!&,\;\;
  \tX{}{\iota} \rightarrow \tX{\tilde}{\iota} &= \prod_{\ell \in +}^\iota   \tilde\tau  ^x_\ell \nonumber
\end{align}
Thus, we are projecting into the $\sigma^z_i \sigma^z_j = 1$ subspace and doing a unitary rotation on the shared-links.
$\sigma^0_i$ (e.g in $\sigma^0_i \sigma^z_j \rightarrow \tilde\sigma^z_\ell$) is an identity operator, and is only written to emphasize that $i$ and $j$ are the sites on the two sides of the link $\ell$.
We will use a tilde to denote operators in this new basis.
In this basis, the operators $\sQ{}{p} \rightarrow \sQ{\tilde}{p}$ and $\sX{}{\iota} \rightarrow \sX{\tilde}{\iota}$ (also given in \figref{fig:toricCode}) in $\tilde{H}_\text{toric}$ exactly reproduce the flux and charge operators in Kitaev's toric code \cite{KitaevToric}.
Therefore, when $t = 0$ and $\max(K_x,K_y) \ll K_z$, our model (\eqnref{eq:H_t}) is in a decoupled phase where every honeycomb lattice on an XY or XZ plane hosts its own gapped 2+1D $Z_2$ topological order.
This is the ``XY\&XZ plane 2+1D $Z_2$ QSL'' phase in \figref{fig:phaseDiagram}.

\subsection{X-Cube Fracton Phase}
\label{sec:X-cube}

In this section, we will explain how X-cube fracton topological order results from our model $H$ (\eqnref{eq:H}) in the limit
\begin{equation}
  \max(K_x,K_y) \ll t \ll K_z \label{eq:X-cube limit}
\end{equation}
More generally, in the \appref{app:X-cube} we show that $\max(K_x,K_y) \ll \min(t, K_z)$ is a sufficient condition for X-cube order.
However, \eqnref{eq:X-cube limit} allows us to apply the results from the previous section and is more closely connected to the previous layer constructions in \refcite{VijayLayer,MaLayers}, and so we will discuss this limit in the main text and consider the more general condition (\eqnref{eq:X-cube condition}) in the appendix.

Thus, we will start from the decoupled $t=0$ limit discussed in the previous section, and then make $t$ very large.
In the toric code basis (\eqnref{eq:toric basis}), $H$ (\eqnref{eq:H}) becomes:
\begin{align}
  H   \rightarrow \tilde{H}   &= \tilde{H}_\text{toric} + \tilde{H}_t \label{eq:H tilde} \\
  H_t \rightarrow \tilde{H}_t &= - \sum_\ell^\text{shared-links} t\, \tilde\sigma^x_\ell \tilde\tau^x_\ell \label{eq:H_t tilde}
\end{align}
  where $\tilde{H}_\text{toric}$ was given in \eqnref{eq:H toric'}.
We see that $\tilde{H}_t$ couples the XY and XZ--plane toric codes together on the shared-links (gray links parallel to x-axis in \figref{fig:cubicLattice}) with a $\tilde\sigma^x_\ell \tilde\tau^x_\ell$ coupling.
When acting on the decoupled $t=0$ ground state, $\tilde\sigma^x_\ell \tilde\tau^x_\ell$ creates two $\sQ{\tilde}{p}$ and two $\tQ{\tilde}{p}$ excitations on the four plaquettes (of the gray cubic lattice in \figref{fig:cubicLattice}) that neighbor the link $\ell$.
When $t$ is large, these excitations condense and the result is X-cube fracton topological order.
(A similar condensation process is described in \refcite{VijayLayer} and \cite{MaLayers}; see \secref{sec:previousCoupledLayer} for connections.)

We will derive this result using degenerate perturbation theory (DPT).
First, we split $H$ (\eqnref{eq:H}) into
\begin{align}
  H   &= H_0 + H_1 \label{eq:DPT from toric} \nonumber\\
  H_0 &= \tilde{H}_t \\
  H_1 &= \tilde{H}_\text{toric} \nonumber
\end{align}
where $H_0$ is the unperturbed Hamiltonian, $H_1$ is the perturbing Hamiltonian, and $\tilde{H}_\text{toric}$ (\eqnref{eq:H toric'}) was derived in the previous section.
\eqnref{eq:X-cube limit} is sufficient to ensure that DPT converges.

The effective Hamiltonian of the low energy Hilbert space is given by (see \appref{app:DPT} for details):
\begin{align}
  H^\text{eff} &= \sum_{n=1}^\infty \scP (H_1 \scD)^{n-1} H_1 \scP + \cdots \label{eq:DPT}\\
  \scP         &= \prod_\ell^\text{shared-links} \frac{1}{2} (1 + \tilde\sigma^x_\ell \tilde\tau^x_\ell) \nonumber\\
  \scD         &= - \frac{1 - \scP}{H_0 - E_0} \nonumber
\end{align}
$\scP$ projects into the low energy Hilbert space where $H_0 = \tilde{H}_t$ is minimized.
$\scD$ projects onto states not in the low energy Hilbert space and penalizes energy excitations by a factor $1/(H_0 - E_0)$
  where $E_0$ is the (degenerate) ground state energy of $H_0$.
``$\cdots$'' contains many terms (e.g. $-\frac{1}{2} \scP H_1 \scD^2 H_1 \scP H_1 \scP$);
  but these more complicated terms will be subleading corrections in this section.
$\sX{\tilde}{\iota}$ and $\tX{\tilde}{\iota}$ (in $H_1 = \tilde{H}_\text{toric}$) commute with $H_t$ (\eqnref{eq:H_t}),
  which implies that they do not create excitations out of the low energy Hilbert space.
However, $\sQ{\tilde}{p}$ and $\tQ{\tilde}{p}$ do not commute with $\tilde{H}_t$,
  since $\tilde{H}_t$ creates two $\sQ{\tilde}{p}$ and two $\tQ{\tilde}{p}$ excitations.
Viewed differently, $\sQ{\tilde}{p}$ and $\tQ{\tilde}{p}$ create $\tilde{H}_t$ excitations, which are not in the low energy Hilbert space.
Therefore, $\sX{\tilde}{\iota}$ and $\tX{\tilde}{\iota}$ will be contained in $H^\text{eff}$ but $\sQ{\tilde}{p}$ and $\tQ{\tilde}{p}$ will not.

\begin{figure}
\includegraphics[width=\columnwidth]{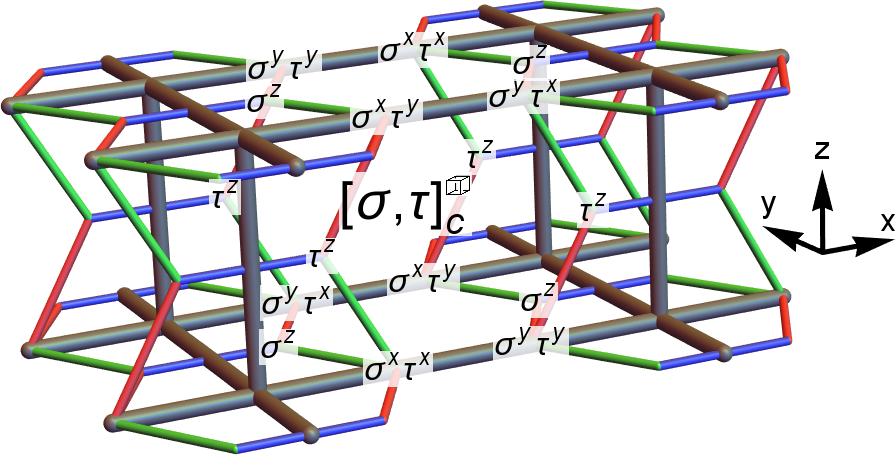}
\includegraphics[width=.5\columnwidth]{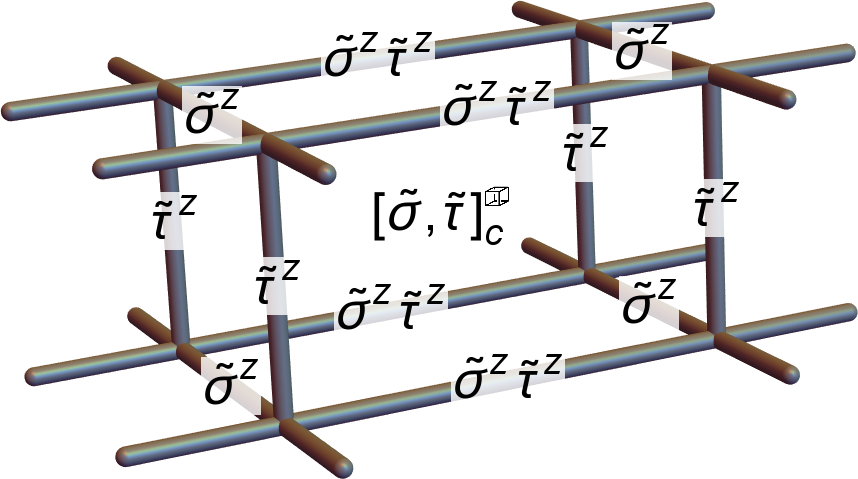}\includegraphics[width=.5\columnwidth]{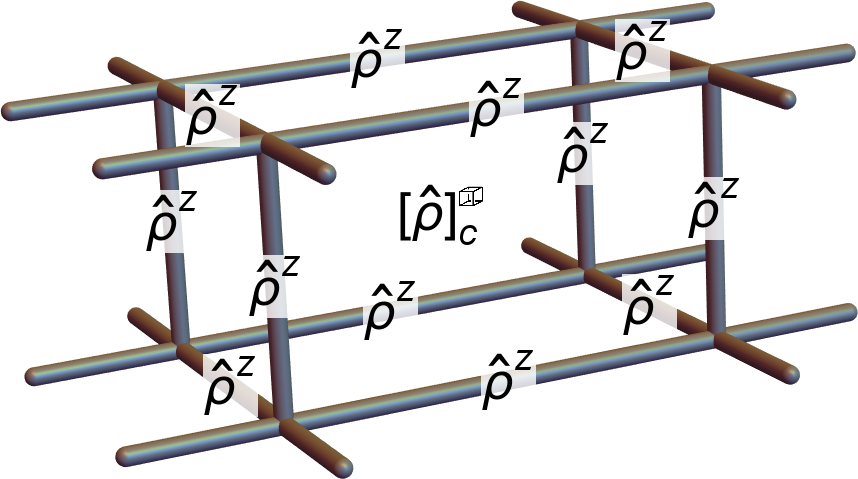}
\caption{
The $Z_2$ fracton number operator generated from degenerate perturbation theory (\eqnref{eq:H X-cube}) when $\max(K_x,K_y) \ll \min(K_z, t)$.
The operator is expressed in three different basis:
  $\rC{}{c}$ is in original basis (\eqnref{eq:H}),
  $\rC{\tilde}{c}$ is in the 2D toric code basis (\eqnref{eq:toric basis}), and
  $\HrC{\bhat}{c}$ is in the X-cube model basis (\eqnref{eq:X-cube basis}).
$\rC{\tilde}{c}$ is the product of two XY-plane and two XZ-plane plaquette/flux operators ($\sQ{\tilde}{p}$ and $\tQ{\tilde}{p}$, \figref{fig:toricCode}) around the cube.
}\label{fig:fractonOperator}
\end{figure}

We must now consider higher order terms ($n \ge 2$) in $H^\text{eff}$ (\eqnref{eq:DPT}) to discover what role $\sQ{\tilde}{p}$ and $\tQ{\tilde}{p}$ end up playing.
The first nonzero term involving $\sQ{\tilde}{p}$ or $\tQ{\tilde}{p}$ is $\EC \rC{\tilde}{c}$, which is obtained at fourth order ($n=4$), and the resulting effective Hamiltonian is
\begin{align}
  H^\text{eff}    &= \tilde{H}_\text{X-cube} + \cdots \nonumber\\
  \tilde{H}_\text{X-cube} &= - \sum_{\iota \is +}^\text{XY planes} E _+\, \sX{\tilde}{\iota}
                             - \sum_{\iota \is +}^\text{XZ planes} E _+\, \tX{\tilde}{\iota} \nonumber\\
                    &\,\quad 
                             - \sum_{c \is \cube} \EC\, \rC{\tilde}{c} \label{eq:H X-cube}\\
  \EC         &\sim \frac{\EQ^4}{t^3} \sim \frac{K_x^8 K_y^8}{K_z^{12} t^3} \quad\text{assuming \eqref{eq:X-cube limit}} \\
  \rC{\tilde}{c} & \quad \text{is defined in \figref{fig:fractonOperator}} \nonumber
\end{align}
$\sum_{c \is \cube}$ sums over the cubes $c$ shown in \figref{fig:cubicLattice}.
$\rC{\tilde}{c}$ is the product of two XY-plane plaquette $\sQ{\tilde}{p}$ and two XZ-plane plaquette $\tQ{\tilde}{p}$ operators around a cube (\figref{fig:fractonOperator}).
Note that $\rC{\tilde}{c}$ commutes with $\tilde{H}_t$, and can therefore appear in the effective Hamiltonian since it preserves the low energy Hilbert space.
The ``$\cdots$'' in $H^\text{eff}$ denotes other subleading terms generated by DPT which are not important.
Another round of DPT on $H^\text{eff}$ with an unperturbed Hamiltonian $H_0' = \tilde{H}_\text{X-cube}$ (and perturbation $H_1' = H^\text{eff} - \tilde{H}_\text{X-cube}$) will eliminate terms in ``$\cdots$'' which anticommute with $\tilde{H}_\text{X-cube}$,
  resulting in an exactly solvable effective Hamiltonian with the same ground state as $\tilde{H}_\text{X-cube}$.
The ``$\sim$'' in the energy expressions means that these expressions are only asymptotically correct (in the $\max(K_x,K_y) \ll t \ll K_z$ limit);
  e.g. factors of two are ignored.

\begin{figure}
\includegraphics[width=.5\columnwidth]{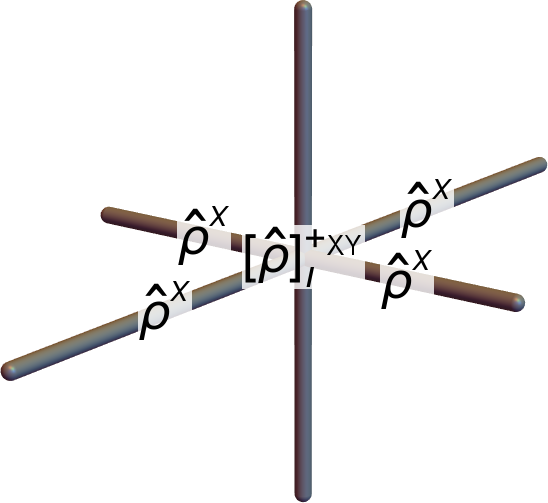}\includegraphics[width=.5\columnwidth]{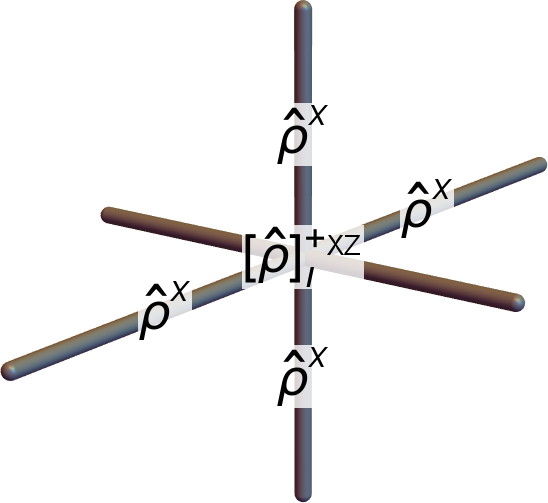}
\includegraphics[width=.5\columnwidth]{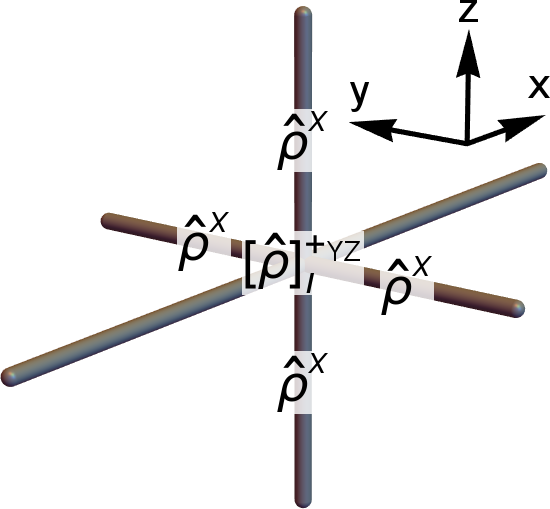}
\caption{
The $\HsX{\bhat}{\iota}$, $\HtX{\bhat}{\iota}$, and $\HrX{\bhat}{\iota}$ operators in the X-cube basis (\eqnref{eq:X-cube basis}).
These operators, along with $\HrC{\bhat}{c}$ (\figref{fig:fractonOperator}), are the commuting operators used to define the X-cube fracton Hamiltonian (\eqnref{eq:H X-cube} after a change of basis \eqref{eq:X-cube basis}).
$\sX{}{\iota}$ was given in the original basis in \figref{fig:toricCode}; $\tX{}{\iota}$ is similar, but with $\sigma \leftrightarrow \tau$ and on a XZ plane instead of a XY plane.
$\rX{}{\iota}$ is given by $\rX{}{\iota} = \sX{}{\iota} \tX{}{\iota}$.
}\label{fig:plusOperators}
\end{figure}

It is again useful to do a change of basis and projection into the low energy Hilbert space where now the energy of $H_0$ is minimized with $\tilde\sigma^x_\ell \tilde\tau^x_\ell = 1$ on every shared-link:
\begin{align}
  \sigma\text{-links:} && \tilde\sigma^\mu_\ell                 &\rightarrow \bhat\rho^\mu_\ell \nonumber\\
  \tau  \text{-links:} && \tilde\tau  ^\mu_\ell                 &\rightarrow \bhat\rho^\mu_\ell \nonumber\\
  \text{shared-links:} && \tilde\sigma^x_\ell \tilde\tau^x_\ell &\rightarrow 1 \label{eq:X-cube basis}\\
                       && \tilde\sigma^x_\ell                   &\rightarrow \bhat\rho^x_\ell \nonumber\\
                       && \tilde\sigma^z_\ell \tilde\tau^z_\ell &\rightarrow \bhat\rho^z_\ell \nonumber
\end{align}
\begin{align}
  \tilde{H}_\text{X-cube} &\rightarrow \bhat{H}_\text{X-cube} = \label{eq:H X-cube'}
\end{align}
\vspace{-.8cm}
\[                       - \sum_{\iota \is +}^\text{XY planes} E _+\, \HsX{\bhat}{\iota}
                         - \sum_{\iota \is +}^\text{XZ planes} E _+\, \HtX{\bhat}{\iota}
                   \quad 
                         - \sum_{c \is \cube} \EC\, \HrC{\bhat}{c} \]
Thus, we are relabeling the $\sigma$ and $\tau$--links, but on the shared-links we are
  projecting into the $\tilde\sigma^x_\ell \tilde\tau^x_\ell = 1$ subspace.
A hat over operators is used to denote operators in this new basis.
See \figref{fig:fractonOperator} and \figref{fig:plusOperators} for descriptions of the operators in $\bhat{H}_\text{X-cube}$.
In this basis, $\bhat{H}_\text{X-cube}$ almost exactly reproduces the X-cube model introduced in \cite{VijayXCube}.
The only difference is that $\bhat{H}_\text{X-cube}$ does not include the $\HrX{\bhat}{\iota} \equiv \HsX{\bhat}{\iota} \HtX{\bhat}{\iota}$ operator.
However, this does not change the phase of the Hamiltonian; it merely adds anisotropy.
Only two of $\HsX{\bhat}{\iota}$, $\HtX{\bhat}{\iota}$, and $\HrX{\bhat}{\iota}$ are necessary for X-cube order, since the third is a product of the other two.
Therefore, when $\max(K_x,K_y) \ll \min(K_z, t)$, our model (\eqnref{eq:H}) has X-cube fracton topological order, as claimed in the phase diagram \figref{fig:phaseDiagram}.
Note that $\bhat{H}_\text{X-cube}$ is actually a rather simple Hamiltonian in the sense that all of its terms commute.
Its ground state is specified by $\HsX{\bhat}{\iota} = \HtX{\bhat}{\iota} = \HrC{\bhat}{c} = 1$ and has degeneracy $2^{6L-3}$ on a 3D torus \cite{VijayXCube,MaLayers}.
It is also interesting to note that the X-cube operators generated by DPT ($\rC{}{p}$, $\sX{}{\iota}$ and $\tX{}{\iota}$)
  all commute with the original Hamiltonian (\eqnref{eq:H}).

$\HrC{\bhat}{c}$ is the fracton number operator.
Its excitations are of particular interest because they can only be created in groups of four \cite{VijayFracton,VijayXCube}.
This can be seen from \figref{fig:fractonOperator} where the application of a single $\bhat\rho^x_\ell$ operator on any link $\ell$ will excite four $\HrC{\bhat}{c}$ fracton operators on the four cubes that share the link $\ell$.
If we have an isolated fracton, then due to the $Z_2$ nature of the fracton operator, the application of a $\bhat\rho^x_\ell$ operator will annihilate the isolated fracton but create three other fractons.
Therefore, an isolated fracton can not move without creating additional excitations.
An isolated fracton is immobile;
  only a pair of fractons can move without creating additional excitations.

\section{Duality and Phase Transitions out of X-cube Order}
\label{sec:dualities}

\subsection{2+1D Duality}
\label{sec:duality}

In this subsection we will discuss a duality (similar to the one in \refcite{VijayLayer}, see \secref{sec:previousCoupledLayer} for connections)
  which describes a (likely first order) phase transition from X-cube fracton order to decoupled XY\&XZ plane 2+1D $Z_2$ topological order.
The duality will be applied to $\tilde{H}$ (\eqnref{eq:H tilde}):
\begin{align}
  \tilde{H} &= \tilde{H}_\text{toric} + \tilde{H}_t \label{eq:H coupled} \\
            &\sim  - \EQ\, (\sQ{\tilde}{p} + \tQ{\tilde}{p})
                   - \EX\, (\sX{\tilde}{\iota} + \tX{\tilde}{\iota}) \nonumber\\
          &\,\quad - t\, \tilde\sigma^x_\ell \tilde\tau^x_\ell \nonumber
\end{align}
The last line above is just a short-hand representation of $\tilde{H}$, and is only shown for convenience.

Recall that $\tilde{H}_\text{toric}$ (\eqnref{eq:H toric'}) describes decoupled stacks of 2+1D toric codes along the XY and XZ planes.
(The tilde over operators is still used to distinguish from the Pauli operators in the original Hamiltonian (\eqnref{eq:H}).)
The XY-plane toric codes are composed of Pauli operators $\tilde\sigma^\mu_\ell$ on the links of the XY-plane square lattices,
  and are described by plaquette/flux operators $\sQ{\tilde}{p}$ and vertex/charge operators $\sX{\tilde}{\iota}$ (\figref{fig:toricCode}).
The XZ-plane toric codes are similar, but are defined using $\tilde\tau$ instead of $\tilde\sigma$.
The XY and XZ--plane toric code square lattices intersect on the links parallel to the x-axis,
  which we have previously referred to as shared-links.
$\tilde{H}_t$ induces X-cube order at large $t$ by coupling these shared-links together via a $t\, \tilde\sigma^x_\ell \tilde\tau^x_\ell$ coupling.

First, note that that $\sX{\tilde}{\iota}$, $\tX{\tilde}{\iota}$, and $\rC{\tilde}{c}$ all commute with $\tilde{H}$.
Recall that $\rC{\tilde}{c}$ is the product of two XY-plane and two XZ-plane flux operators ($\sQ{\tilde}{p}$ and $\tQ{\tilde}{p}$) that boarder the cube $c$.
(Note, there are no YZ-plane flux operators on the other two faces of the cube.)
Therefore, we can work in the restricted Hilbert space where
\begin{equation}
  \sX{\tilde}{\iota} \ket{\Psi} = \tX{\tilde}{\iota} \ket{\Psi} = \rC{\tilde}{c} \ket{\Psi} = \ket{\Psi} \label{eq:duality restriction}
\end{equation}
This is similar to imposing a charge-free gauge constraint when one considers dualities of pure $Z_2$ gauge theory.
For example, our restricted Hilbert space does not allow any fracton excitations,
  which are excitations of $\rC{\tilde}{c}$.

The dual theory will consist of a stack of decoupled YZ-plane square lattices with vertices $\hat\iota$ at the center of the x-axis links $\ell$ of the original cubic lattice.
We introduce a new Pauli operator $\eta^\mu_{\hat\iota}$ which lives on the dual lattice vertices (\figref{fig:duality} (b)).
$\eta^\mu_{\hat\iota}$ will only be coupled along YZ planes; different YZ planes will be completely decoupled from each other.
The duality is given by
\begin{align}
  \sQ{\tilde}{p}                        &\leftrightarrow \eta^z_{\hat\iota} \eta^z_{\hat\iota'} \nonumber\\
  \tQ{\tilde}{p}                        &\leftrightarrow \eta^z_{\hat\iota} \eta^z_{\hat\iota'} \quad \text{where $\hat\iota$ and $\hat\iota'$ boarder $p$} \nonumber\\
  \tilde\sigma^x_\ell \tilde\tau^x_\ell &\leftrightarrow \eta^x_{\hat\iota=\ell} \label{eq:duality}\\
  \sX{\tilde}{\iota}                    &\leftrightarrow 1 \nonumber\\
  \tX{\tilde}{\iota}                    &\leftrightarrow 1 \nonumber\\
  \rC{\tilde}{c}                        &\leftrightarrow 1 \nonumber
\end{align}
Therefore, the flux (XY plane $\sQ{\tilde}{p}$ or XZ plane $\tQ{\tilde}{p}$) through the plaquette $p$ is mapped to $\eta^z_{\hat\iota} \eta^z_{\hat\iota'}$ where $\hat\iota$ and $\hat\iota'$ reside at the center of the two x-axis links of the plaquette $p$.
The coupling $\tilde\sigma^x_\ell \tilde\tau^x_\ell$ (on an x-axis link) is mapped to $\eta^x_{\hat\iota}$, which resides at the center of the link $\ell$.
$\sX{\tilde}{\iota}$, $\tX{\tilde}{\iota}$, and $\rC{\tilde}{c}$ are mapped to the identity, which solves the Hilbert space constraint \eqnref{eq:duality restriction}.

Note that the algebra (e.g. commutation relations) of the operators on the left hand side (LHS) and right hand side (RHS) of \eqnref{eq:duality} are the same.
Furthermore, the LHS and RHS both have exactly one effective qubit per unit cell.
The LHS originally has 4 qubits (two $\tilde\sigma$ and two $\tilde\tau$), but this gets subtracted by 3 qubits due to the 3 restricted Hilbert space conditions (\eqnref{eq:duality restriction}).
Therefore, the transformation (\eqnref{eq:duality}) is an algebra isomorphism (on an infinite lattice) between the algebra of $\tilde\sigma^\mu_\ell$ and $\tilde\tau^\mu_\ell$ modulo the Hilbert space restriction (\eqnref{eq:duality restriction}), and the algebra of $\eta^\mu_{\hat\iota}$.
Thus, \eqnref{eq:duality} is a valid duality transformation.

The dual Hamiltonian is described by decoupled 2+1D transverse field Ising models on the dual YZ-plane square lattices:
\begin{equation}
  \tilde{H}^\text{dual} = - \sum_{\langle\hat\iota \hat\iota' \rangle}^\text{YZ planes} \EQ\, \eta^z_{\hat\iota} \eta^z_{\hat\iota'} - \sum_{\hat\iota} t\, \eta^x_{\hat\iota} \label{eq:H dual}
\end{equation}
$\sum_{\langle\hat\iota \hat\iota' \rangle}^\text{YZ planes}$ sums over nearest-neighbor dual lattice vertices that lie on the same YZ plane.
We see that the decoupled phase (XY \& XZ plane $Z_2$ topological order) with small coupling $t$ is dual to the ordered phase of $\tilde{H}^\text{dual}$,
  and the X-cube phase with large $t$ is dual to the disordered phase of $\tilde{H}^\text{dual}$.

In the original model $\tilde{H}$ (\eqnref{eq:H coupled}), the $t$ coupling, which couples together orthogonal layers of toric code, creates a YZ-plane loop of four fluxes (two XY and two XZ--plane fluxes: \figref{fig:duality} (b)).
In the previous subsection (\secref{sec:X-cube}) we use degenerate perturbation theory to show that at strong coupling, this YZ-plane composite flux loop condenses, and X-cube order results.
Similar to the explanations in \refcite{VijayLayer,MaLayers}, this occurs because the XY and XZ--plane toric code charge excitations ($\sX{\tilde}{\iota}$ and $\tX{\tilde}{\iota}$) can no longer move in the x direction since they have nontrivial braiding statistics with the YZ-plane composite flux loop.
Only a bound state of an XY and an XZ--plane toric code charge excitation can move in the x direction;
  but this composite particle can no longer move in any other direction.
One can also consider a pair of open YZ-plane flux loop excitations, which have immobile fracton excitations $\rC{\tilde}{c}$ (\eqnref{fig:fractonOperator}) at the open endpoints.
Before the composite flux condensation, open flux loops have energy proportional to their length;
  however, after condensation, the bulk of the open flux loop is indistinguishable from the bulk of a closed flux loop, which carries no energy.
Thus, arbitrarily long open YZ-plane flux loops only carry a finite energy cost at their boundaries.
We see that in $\tilde{H}^\text{dual}$, the $t$ coupling flips a $\eta^z$ spin, which will either create or deform a domain wall on a YZ plane.
Thus, the condensation of YZ-plane flux loops is dual to a proliferation of YZ-plane domain walls in 2+1D YZ-plane Ising models.

Although the spontaneous symmetry breaking phase transition of a single layer of the 2+1D transverse-field Ising model is continuous,
  the same transition for a stack of Ising models (i.e. $\tilde{H}^\text{dual}$) appears to be first order once they are weakly coupled.
This is because a $\phi_L^2 \phi_{L'}^2$ term (in a stack of $\phi^4$ field theories) or $\sigma_{LI}^z \sigma_{LJ}^z \sigma_{L'I}^z \sigma_{L'J}^z$ term
  (in a stack of transverse-field Ising models) is allowed by the subsystem symmetry,
  where $L$ and $L'$ denote different layers, and $I$ and $J$ are nearest-neighbor sites on a layer.
($\phi_L \phi_{L'}$ or  $\sigma_{LI}^z \sigma_{L'I}^z$ are odd under subdimensional symmetry and thus not allowed.)
$\phi^2$ has scaling dimension $\sim1.4$, and thus $\phi_L^2 \phi_{L'}^2$ can be expected to have dimension $\sim 1.4 \times 2 \sim 2.8$,
  which is less than the dimension (2+1) of the critical Ising layers.
Thus $\phi_L^2 \phi_{L'}^2$ is likely to be relevant under RG at the critical point,
  which implies that the transition of weakly coupled 2+1D Ising layers is not described by the decoupled transition.
According to the duality mapping (\eqnref{eq:duality}), the relevant operator corresponds to a product of two $\sQ{\tilde}{p}$ or $\tQ{\tilde}{p}$ operators.
In \appref{app:MC} we provide classical Monte Carlo evidence that these symmetry allowed perturbations are relevant under RG,
  and that this transition and its dual are likely to be discontinuous and first order.

\subsection{1+1D Duality}
\label{sec:duality'}

In this subsection we will discuss a duality describing a (likely first order) phase transition from the X-cube phase to a phase of decoupled YZ-plane 2+1D $Z_2$ topological order.
To do this, we will consider the X-cube model, but with a dimension-1 particle hopping term (and a perturbation):
\begin{align}
  \bhat{H}_\text{hop} &= \bhat{H}_\text{X-cube} - \sum_\ell^\text{x-axis links} h\, \bhat\rho^z_\ell
                       - \sum_{p \is \square}^\text{YZ planes} \EQyz\, \HrQ{\bhat}{p} \label{eq:H hop'}\\
  \bhat{H}_\text{X-cube} &= - \sum_{\iota \is +}^\text{XY planes} E _+\, \HsX{\bhat}{\iota}
                            - \sum_{\iota \is +}^\text{XZ planes} E _+\, \HtX{\bhat}{\iota} \nonumber\\
                   &\,\quad - \sum_{c \is \cube} \EC\, \HrC{\bhat}{c} \label{eq:X-cube}
\end{align}
where $\sum_\ell^\text{x-axis links}$ sums over all links $\ell$ which are parallel to the x-axis.
The hopping term $h$ (which we also consider in \secref{sec:extended}) hops the $\HsX{\bhat}{\iota}$ and $\HtX{\bhat}{\iota}$ excitations (\figref{fig:plusOperators}) along the x-axis,
  which will condense the X-cube ordered phase into YZ-plane 2+1D $Z_2$ topological order.
A small $\EQyz\, \HrQ{\bhat}{p}$ perturbation is included to prevent fine tuning,
  where $\HrQ{\bhat}{p}$ is a YZ-plane toric code flux operator (\figref{fig:3+1D}).
Similar to $H_{h'}$ (\eqnref{eq:H_h'}) in a later section, without the $\EQyz$ term, the large $h$ phase would have an unphysical degeneracy $\sim 2^{L^2}$ due to accidental symmetries which prevent a $ \HrQ{\bhat}{p}$ flux term from being generated.
Terms other than $\EQyz\, \HrQ{\bhat}{p}$ can also work;
  however, this is the simplest choice for understanding the duality.

$\bhat{H}_\text{X-cube}$ describes the X-cube model \cite{VijayXCube} with $\bhat\rho^\mu$ operators on the links of a cubic lattice.
(The hat over operators is used to distinguish them from the Pauli operators in the original Hamiltonian (\eqnref{eq:H} and previous section).)
$\HrC{\bhat}{c}$ (\figref{fig:fractonOperator}) measures the fracton number and is a product of the twelve $\bhat\rho^z_\ell$ operators on the edges of the cube $c$.
$\HsX{\bhat}{\iota}$, $\HtX{\bhat}{\iota}$, and $\HrX{\bhat}{\iota} = \HsX{\bhat}{\iota} \HtX{\bhat}{\iota}$ (\figref{fig:plusOperators}) are products of four $\bhat\rho^x_\ell$ operators and correspond to dimension-1 particle excitations,
  which can only move along 1-dimensional lines.
(The absence of $\HrX{\bhat}{\iota}$ in $\bhat{H}_\text{X-cube}$ is not important.)

$\HrX{\bhat}{\iota}$ and $\HrC{\bhat}{c}$ commute with $\bhat{H}_\text{hop}$ (\eqnref{eq:H hop'}),
  and therefore we can work in a restricted Hilbert space where
\begin{equation}
  \HrX{\bhat}{\iota} \ket{\Psi} = \HrC{\bhat}{c} \ket{\Psi} = \ket{\Psi} \label{eq:duality restriction'}
\end{equation}

The dual theory will consist of a grid of weakly coupled Ising chains along the x-axis with the same lattice sites $\iota$ as the original cubic lattice.
We introduce a new Pauli operator $\mu^\alpha_\iota$ which lives on the vertices of the cubic lattice.
The duality is given by
\begin{align}
  \HsX{\bhat}{\iota} &\leftrightarrow \mu^x_\iota \nonumber\\
  \HtX{\bhat}{\iota} &\leftrightarrow \mu^x_\iota \nonumber\\
  \HrX{\bhat}{\iota} &\leftrightarrow 1 \label{eq:duality'}\\
  \bhat\rho^z_\ell   &\leftrightarrow \mu^z_\iota \mu^z_{\iota'} \quad \text{for x-axis link $\ell = \langle\iota \iota' \rangle$} \nonumber\\
  \HrQ{\bhat}{p}     &\leftrightarrow \prod_{\iota \in \square}^p \mu^z_\iota \nonumber\\
  \HrC{\bhat}{c}     &\leftrightarrow 1 \nonumber
\end{align}
Thus, the dimension-1 particle ($\HsX{\bhat}{\iota}$ and $\HtX{\bhat}{\iota}$) number operator is mapped to a spin flip operator $\mu^x_\iota$;
 and the dimension-1 particle hopping $\bhat\rho^z_\ell$ is mapped to a domain wall number operator $\mu^z_\iota \mu^z_{\iota'}$.
$\HrX{\bhat}{\iota}$ and $\rC{\tilde}{c}$ are mapped to the identity, which solves the Hilbert space constraint \eqnref{eq:duality restriction'}.
The YZ-plane flux operator $\HrQ{\bhat}{p}$ is mapped to a product of four $\mu^z_\iota$ operators at the corners of the plaquette $p$.
The duality mapping forms an algebra isomorphism (on an infinite lattice) of the left hand side (LHS) of \eqnref{eq:duality'} (modulo the Hilbert space restriction (\eqnref{eq:duality restriction'})), and the right hand side (RHS).
For example, both the LHS and RHS have one effective qubit per unit cell.
The LHS originally has 3 qubits ($\bhat\rho^\mu_\ell$ along the three different directions), but this gets subtracted by 2 qubits due to the 2 Hilbert space conditions (\eqnref{eq:duality restriction'}).

The dual Hamiltonian is described by a grid of 1+1D transverse field Ising models parallel to the x-axis (\figref{fig:duality} (c)):
\begin{align}
  \bhat{H}_\text{hop}^\text{dual}
    &= - \sum_\iota \left( 2 \EX  \, \mu^x_\iota + h\, \mu^z_\iota \mu^z_{\iota + \hat x} \right. \nonumber\\
    &\,\quad        \left. + \EQyz\, \mu^z_\iota \mu^z_{\iota + \hat y} \mu^z_{\iota + \hat z}\mu^z_{\iota + \hat y + \hat z}  \right) \label{eq:H dual'}
\end{align}
We see that the X-cube phase with small hopping $h$ is dual to the disordered phase of $\bhat{H}_\text{hop}^\text{dual}$,
  and the YZ-plane topological order phase with large $h$ is dual to the ordered phase of $\bhat{H}_\text{hop}^\text{dual}$.

In the original model $\bhat{H}_\text{hop}$ (\eqnref{eq:H hop'}), the $h$ hopping hops dimension-1 particles (a composite excitation of $\HsX{\bhat}{\iota}$ and $\HtX{\bhat}{\iota}$) parallel to the x-axis.
For large hopping strength $h$, these x-axis particles condense, and YZ-plane 2+1D $Z_2$ topological order results.
This is shown in a latter section (\secref{sec:YZ}) using degenerate perturbation theory with a more general model including a $t$ term, which is taken to be zero in this section.
We now give a physical explanation.
The x-axis particle condensation confines fracton excitations which are created by YZ-plane membrane operators,
  since these membrane operators have nontrivial statistics with the x-axis particle.
This condensation also allows the y-axis particle $\HsX{\bhat}{\iota}$ to move along the z direction,
  since it can do this by splitting up into a z-axis particle (which can move in the z direction) and an x-axis particle, which costs no energy after its condensation.
Thus, the y-axis and z-axis particles become YZ-plane toric code charge excitations.
Therefore, after the x-axis particle condensation, there are no dimension-1 particles,
  which implies that the fracton excitations created by XY and XZ--plane membrane operators are no longer finite energy excitations.
To understand this, note that in the X-cube phase, there is no excitation (with a nontrivial braiding statistic) that can be braided through the membrane along a small closed loop.
However, after the x-axis particles are condensed and YZ-plane toric code charge excitations are permitted,
  the YZ-plane charge excitations can braid through an XY-plane membrane.
As a result, an XY-plane (or XZ-plane) membrane will have an energy cost proportional to the number of YZ-plane charges that can braid around it, which is proportional to the length of the membrane in the x direction.
Mathematically, this energy cost results from the YZ-plane toric code flux term $\EQyz\, \HrQ{\bhat}{p}$ in $\bhat{H}_\text{hop}$ (\eqnref{eq:H hop'}).
(The same physics explains why flux loops in 3+1D toric code have energy cost proportional to their length.)
However, an XY-plane (or XZ-plane) membrane operator with small length (e.g. =1) in the x direction will have a finite energy cost;
  these operators create composite fracton excitations which are free to move along a YZ plane.
After the x-axis particles condense, this composite fracton excitation becomes the YZ-plane toric code flux excitation.

In the dual theory $\bhat{H}_\text{hop}^\text{dual}$, $h$ hops excitations of the spin flip operator $\mu^x_\iota$,
  which spontaneously breaks $Z_2$ Ising symmetry when condensed since these excitations are conserved modulo two due to the $Z_2$ Ising symmetry.
The $\EQyz$ term couples four neighboring 1+1D chains together,
  which is a marginal under RG operator at the Ising phase transition.
This can be understood by mapping the Ising chains to free Majorana chains using a Jordan-Wigner transformation;
  the $\EQyz$ term is then mapped to a mass-mass term $(\chi_l^T \sigma^2 \chi_l) (\chi_{l'}^T \sigma^2 \chi_{l'})$.
Since $\chi_l^T \sigma^2 \chi_l$ has scaling dimension $1$, the $\EQyz$ term has scaling dimension $2 \times 1$, which is equal to the number of space-time dimensions of each chain (1+1).
Thus, the $\EQyz$ term is marginal under RG at the fixed point of critical decoupled Ising chains.
In \appref{app:MC} we show classical Monte Carlo evidence that these symmetry allowed perturbations are likely to be relevant under RG,
  and that this transition and its dual are likely to be discontinuous and first order.

In the next section, we introduce an extended honeycomb-based model $H'$ (\eqnref{eq:H'}) which also exhibits this phase transition with a similar duality description at low energies.

\section{Extended Model with 3+1D \ZTwo Topological Order}
\label{sec:extended}

\begin{figure}
\includegraphics[width=\columnwidth]{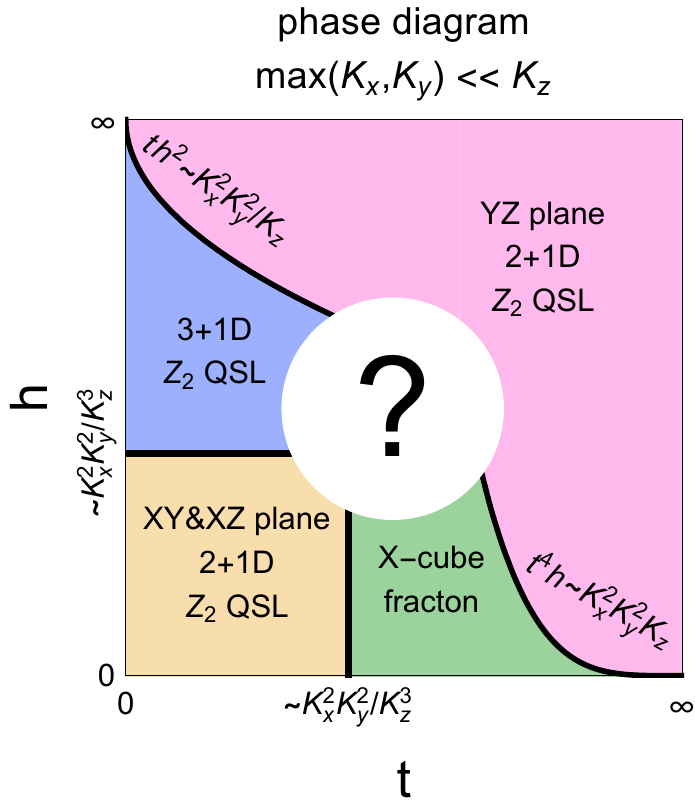}
\caption{
Rough phase diagram of $H'$ (\eqnref{eq:H'}) as a function of $t$ (horizontal axis) and $h$ (vertical axis) in the limit $\max(K_x,K_y) \ll K_z$
  with $\max(K_x,K_y)/K_z$ small but finite and $h'$ infinitesimally small (to break accidental degeneracies).
In \figref{fig:largeKzDiagram} we utilize infinitesimal $\max(K_x,K_y)/K_z$ in order display more precise claims regarding phase boundaries.
Rough phase boundaries are computed in \appref{app:Kz DPT}.
Four phase transitions occur around the edge of the phase diagram.
In \secref{sec:dualities} and \secref{sec:global duality} we present dual descriptions of these four (likely first order) phase transitions.
We do not know what phase(s) occur in the white region;
  this ``region of ignorance'' grows as we move away from the $\max(K_x,K_y) \ll K_z$ limit.
In \appref{app:MC} we find that (depending on the sign of certain perturbations) topologically ordered intermediate phases are also possible in between all of the above phase transitions.
}\label{fig:extendedPhaseDiagram}
\end{figure}

To induce 3+1D $Z_2$ topological order instead of fracton order, we can consider adding to $H = H_K + H_t$ (\eqnref{eq:H}) a coupling $H_h$ and a small perturbation, for which we will choose $H_{\h}$ as an example:
\begin{align}
  H'       & = H_K + H_t + H_h + H_{\h} \label{eq:H'}\\
  H_h      &= - \sum_{\left< i j \right>}^\text{shared-links} \frac{h}{4}\, (\sigma^x_i \sigma^y_j + \sigma^y_i \sigma^x_j) (\tau^x_i \tau^y_j + \tau^y_i \tau^x_j) \label{eq:H_h}\\
  H_{\h}  &= - \sum_i^{\sigma\text{-sites}} \frac{\h}{2} \sigma^z_i
             - \sum_i^{\tau  \text{-sites}} \frac{\h}{2} \tau  ^z_i \label{eq:H_h'}
\end{align}
A rough phase diagram for this model is given in \figref{fig:extendedPhaseDiagram}.
See \secref{sec:previousCoupledLayer} for connections to previous work.

$H_h$ is an alternative coupling of the 2D honeycomb lattices.
$\sum_{\left< i j \right>}^\text{shared-links}$ sums over all shared links.
As we will explain in \secref{sec:3+1D}, this term will cause a composite charge excitation to condense, resulting in a phase which is fine tuned (with unphysical degeneracy $\sim 2^{L^2}$ on a torus).
When generic perturbations are added (e.g. $\h$), the resulting phase is 3+1D $Z_2$ topological order.

$H_{\h}$ will be considered as an example of a generic perturbation to break the unphysical degeneracy mentioned in the previous paragraph.
The particular choice of $H_{\h}$ is not important.


In the tilde basis (which includes a low energy Hilbert space projection from degenerate perturbation theory), $H'$ is given by (ignoring subleading terms)
\begin{align}
  H'     \rightarrow \tilde{H}'     &= \tilde{H}_\text{toric} + \tilde{H}_t + \tilde{H}_h + \tilde{H}_{\h} \label{eq:H' tilde}\\
  H_h    \rightarrow \tilde{H}_h    &= - \sum_\ell^ \text{shared-links}   h\, \tilde\sigma^z_\ell \tilde\tau^z_\ell \label{eq:H_h tilde}\\
  H_{\h} \rightarrow \tilde{H}_{\h} &= - \sum_\ell^{\sigma\text{-links}} \h\, \tilde\sigma^z_\ell
                                       - \sum_\ell^{\tau  \text{-links}} \h\, \tilde\tau  ^z_\ell \label{eq:H_h' tilde}
\end{align}
$\tilde{H}_\text{toric}$ (\eqnref{eq:H toric'}) and $\tilde{H}_t$ (\eqnref{eq:H_t tilde}) were discussed previously.
We see that $\tilde{H}_h$ couples the XY and XZ--plane toric codes together on the shared-links with a $\tilde\sigma^z_\ell \tilde\tau^z_\ell$ coupling.
If $\ell = \langle\iota \iota' \rangle$ is a cubic lattice link between vertices $\iota$ and $\iota'$,
  then when acting on the decoupled $t=h=0$ ground state, $\tilde\sigma^z_\ell \tilde\tau^z_\ell$ creates the following four excitations: $\sX{\tilde}{\iota}$, $\sX{\tilde}{\iota'}$, $\tX{\tilde}{\iota}$, $\tX{\tilde}{\iota'}$.
When $t$ is small and $h$ is large (and $\h$ is small but finite), these excitations condense and the result is 3+1D $Z_2$ topological order.

\subsection{3+1D \ZTwo Topological Order}
\label{sec:3+1D}

In this section, we will explain how 3+1D $Z_2$ topological order results from $H'$ (\eqnref{eq:H'}) in the limit
\begin{equation}
  \max(K_x,K_y) \ll h \ll K_z \text{ and } t=0 \label{eq:3+1D limit}
\end{equation}
  and with infinitesimal $\h$ (to break accidental degeneracies).
In \appref{app:3+1D} we will consider more general limits.
However, $\max(K_x,K_y) \ll h \ll K_z$ allows us to apply the results from \secref{sec:decoupled} and is closely connected to the previous layer constructions in \refcite{VijayLayer,MaLayers}, and so we will discuss this limit here.

\begin{figure}
\includegraphics[width=\columnwidth]{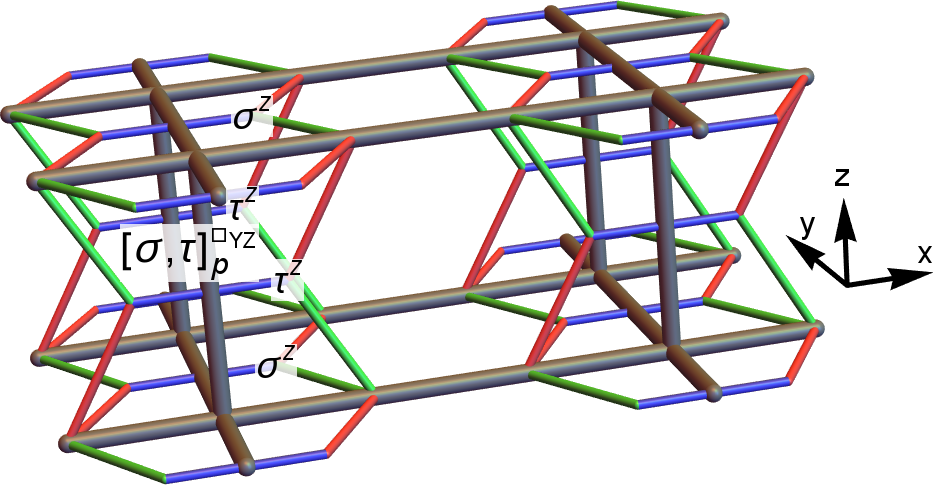}
\includegraphics[width=.49\columnwidth]{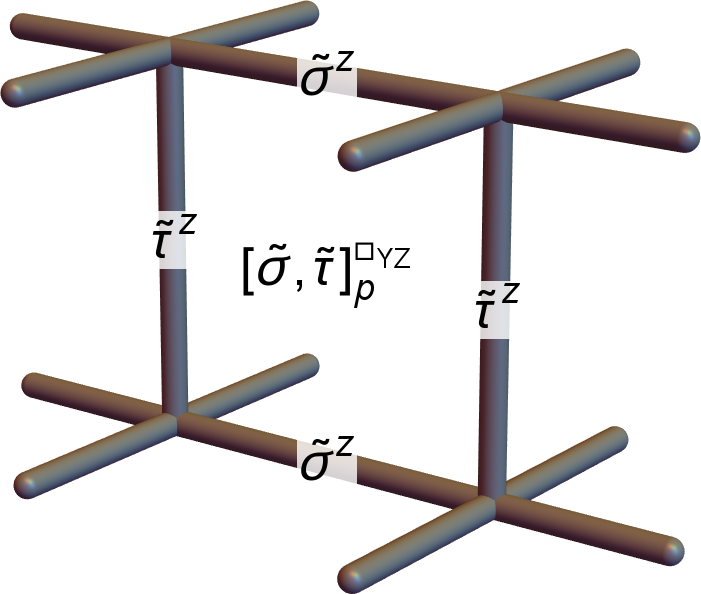}
\includegraphics[width=.49\columnwidth]{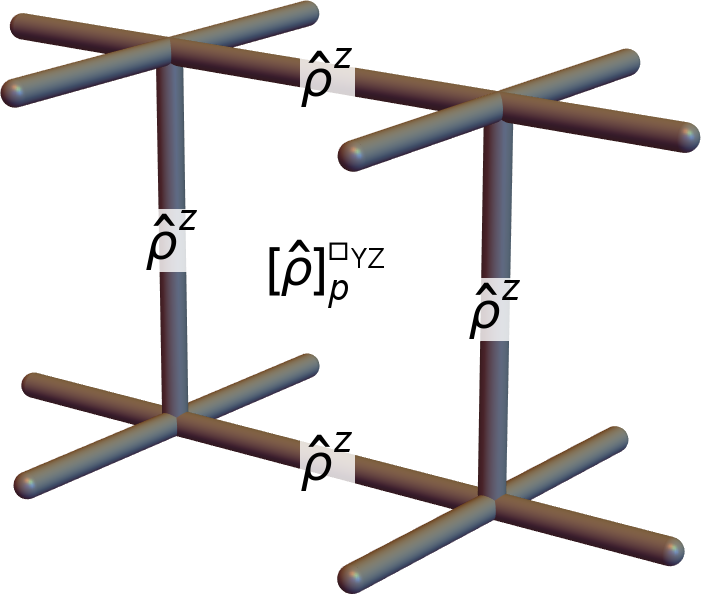}
\includegraphics[width=.5\columnwidth]{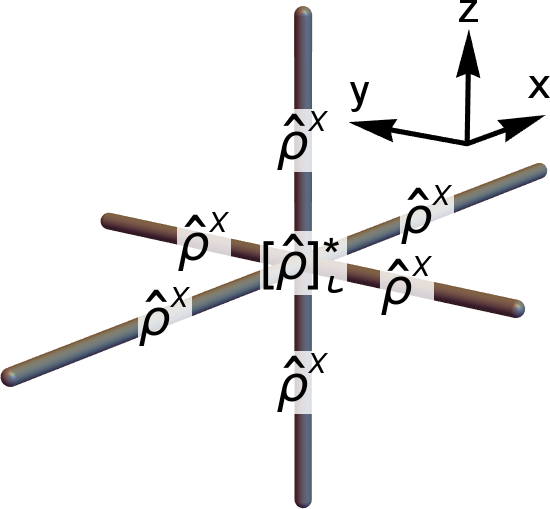}
\caption{
The YZ-plane toric code plaquette/flux operator expressed in three different basis:
  $\rQ{}{p}$ (top) is in the original basis (\eqnref{eq:H}),
  $\rQ{\tilde}{p}$ (mid-left) is in the 2D toric code basis (\eqnref{eq:toric basis}), and
  $\HrQ{\bbar}{p}$ (mid-right) is the X-cube basis (\eqnref{eq:X-cube basis}).
$\HrS{\bbar}{\iota}$ (bottom) is the vertex/charge operator in 3+1D toric code  (\eqnref{eq:H 3D toric}).
}\label{fig:3+1D}
\end{figure}

Again, we will derive this result using degenerate perturbation theory (DPT).
First, we split $H'$ (\eqnref{eq:H'}) into
\begin{align}
  H'  &\approx H_0 + H_1 \nonumber\\ 
  H_0 &= \tilde{H}_h \\
  H_1 &= \tilde{H}_\text{toric} + \tilde{H}_{\h} \nonumber
\end{align}
where $\tilde{H}_\text{toric}$ (\eqnref{eq:H toric'}) describes the decoupled toric codes on XY and XZ planes.
\eqnref{eq:3+1D limit} is sufficient to ensure that DPT converges.
$\sQ{\tilde}{p}$ and $\tQ{\tilde}{p}$ (in $\tilde{H}_\text{toric}$) commute with $H_0 = \tilde{H}_h$, and will therefore appear in the effective Hamiltonian $H^\text{eff}$.
These operators will play the role of flux operators on XY and XZ planes in 3+1D toric code.
However, $\sX{\tilde}{\iota}$ and $\tX{\tilde}{\iota}$ will not appear in $H^\text{eff}$ since they do not commute with $H_0$,
  which means that they create excitations out of the low energy Hilbert space.
However, their product $\rS{\tilde}{\iota} \equiv \sX{\tilde}{\iota} \tX{\tilde}{\iota}$ does commute with $H_0$ and will indeed appear in $H^\text{eff}$.
In a new basis, $\rS{\tilde}{\iota}$ will take the form of a charge operator in 3+1D toric code.
Thus, DPT will produce the following effective Hamiltonian
\begin{align}
  H^\text{eff} &= \tilde{H}_\text{3D toric} + \cdots \nonumber\\
  \tilde{H}_\text{3D toric} &= - \sum_{p \is \square}^\text{XY planes} \EQ\,   \sQ{\tilde}{p}
                               - \sum_{p \is \square}^\text{XZ planes} \EQ\,   \tQ{\tilde}{p} \nonumber\\
                      &\,\quad - \sum_{p \is \square}^\text{YZ planes} \EQyz\, \rQ{\tilde}{p}
                               - \sum_{\iota \is *} \ES\, \rS{\tilde}{\iota} \label{eq:H 3D toric}\\
  \ES &\sim \frac{\EX^2}{h} \sim \frac{K_x^4 K_y^4}{K_z^6 h} \quad\text{(assuming \eqref{eq:3+1D limit})} \nonumber\\
  \rS{\tilde}{\iota} &\equiv \sX{\tilde}{\iota} \tX{\tilde}{\iota} \label{eq:rho*}\\
  \rQ{\tilde}{p} & \quad \text{is defined in \figref{fig:3+1D}} \nonumber
\end{align}
(Actually, two applications of DPT are necessary to generate $\EQyz\, \rQ{\tilde}{p}$, as is done in \appref{app:3+1D};
  this term will be explained soon.)
Again, we perform a change of basis and projection into the low energy Hilbert space where now the energy of $H_0$ is minimized with $\tilde\sigma^z_\ell \tilde\tau^z_\ell = 1$ on every shared-link:
\begin{align}
  \sigma\text{-links:} && \tilde\sigma^\mu_\ell                 &\rightarrow \bbar\rho^\mu_\ell \nonumber\\
  \tau  \text{-links:} && \tilde\tau  ^\mu_\ell                 &\rightarrow \bbar\rho^\mu_\ell \nonumber\\
  \text{shared-links:} && \tilde\sigma^z_\ell \tilde\tau^z_\ell &\rightarrow 1 \label{eq:3D basis}\\
                       && \tilde\sigma^z_\ell                   &\rightarrow \bbar\rho^z_\ell \nonumber\\
                       && \tilde\sigma^x_\ell \tilde\tau^x_\ell &\rightarrow \bbar\rho^x_\ell \nonumber
\end{align}
\begin{align}
  \tilde{H}_\text{3D toric} &\rightarrow \bbar{H}_\text{3D toric} \label{eq:H 3D toric'} \\
                            &= - \sum_{p \is \square}^\text{XY planes} \EQ\,   \HsQ{\bhat}{p}
                               - \sum_{p \is \square}^\text{XZ planes} \EQ\,   \HtQ{\bhat}{p} \nonumber\\
                      &\,\quad - \sum_{p \is \square}^\text{YZ planes} \EQyz\, \HrQ{\bhat}{p}
                               - \sum_{\iota \is *} \ES\, \HrS{\bhat}{\iota} \nonumber
\end{align}
A hat over operators is also used to denote operators in this basis.
In this basis, $\bbar{H}_\text{3D toric}$ has the exact same form as 3+1D toric code, which has 3+1D $Z_2$ topological order.

However, we have not yet discussed the $\EQyz \rQ{\tilde}{p}$ term in $\tilde{H}_\text{3D toric}$.
In the new basis (\eqnref{eq:3D basis}), $\HrQ{\bbar}{p}$ takes the form of a flux operator on YZ planes in 3+1D toric code.
This term is necessary in 3+1D toric code; without it, $\bbar{H}_\text{3D toric}$ would have an unphysical degeneracy $\sim 2^{L^2}$ on a 3D torus.
But without $H_{\h}$, $\rQ{\tilde}{p}$ can not be generated by DPT (due to accidental symmetries). 
Therefore, we add an infinitesimal perturbation $H_{\h}$ to break this degeneracy.
The choice of perturbation is not important, and we only choose $H_{\h}$ as a simple example.
Most sufficiently general and sufficiently small perturbations will generate $\rQ{\tilde}{p}$ and not affect the resulting phase.

\subsection{YZ Plane 2+1D \ZTwo Topological Order}
\label{sec:YZ}

In this section, we will explain why 2+1D $Z_2$ topological order on YZ planes results from $H'$ (\eqnref{eq:H'}) in the limit
\begin{equation}
 \max(K_x,K_y) \ll \min(t,h) \text{ and } \max(t, h) \ll K_z \label{eq:YZ limit}
\end{equation}
with infinitesimal $\h$ (to break accidental degeneracies).
This phase is the same as a stack of decoupled toric codes on YZ planes.
In the \appref{app:YZ} we will consider limits more general than \eqnref{eq:YZ limit}.
Starting again from the decoupled $t=h=0$ limit discussed in \secref{sec:decoupled}, we make both $h$ and $t$ very large and perform DPT on $\tilde{H}'$ (\eqnref{eq:H' tilde}).
$\tilde{H}_t$ (\eqnref{eq:H_t tilde}) and $\tilde{H}_h$ (\eqnref{eq:H_h tilde}) couple the XY and XZ--plane toric codes together on the shared-links with $\tilde\sigma^x_\ell \tilde\tau^x_\ell$ and $\tilde\sigma^z_\ell \tilde\tau^z_\ell$ couplings.
When acting on the decoupled $t=h=0$ ground state, these couplings create toric code charge and flux excitations, and when $t$ and $h$ are large, these excitations condense and the result is 2+1D $Z_2$ topological order on YZ planes.

To understand this, we again apply degenerate perturbation theory (DPT) by splitting $H'$ (\eqnref{eq:H'}) into
\begin{align}
  H'  &\approx H_0 + H_1 \nonumber\\
  H_0 &= \tilde{H}_t + \tilde{H}_h \\
  H_1 &= \tilde{H}_\text{toric} + H_{\h} \nonumber
\end{align}
\eqnref{eq:YZ limit} is sufficient to ensure that DPT converges.
No operator in $\tilde{H}_\text{toric}$ (\eqnref{eq:H toric'}) will appear in $H^\text{eff}$ since they do not commute with $H_0$;
  i.e. they create excitations out of the low energy Hilbert space.
However, $\rC{\tilde}{c}$ and $\rX{\tilde}{\iota}$ ($= \rS{\tilde}{\iota}$ from \eqnref{eq:rho*}) are generated by DPT, and the effective Hamiltonian is
\begin{align}
  H^\text{eff} &= \tilde{H}^\text{YZ}_\text{toric}
                - \sum_{c \is \cube} \EC\, \rC{\tilde}{c}
                + \cdots \nonumber\\
  \tilde{H}^\text{YZ}_\text{toric} &= - \sum_{p \is \square}^\text{YZ planes} \EQyz\, \rQ{\tilde}{p} \nonumber\\
                             &\,\quad - \sum_{\iota \is +  }^\text{YZ planes} \EXyz\, \rX{\tilde}{\iota} \label{eq:H YZ toric}\\
  \EXyz       &\sim \frac{\EX^2}{h} \sim \frac{K_x^4 K_y^4}{K_z^6 h} \quad\text{assuming \eqref{eq:YZ limit}} \nonumber\\
  \EC    &\sim \frac{\EQ^4}{t^3} \sim \frac{K_x^8 K_y^8}{K_z^{12} t^3} \quad\text{assuming \eqref{eq:YZ limit}} \nonumber\\
  \rX{\tilde}{\iota} &\equiv \sX{\tilde}{\iota} \tX{\tilde}{\iota}
\end{align}
(Actually, two applications of DPT are necessary to generate $\EXyz\, \rQ{\tilde}{p}$, as is done in \appref{app:YZ}.)
We perform a change of variables and projection into the low energy Hilbert space with $\tilde\sigma^x_\ell \tilde\tau^x_\ell = \tilde\sigma^z_\ell \tilde\tau^z_\ell = 1$ on every shared-link:
\begin{align}
  \sigma\text{-links:} && \tilde\sigma^\mu_\ell                 &\rightarrow \bdot\rho^\mu_\ell \nonumber\\
  \tau  \text{-links:} && \tilde\tau  ^\mu_\ell                 &\rightarrow \bdot\rho^\mu_\ell \nonumber\\
  \text{shared-links:} && \tilde\sigma^x_\ell \tilde\tau^x_\ell &\rightarrow 1 \label{eq:YZ basis}\\
                       && \tilde\sigma^z_\ell \tilde\tau^z_\ell &\rightarrow 1 \nonumber
\end{align}
\begin{align}
  \tilde{H}^\text{YZ}_\text{toric} &\rightarrow \bdot{H}^\text{YZ}_\text{toric} \label{eq:H YZ toric'}
\end{align}
A hat over operators is also used to denote operators in this basis.
In this basis, $\bdot{H}^\text{YZ}_\text{toric}$ has the exact same form as decoupled 2+1D toric codes on YZ planes.
$\HrX{\bdot}{\iota}$ and $\HrQ{\bdot}{p}$ are charge and flux operators on YZ planes in 2+1D toric code, respectively.

But just like $\rQ{\tilde}{p}$ (\eqnref{eq:H 3D toric}) in the previous section,
  $\rQ{\tilde}{p}$ in \eqnref{eq:H YZ toric} can not be generated from $H'$ (\eqnref{eq:H'}) with $\h = 0$.
However, $H'$ with $\h = 0$ does generate $\rC{\tilde}{c}$,
  which corresponds to a product of flux operators on neighboring YZ planes in the new basis (\eqnref{eq:H YZ toric'}); that is,
  $\rC{\tilde}{c} \rightarrow \HrC{\bdot}{c} = \HrQ{\bdot}{p} \HrQ{\bdot}{p'}$ where YZ-plane plaquettes $p$ and $p'$ lie on the boundary of the cube $c$.
Without perturbations such as $H_\h$, $H^\text{eff}$ in \eqnref{eq:H YZ toric} (and therefore also $H'$) would have an unphysical degeneracy $\sim 2^{L^2}$ on a torus.
Again, the choice of perturbation $H_{\h}$ is not important.

\subsection{Global Duality of the Extended Model}
\label{sec:global duality}

\begin{figure}
\includegraphics[width=.8\columnwidth]{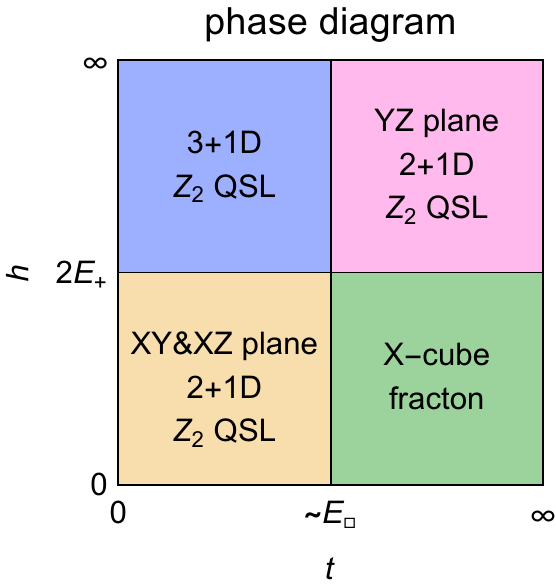}
\caption{
Phase diagram of $\tilde{H}_\text{global}$ (\eqnref{eq:H global}) for $\EQyz \ll \EX$.
$\tilde{H}_\text{global}$ is used to derive dual descriptions of the phase transitions between the above phases,
  which occur on the lines $t \sim \EQ$ and $h = 2\EX$.
The phase boundaries in \figref{fig:extendedPhaseDiagram} are different because that phase diagram describes a model $H'$ (\eqnref{eq:H'}) where the effective $\EQ$ and $\EX$ depend on $t$ and $h$ as shown in \eqnref{eq:energies}.
}\label{fig:globalPhaseDiagram}
\end{figure}

In this subsection we discuss a global duality which describes the four (likely first order) phase transitions along the edge of \figref{fig:extendedPhaseDiagram}.
To do this, we merge $\tilde{H}$ (\eqnref{eq:H coupled}) and $\bhat{H}_\text{hop}$ (\eqnref{eq:H hop'}) from \secref{sec:dualities}:
\begin{align}
   \tilde{H}_\text{global}
     &= \tilde{H}_\text{toric} + \tilde{H}_t + \tilde{H}_h - \sum_{p \is \square}^\text{YZ planes} \EQyz\, \rQ{\tilde}{p} \label{eq:H global}\\
     &\sim  - \EQ\, (\sQ{\tilde}{p}     + \tQ{\tilde}{p})
            - \EX\, (\sX{\tilde}{\iota} + \tX{\tilde}{\iota}) \nonumber\\
   &\,\quad - t\, \tilde\sigma^x_\ell \tilde\tau^x_\ell - h\, \tilde\sigma^z_\ell \tilde\sigma^z_\ell - \EQyz\, \rQ{\tilde}{p} \nonumber
\end{align}
$\tilde{H}_\text{global}$ is a model of XY and XZ--plane toric codes $\tilde{H}_\text{toric}$ (\eqnref{eq:H toric'})
  which are coupled together by $\tilde{H}_t$ (\eqnref{eq:H_t tilde}) and $\tilde{H}_h$ (\eqnref{eq:H_h'}),
  with an infinitesimal perturbation $\EQyz\, \rQ{\tilde}{p}$ (\figref{fig:3+1D}) to break accidental degeneracies (discussed in \secref{sec:duality'}, \ref{sec:3+1D}, and \ref{sec:YZ}).
$\tilde{H}_\text{toric}$ and $\tilde{H}_t$ were discussed in \secref{sec:duality}.
$\tilde{H}_h$ and $\EQyz\, \rQ{\tilde}{p}$ were discussed in \secref{sec:duality'} in the X-cube basis (\eqnref{eq:X-cube basis});
  here we use the decoupled toric code basis (\eqnref{eq:toric basis}).
$\tilde{H}_\text{global}$ is very similar to $\tilde{H}'$ (\eqnref{eq:H' tilde});
  the only difference is that $\tilde{H}_\text{global}$ replaces $\tilde{H}_{\h}$ (\eqnref{eq:H_h' tilde}) with the $\EQyz$ term.
(Recall that the choice of $\tilde{H}_{\h}$ in $\tilde{H}'$ was not important.)
$\sQ{\tilde}{p}$, $\sX{\tilde}{\iota}$ (\figref{fig:toricCode}), and $\tilde\sigma^x_\ell$ are defined on the XY plane toric codes,
  while $\tilde\tau$ operators are defined on XZ plane toric codes.
The second line in \eqnref{eq:H global} is just a short-hand representation of the included terms, and is only shown for convenience.
A phase diagram is given in \figref{fig:globalPhaseDiagram}.
We will work in a restricted Hilbert space where
\begin{equation}
  \sX{\tilde}{\iota} \tX{\tilde}{\iota} \ket{\Psi} = \rC{\tilde}{c} \ket{\Psi} = \ket{\Psi} \label{eq:duality restriction global}
\end{equation}
The dual theory looks like a decoupled combination of the two dual theories discussed in \secref{sec:dualities}.
Thus, the dual theory consists of a stack of decoupled YZ-plane square lattices with Pauli operators $\eta^\alpha_{\hat\iota}$ at vertices $\hat\iota$ centered on x-axis links $\ell$ of the original cubic lattice,
  and a grid of weakly coupled Ising chains along the x-axis with Pauli operators $\mu^\alpha_\iota$ on the same lattice sites $\iota$ as the original cubic lattice.
$\eta^\alpha_{\hat\iota}$ and $\mu^\alpha_\iota$ will be completely decoupled.
The duality is given by
\begin{align}
  \sQ{\tilde}{p}                          &\leftrightarrow \eta^z_{\hat\iota} \eta^z_{\hat\iota'} \nonumber\\
  \tQ{\tilde}{p}                          &\leftrightarrow \eta^z_{\hat\iota} \eta^z_{\hat\iota'} \quad \text{where $\hat\iota$ and $\hat\iota'$ boarder $p$} \nonumber\\
  \tilde\sigma^x_\ell \tilde\tau^x_\ell   &\leftrightarrow \eta^x_{\hat\iota=\ell} \nonumber\\
  \sX{\tilde}{\iota}                      &\leftrightarrow \mu^x_\iota \label{eq:duality global}\\
  \tX{\tilde}{\iota}                      &\leftrightarrow \mu^x_\iota \nonumber\\
  \tilde\sigma^z_\ell \tilde\sigma^z_\ell &\leftrightarrow \mu^z_\iota \mu^z_{\iota'} \quad \text{for x-axis link $\ell = \langle\iota \iota' \rangle$} \nonumber\\
  \rQ{\tilde}{p}                          &\leftrightarrow \prod_{\iota \in \square}^p \mu^z_\iota \nonumber\\
  \rC{\tilde}{c}                          &\leftrightarrow 1 \nonumber
\end{align}
The mapping is essentially a combination of the duality mappings in \secref{sec:dualities}: \eqnref{eq:duality} and \eqref{eq:duality'}.
Both the original theory (modulo the Hilbert space restriction (\eqnref{eq:duality restriction global})) and the dual theory have two qubits per site.
The dual Hamiltonian is simply the decoupled sum of the two dual Hamiltonians in \secref{sec:dualities} (\eqnref{eq:H dual} and \eqref{eq:H dual'}):
\begin{equation}
  \tilde{H}_\text{global}^\text{dual} = \tilde{H}^\text{dual} + \bhat{H}_\text{hop}^\text{dual} \label{eq:H dual global}
\end{equation}
In \secref{sec:dualities} we described two phase transitions out of the X-cube order phase.
$\tilde{H}_\text{global}^\text{dual}$ describes these transitions in the same way.
We now explain how this duality also describes two transitions out of 3+1D $Z_2$ topological order.

Let us start in the XY \& XZ plane $Z_2$ topological order phase with small $t$ and $h$, and then increase $h$ to transition into the 3+1D $Z_2$ topological order phase (\figref{fig:globalPhaseDiagram}).
In the original model $\tilde{H}_\text{global}$ (\eqnref{eq:H global}), the $h$ coupling hops a composite excitation of XY and XZ--plane toric code charge excitations ($\sX{\tilde}{\iota}$ and $\tX{\tilde}{\iota}$, see \figref{fig:duality} (c)).
In \secref{sec:3+1D} we used degenerate perturbation theory to show that for large hopping strength $h$, this composite charge excitation condenses,
  and 3+1D $Z_2$ topological order results.
We now give a physical explanation.
After the composite charge condenses, the XY and XZ--plane 2+1D toric code charge excitations are no longer bound to the XY and XZ planes (respectively)
  since an XY plane charge can split into a XZ plane charge and a composite charge excitation,
  which means that the XY and XZ--plane charges can freely transform into the other.
A XY or XZ--plane charge excitation thus becomes a 3+1D toric code charge excitation after the composite charge condenses.
Once the charges are no longer bound to planes, the XY and XZ--plane 2+1D toric code flux excitations are confined since they no longer have well defined braiding statistics with the charge excitations.
That is, a pair of XY or XZ--plane flux excitations will incur an energy cost proportional the their distance
  due to two strings of YZ-plane toric code flux terms $\EQyz\, \rQ{\tilde}{p}$ which will also be excited.
Thus, the XY and XZ--plane 2+1D toric code flux excitations become 3+1D toric code flux loop excitations when the composite charge condenses.

The dual theory is described by $\tilde{H}_\text{global}^\text{dual}$ (\eqnref{eq:H dual global}).
This transition is driven by increasing $h$ (see \figref{fig:globalPhaseDiagram}) while keeping $t$ constant.
Since $\tilde{H}^\text{dual}$ does not depend on $h$, and is completely decoupled from $\bhat{H}_\text{hop}^\text{dual}$, the phase transition is just described by $\bhat{H}_\text{hop}^\text{dual}$: i.e. decoupled 1+1D Ising chains.
The composite charge excitation is dual to an excitation of the 1+1D Ising paramagnet (\figref{fig:duality} (c)).
Just like in \secref{sec:duality'}, where a similar duality is discussed, $\EQyz\, \rQ{\tilde}{p}$ appears to be marginally relevant under RG (see \appref{app:MC}) at the phase transition,
  and the phase transition is likely first order.
The extended honeycomb-based model (\eqnref{eq:H'}) also exhibits this phase transition since $\tilde{H}_\text{global}$ (\eqnref{eq:H global}) is the same as the effective Hamiltonian $\tilde{H}'$ (\eqnref{eq:H' tilde}) of the honeycomb-based model in the $\max(K_x,K_y,t,h) \ll K_z$ limit (ignoring subleading terms),
  with the unimportant difference that $\tilde{H}_\text{global}$ includes $\EQyz\, \rQ{\tilde}{p}$ as a perturbation,
  while $\tilde{H}'$ instead includes $\tilde{H}_{\h}$ (\eqnref{eq:H_h' tilde}).

Let us now consider the 3+1D $Z_2$ topological order phase with small $t$ and large $h$, and then increase $t$ to transition into the YZ-plane $Z_2$ topological order phase (\figref{fig:globalPhaseDiagram}).
In the original model $\tilde{H}_\text{global}$ (\eqnref{eq:H global}), the $t$ coupling creates a YZ-plane loop of four 3+1D toric code fluxes (\figref{fig:duality} (b)).
In \secref{sec:YZ} we used degenerate perturbation theory to show that for strong coupling $t$, the composite flux loops condense,
  and YZ-plane topological order results.
This occurs because when the YZ-plane flux loops condense, the 3+1D toric code charge excitations can no longer move in the x direction,
  since this would incur a nontrivial braiding statistic with the condensed YZ-plane flux loops.
Now consider an open flux loop excitation of 3+1D toric code that is long in the y and z directions, but very short in the x direction.
Before the YZ-plane flux loops condense, this excitation has a large energy proportional to its length.
However, after the YZ-plane flux loops condense, its bulk is indistinguishable to part of a YZ-plane flux loop,
  and therefore only a finite energy cost is incurred.
Thus, the YZ-plane flux loop condensation allows 2+1D toric code flux excitations (at the end of the open flux loop) to become deconfined along the YZ plane.

The dual theory is described by decoupled YZ-plane 2+1D Ising models $\tilde{H}^\text{dual}$ (\eqnref{eq:H dual global}) since $\bhat{H}_\text{hop}^\text{dual}$ does not depend on $t$.
The YZ-plane flux loop excitation of $\tilde{H}_\text{global}$ is dual to a domain wall in a YZ-plane 2+1D Ising model (\figref{fig:duality} (b)).
Just like in $\secref{sec:duality}$, where a similar duality is discussed, this phase transition is also likely first order.
The extended honeycomb-based model (\eqnref{eq:H'}) also exhibits this phase transition since $\tilde{H}_\text{global}$ (\eqnref{eq:H global}) is the same as
  the effective Hamiltonian $\tilde{H}'$ (\eqnref{eq:H' tilde}) of the honeycomb-based model in the $\max(K_x,K_y,t,h) \ll K_z$ limit (ignoring subleading terms).
This $\max(K_x,K_y,t,h) \ll K_z$ limit describes the phase transition out of 3+1D topological order across the vertical red line in \ref{fig:largeKzDiagram}.
However, this $\max(K_x,K_y,t,h) \ll K_z$ limit does not correspond to the top-left corner of \figref{fig:extendedPhaseDiagram} for which $\max(K_x,K_y,t) \ll \min(K_z,h)$,
  which instead corresponds to the diagonal red on the top-left of \ref{fig:largeKzDiagram}.
To describe the $\max(K_x,K_y,t) \ll \min(K_z,h)$ limit, we could apply a similar duality to the following Hamiltonian,
  which describes the extended honeycomb-based model (\eqnref{eq:H'}) when $\max(K_x,K_y,t) \ll \min(K_z,h)$:
\begin{equation}
  \bhat{H}_\text{coupled} = \bhat{H}_\text{3D toric} - \sum_\ell^\text{x-axis links} t\, \bhat\rho^x_\ell
\end{equation}
where $\bhat{H}_\text{3D toric}$ is given in \eqnref{eq:H 3D toric'}.

\subsection{Connections to Previous Coupled Layer Constructions}
\label{sec:previousCoupledLayer}

$\tilde{H}_\text{toric} + \tilde{H}_t + \tilde{H}_h$ (\eqnref{eq:H toric'}, \eqref{eq:H_t tilde}, and \eqref{eq:H_h' tilde}) is very similar to the coupled layer constructions considered in \refcite{VijayLayer} and \cite{MaLayers}.
The primary difference is that we have only XY and XZ--plane toric codes, while they also had YZ planes.
This difference was motivated in our work by the desire to have the simplest possible lattice model.
Although the lack of YZ planes in $\tilde{H}_\text{toric}$ adds some anisotropy to the resulting X-cube Hamiltonian (\eqnref{eq:H X-cube'}) and 3+1D toric code Hamiltonian (\eqnref{eq:H 3D toric'}),
  these details do not affect the universality class of these phases.
However, when both $t$ and $h$ are large, the lack of YZ-plane toric codes in our model results in radically different physics.
Specifically, in \secref{sec:YZ} we showed that XY and XZ--plane toric codes with large $t$ and $h$ coupling leads to a phase described by decoupled YZ-plane toric codes.
However, with XY, XZ, and YZ planes, \refcite{VijayLayer} showed that large $t$ and $h$ coupling drives the system into a trivial phase with no topological order.
It is interesting that large $t$ and $h$ coupling seems to have a $Z_2$ action on the presence of topological order on YZ planes.

\section{Conclusion}
\label{sec:conclusion}

In this work we strive to bring the recently discovered fracton topological order closer to an experimental realization.
We do this by presenting a simple model with X-cube fracton topological order which only requires nearest-neighbor 2-spin interactions.
This is a significant improvement over the original X-cube model with 12-spin interactions \cite{VijayXCube},
  or previous coupled layer constructions with 4-spin interactions.
Unfortunately however, it is likely that much future work will be necessary to realize fracton order in an experiment.

Perhaps there exists a material for which our model could be a useful description.
If this is not the case, we hope that this work could be useful inspiration for deriving more realistic models of fracton order with possible experimental realizations.
Additionally, synthetic quantum matter, such as cold atom experiments, provide another possible route to experimental realization.
Perhaps previous theoretical work \cite{TanamotoSpinPulse,LiuColdAtoms} on synthetic quantum matter realizations of Kitaev's honeycomb model \cite{KitaevHoneycomb} could be generalized to apply to our model or to similar models.
It would also be interesting to discover realistic models of other kinds of fracton order.
In particular, it would be very interesting to realize fractal-like (type II in \refcite{VijayXCube}) fracton topological order (e.g. Haah's code \cite{HaahCode}) via a coupled layer construction or via a two-spin Hamiltonian.

In this work, we found two first order phase transitions out of the X-cube phase.
\refcite{VijayLayer} also showed that their isotropic coupled layer construction admits dualities describing two first order phase transition out of X-cube fracton topological order.
Additionally, \refcite{VijayXCube} gave dualities describing phase transitions out of an entire class of fracton phases;
  however, the nature of these phase transitions is currently unknown.
It remains an open and interesting question if it is possible to have a continuous phase transition out of fracton topological order.

In \figref{fig:extendedPhaseDiagram} and especially in \figref{fig:phaseDiagram}, there are regions of the phase diagrams which could not be described by degenerate perturbation theory or the dualities that we found.
In particular, it is not clear what occurs when $K_x$ and $K_y$ are increased above the $\max(K_x,K_y) \ll K_z$ limit in \figref{fig:phaseDiagram}.
A gapless version of X-cube fracton order would be an interesting possibility;
  however, such a phase is likely to be unstable to a gapped phase \cite{VijayLayer,Xu2008}.
It is also natural to ask what occurs near the center of \figref{fig:extendedPhaseDiagram}.

It is also interesting to note that the X-cube operators that we generated using degenerate perturbation theory ($\rC{}{p}$ (\figref{fig:fractonOperator}), and $\sX{}{\iota}$ and $\tX{}{\iota}$ (\figref{fig:toricCode}))
  all commute with the original Hamiltonian (\eqnref{eq:H}).
The original Hamiltonian has 8 qubit degrees of freedom for each unit cell; and we have found 3 qubit integrals of motion.
In Kitaev's exactly solvable honeycomb model \cite{KitaevHoneycomb}, every unit cell is comprised of 2 qubit degrees of freedom which are split into a single integral of motion given by a hexagon/vison operator (\figref{fig:toricCode}), and two Majorana fermions with only quadratic interactions for a given static configuration of hexagon/vison excitations.
Perhaps our model, or a similar model, admits a similar exact solution, possibly with connections to the generalized parton constructions in \refcite{HsiehPartons}.

\acknowledgments

\emph{Acknowledgements} ---
We thank Leon Balents for pointing out the relevance of the layer couplings discussed in \appref{app:MC}.
This work was supported by the NSERC of Canada and the Center for Quantum Materials at the University of Toronto.
Computations were performed on the gpc supercomputer at the SciNet HPC Consortium. SciNet is funded by: the Canada Foundation for Innovation under the auspices of Compute Canada; the Government of Ontario; Ontario Research Fund - Research Excellence; and the University of Toronto. \cite{scinet}

\bibliography{fracton2}

\newpage
\appendix

\section{Large \texorpdfstring{$K_z$}{Kz} Degenerate Perturbation Theory}
\label{app:Kz DPT}

\begin{figure}
\includegraphics[width=\columnwidth]{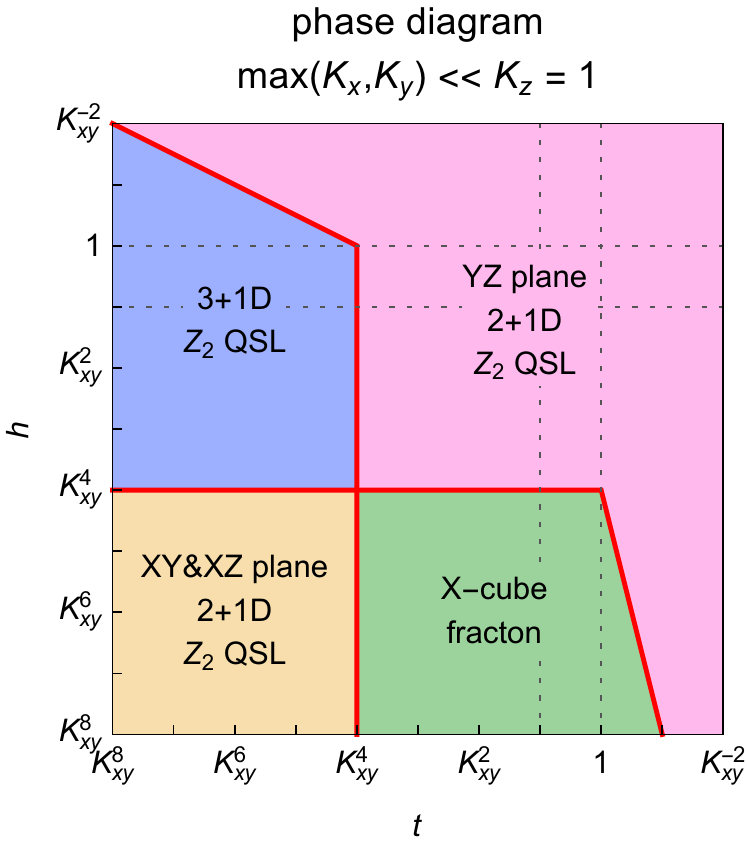}
\caption{
Phase diagram of $H'$ (\eqnref{eq:H'}) in the limit $\max(K_x,K_y) \ll K_z = 1$ with $\max(K_x,K_y)/K_z$ infinitesimally small.
Finite $\max(K_x,K_y)/K_z$ is shown in \figref{fig:phaseDiagram} and \ref{fig:extendedPhaseDiagram}.
$Z_2$ QSL is short for quantum spin liquid with $Z_2$ topological order.
The 2+1D QSL phases are in the same phase as parallel layers of 2+1D toric code along certain orthogonal planes.
The phase boundaries are calculated in \appref{app:Kz DPT} and given by \eqnref{eq:decoupled condition}, \eqref{eq:X-cube condition}, \eqref{eq:3+1D condition}, and \eqref{eq:YZ QSL condition}.
Degenerate perturbation theory does not determine the nature of the phase transitions, nor the possible existence of intermediate phases.
However, duality arguments can be used to describe these transitions (\secref{sec:dualities} and \secref{sec:global duality}).
The dotted gray lines separate overlapping regions where different choices of unperturbed Hamiltonian $H_0$ were used in degenerate perturbation theory in \appref{app:Kz DPT}.
The diagram is a log-log plot where the axes have log base $K_{xy}^{-1}$ where $K_{xy} = \sqrt{K_x K_y}$.
(The axes are labeled by powers of $K_{xy}$, which is to be thought of as a very small number $\epsilon$.)
Note that the point set topology of the log-log coordinates in the $K_{xy} \ll K_z$ limit can be surprising.
Specifically, for finite $K_{xy} / K_z$, intermediate phases are possible (see \appref{app:MC}), but in the $K_{xy} \ll K_z$ limit, these intermediate phases shrink into the red lines.
}\label{fig:largeKzDiagram}
\end{figure}

In this section, we will apply degenerate perturbation theory (DPT) (\appref{app:DPT}) to the extended model $H'$ (\eqnref{eq:H'}) and produce a phase diagram (\figref{fig:largeKzDiagram}) in the limit when $K_z \gg \max(K_x,K_y)$.
To reduce verbosity, throughout this section we will often refer to operators by their coefficient.
For example, when we say that $\EQ$ and $t$ anticommute, we mean that an operator with coefficient $\EQ$ anticommutes with an operator with coefficient $t$.
We will perform DPT by splitting the Hamiltonian in \eqnref{eq:H'} into $H' = H_0 + H_1$ where $|H_0| \gg |H_1|$ (where any operator norm can be chosen).
$K_z$ will always be included in $H_0$,
  and $K_x$ and $K_y$ will always be in $H_1$.
Therefore, $t$ can only be in $H_0$ if $t \gg \max(K_x,K_y)$;
and        $t$ can only be in $H_1$ if $t \ll K_z$.
Similarly, $h$ can only be in $H_0$ if $h \gg \max(K_x,K_y)$;
and        $h$ can only be in $H_1$ if $h \ll K_z$.
These (overlapping) conditions partition (via the dotted gray lines) the phase diagram (\figref{fig:largeKzDiagram}) into four overlapping quadrants.

DPT will generate many different terms.
But how do we determine the resulting phase?
Instead of trying to answer this question direction, we will instead consider one target phase at a time (e.g. X-cube fracton order), and then attempt to calculate a sufficient condition for that phase
  (e.g. $\max(K_x,K_y) \ll t \ll K_z$).
In our model, doing this is simplified because all terms in the Hamiltonian that we consider are products of Pauli operators, which we will refer to as Pauli strings.
Furthermore, all of the phases that we're interested in are gapped and can be described by a stabilizer Hamiltonian.
By stabilizer Hamiltonian, we will mean a Hamiltonian that is a sum of Pauli strings which all commute with each other.
The ground state of the stabilizer Hamiltonian will also be the ground state of an extensive number of Pauli string terms in the stabilizer Hamiltonian;
  we will refer to these Pauli strings as stabilizers.
Examples of stabilizer Hamiltonians are:
  $H=\sum_i X_i$ which stabilizes a paramagnet,
  $H=\sum_{\langle i j \rangle} Z_i Z_j$ which stabilizes a ferromagnet,
  or the toric code $H = \sum_\square \prod_{i \in \square} Z_i + \sum_+ \prod_{i \in +} X_i$ which stabilizes $Z_2$ topological order.
In certain limits, such as $\max(K_x,K_y) \ll t \ll K_z$ in \eqnref{eq:X-cube limit}, we will find that (multiple applications of) DPT produces a Hamiltonian that is a stabilizer Hamiltonian.
If we obtain the right stabilizers, then we know that the limits considered are a sufficient condition for the target phase.

As a generic example, suppose we want to find sufficient conditions for a target phase with stabilizers: $A_i$, $B_i$, $C_i$, and $D_i$.
If at least one of these operators appears in $H$, then we can apply DPT with at least one of them appearing in $H_0$.
Suppose only $A_i$ appears in $H$ with energy $E_A$.
Then we must choose $H_0 = - \sum_i E_A A_i$.
In order for DPT to converge, we will have to assume that $E_A$ is much larger than all energies in $H_1$.
This will thus be part of the sufficient condition for the target phase.
If we assume this, and generate an effective Hamiltonian $H_\text{eff}$, then $A_i$ will be a stabilizer (after more applications of DPT) of the ground state since $H_\text{eff}$ resides in the degenerate ground state manifold of $H_0$.
Now suppose that $\sum_i E_B B_i$ is a term in $H_\text{eff}$, but neither $C_i$ nor $D_i$ appear.
Then we will want to apply DPT to $H_\text{eff}$ with $H_0' = - \sum_i E_B B_i$ and $H_1' = H_\text{eff} - H_0'$ to generate a new effective Hamiltonian $H_\text{eff}$ in hope of eventually generating $C_i$ and $D_i$.
In order for DPT to converge, we will have to assume that $E_B$ is larger than all other terms in $H_\text{eff}$.
In particular, we will be most worried about terms which anticommute with $B_i$, since these terms will create excitations of $B_i$ and will therefore try to prevent $B_i$ from being a stabilizer.
Suppose the new effective Hamiltonian $H_\text{eff}'$ does include $C_i$ and $D_i$ with energies $E_C$ and $E_D$, respectively.
Then if $E_C$ and $E_D$ are sufficiently large, we can apply DPT once more with $H_0'' = - \sum_i (E_C C_i + E_D D_i)$, and we will be left with a stabilizer Hamiltonian of the desired phase.
Note that terms like $\sum_{\langle i j \rangle} E_{CC} C_i C_j$ may also be generated.
However, as long as $E_{CC} \ll E_C$, these terms don't affect the ground state, and so they will be ignored since $C_i C_j$ is not an independent stabilizer from $C_i$
  (since it's the product of $C_i$ and $C_j$).
Thus, a sufficient condition for the target phase would require that $E_A$, $E_B$, $E_C$, and $E_D$ are sufficiently large.

In \figref{fig:largeKzDiagram} we show the results from this approach.
The sufficient conditions that we generate for each phase are not necessary conditions.
Thus, we can not fill in the entire phase diagram (as seen in \figref{fig:extendedPhaseDiagram}).
However, \figref{fig:largeKzDiagram} is drawn in a subspace of couplings ($K_\mu,t,h$) where we can identify the phase at almost every point.

Below we list the energies (calculated in \appref{app:energies}) for the important terms generated by degenerate perturbation theory (DPT) that are required for drawing the phase diagram (\figref{fig:largeKzDiagram}).
\begin{align}
  \EQ     &\sim \frac{K_x^2 K_y^2}{K_z \max(K_z,h)^2} \nonumber\\
  \EX           &\sim \frac{K_x^2 K_y^2}{\max(K_z,t)^4/K_z} \label{eq:energies}
\end{align}
$\EQ$  (introduced in \eqnref{eq:H toric}) corresponds to a toric code flux operator (\figref{fig:toricCode}) on an XY or XZ plane.
$\EX$ (\eqnref{eq:H toric}) corresponds to a 2+1D toric code charge operator (\figref{fig:toricCode}) on an XY or XZ plane.
The ``$\sim$'' in the expressions means that these expressions are only asymptotically correct;
  e.g. factors of two are ignored which allows simplifications such as $K_z + t \sim \max(K_z,t)$.

Energies for other terms generated by DPT will also be mentioned.
These energies can be difficult to calculate in some cases due to the high order of DPT required to calculate them;
  but fortunately their values will not be important in this work.
These energies are:
\vbox{ 
\begin{itemize}
\item $\EQyz$ (in \eqnref{eq:H 3D toric} and \eqref{eq:H YZ toric}), corresponding to a toric code flux operator (\figref{fig:3+1D}) on a YZ plane
\item $\EC$ (in \eqnref{eq:H X-cube}), corresponding to an X-cube fracton number operator (\figref{fig:fractonOperator})
\item $\ES$ (in \eqnref{eq:H 3D toric}), corresponding to a 3+1D toric code charge operator (\figref{fig:3+1D})
\item $\EXyz$ (in \eqnref{eq:H YZ toric}), corresponding to the same operator as $\ES$,
  but we instead view it as a YZ-plane toric code charge operator
\end{itemize}}

We now apply DPT to $H'$ (\eqnref{eq:H'}) in various limits.
$H_{\h}$ (\eqnref{eq:H_h'}) will be considered as an example of a generic perturbation that can be added to break unphysical degeneracy which results in some cases;
  it's coefficient $\h$ will always be assumed to be infinitesimally small.
Then we will calculate $\EQ$ and $\EX$ (\eqnref{eq:energies}) in \appref{app:energies}.

\subsection{XY\&XZ plane 2+1D \ZTwo Topological Order}
\label{app:XY-XZ}

This phase occurs when $K_z$, $\EQ$, and $\EX$ are the stabilizers.
Let us determine sufficient conditions for this.
We'll need to do DPT in order to produce $\EQ$ and $\EX$.
We'll do DPT with $H_0 = - K_z\, (\cdots)$.
To ensure that DPT converges, we'll need to assume $K_z \gg \max(K_x, K_y, t, h)$.
$\EQ$ and $\EX$ are the highest energy operators generated by DPT.
But $\EQ$ anticommutes with $t$, and can therefore only be a stabilizer if $\EQ \gg t$.
And $\EX$ anticommutes with $h$, and can therefore only be a stabilizer if $\EX \gg h$.
Therefore, if
\begin{equation}
  K_z \gg \max(K_x, K_y, t, h) \text{ and } \EQ \gg t \text{ and } \EX \gg h \label{eq:decoupled condition}
\end{equation}
then another application of DPT with $H_0' = - K_z\, (\cdots) - \EQ\, (\cdots) - \EX\, (\cdots)$ will produce a stabilizer Hamiltonian with
  2+1D $Z_2$ topological order along the original XY and XZ planes.
The above conditions are already sufficient to ensure that DPT converges.
This is the decoupled phase.

\subsection{X-cube Fracton Topological Order}
\label{app:X-cube}

This phase occurs when $K_z$, $t$, $\EC$, and $\EX$ are stabilizers.
Let us determine sufficient conditions for this.
We'll need to do DPT in order to produce $\EC$ and $\EX$.

Let us first try DPT with $H_0 = - K_z\, (\cdots) - t\, (\cdots)$.
To ensure that DPT converges, we'll need to assume $\min(K_z, t) \gg \max(K_x, K_y, h)$.
$E_{+\Et}$ is the highest energy operator generated by DPT.
But $E_{+\Et}$ anticommutes with $h$, and can therefore only be a stabilizer if $E_{+\Et} \gg h$.
$\EC$ is the next highest energy operator generated by DPT that's independent from and commutes with the previous (candidate) stabilizers: $K_z$, $t$, and $E_{+\Et}$.
(The product of two $\EX$ terms may be generated with a higher energy, but such terms will not affect the ground state since their energy is much smaller than $\EX$.)
$\EC$ can therefore be a stabilizer.
In fact, if $K_z$, $t$, and $\EX$ are stabilizers, then $\EC$ will have to be a stabilizer if we require the ground state to be the ground state of a stabilizer Hamiltonian which is stable to arbitrary perturbations.
There simply is no other Pauli string that commutes with  $K_z$, $t$, and $\EX$, and isn't a product of $K_z$, $t$, $\EX$, and $\EC$.
Therefore, if $\min(K_z, t) \gg \max(K_x, K_y, h)$ and $E_{+\Et} \gg h$, then X-cube fracton topological order will be stabilized.
That is, applying DPT again with $H_0' =  - K_z\, (\cdots) - t\, (\cdots) - \EC\, (\cdots) - \EX\, (\cdots)$ will result in a stabilizer Hamiltonian with X-cube fracton order;
  the above assumptions are already sufficient to guarantee that DPT converges.

But what if it isn't the case that $t \gg \max(K_x, K_y, h)$?
To cover this case, we'll also try DPT with $H_0 = - K_z\, (\cdots)$.
To ensure that DPT converges, we'll need to assume $K_z \gg \max(K_x, K_y, t, h)$.
(This assumption will overlap with the $\min(K_z, t) \gg \max(K_x, K_y, h)$ assumption in the previous paragraph, but the results in this overlapping region are the same.)
$\EQ$ and $\EX$ are the highest energy operators generated by DPT.
$t$ anticommutes with $\EQ$, and can therefore only be a stabilizer if $t \gg \EQ$.
We can do DPT again with $H_0' = - t\, (\cdots)$ and $H_1' = - \EQ\, (\cdots) - \EX\, (\cdots) - h\, (\cdots) + \cdots$
  where the last ``$\cdots$'' denotes unimportant subleading terms.
$t \gg \EQ$ (note $\EQ \sim \EX$) is already sufficient to ensure that DPT converges.
$\EX$ anticommutes with $h$, and can therefore only be a stabilizer if $\EX \gg h$.
$\EC$ is the highest energy operator generated by DPT that's independent from and commutes with the previous (candidate) stabilizers: $K_z$, $t$ and $\EX$.
$\EC$ can therefore be a stabilizer.
Therefore, if $K_z \gg \max(K_x, K_y, t, h)$ and $t \gg \EQ$ and $\EX \gg h$, then X-cube fracton topological order will be stabilized.

In summary, a sufficient condition for X-cube fracton topological order is
\begin{align}
  &\left[ \min(K_z, t) \gg \max(K_x, K_y, h) \text{ and } E_{+\Et} \gg h \right] \text{ or} \nonumber\\
  &\left[ K_z \gg \max(K_x, K_y, t, h) \text{ and } t \gg \EQ \text{ and } \EX \gg h \right] \label{eq:X-cube condition}
\end{align}

\subsection{3+1D \ZTwo Topological Order}
\label{app:3+1D}

This phase occurs when $K_z$, $h$, $\EQ$, $\EQyz$, and $\ES$ are the stabilizers.
Let us determine sufficient conditions for this.
We'll need to do DPT in order to produce $\EQ$, $\EQyz$, and $\ES$.

Let us first try DPT with $H_0 = - K_z\, (\cdots) - h\, (\cdots)$.
To ensure that DPT converges, we'll need to assume $\min(K_z, h) \gg \max(K_x, K_y, t)$.
$\EQ$ is the highest energy operator generated by DPT.
But $\EQ$ anticommutes with $t$, and can therefore only be a stabilizer if $\EQ \gg t$.
$\EC$ and $E_{*\Eh}$ are the next highest energy operators generated by DPT that are independent from and commute with the previous (candidate) stabilizers: $K_z$, $h$, and $\EQ$.
$\EC$ and $E_{*\Eh}$ can therefore be stabilizers since they commute.
(If they anticommuted, then they couldn't both be stabilizers.)
The current stabilizers ($K_z$, $h$, $\EQ$, $\ES$, and $\EC$) describe a phase corresponding to 3+1D toric code, but without YZ-plane plaquette/flux operators.
This phase is unstable to perturbations
  and has unphysical degeneracy $\sim 2^{L^2}$ on a torus due to the $L^2$ missing $\EQyz$ operators on any single YZ plane.
However, with generic perturbations (for which we choose $H_{\h}$ (\eqnref{eq:H_h'}) as an example),
  we can do DPT again with $H_0' = - \EQ\, (\cdots) - \ES\, (\cdots) - \EC\, (\cdots)$
  and $H_1' = H_{\h} + \cdots$
  where the last ``$\cdots$'' denotes unimportant subleading terms.
$\EQyz$ is the highest energy operator generated by DPT.
$\EQyz$ can therefore be a stabilizer (which makes $\EC$ a redundant stabilizer since it's the product of two $\EQyz$ operators, modulo $K_z$ and $h$ operators).
Therefore, if $\min(K_z, h) \gg \max(K_x, K_y, t)$ and $\EQ \gg t$, then 3+1D $Z_2$ topological order will be stabilized.

But what if it isn't the case that $h \gg \max(K_x, K_y, t)$?
To cover this case, we'll also try DPT with $H_0 = - K_z\, (\cdots)$.
To ensure that DPT converges, we'll need to assume $K_z \gg \max(K_x, K_y, t, h)$.
(This assumption will overlap with the $\min(K_z, h) \gg \max(K_x, K_y, t)$ assumption in the previous paragraph, but the results in this overlapping region are the same.)
$\EQ$ and $\EX$ are the highest energy operators generated by DPT.
But $h$ anticommutes with $\EX$, and can therefore only be a stabilizer if $h \gg \EX$.
We can do DPT again with $H_0' = - h\, (\cdots)$ and $H_1' = - \EQ\, (\cdots) - \EX\, (\cdots) - t\, (\cdots) + \cdots$.
$h \gg \EX$ is already sufficient to ensure that DPT converges.
$\EQ$ anticommutes with $t$, and can therefore only be a stabilizer if $\EQ \gg t$.
$\EC$ and $\ES$ are the highest energy operators generated by DPT that are independent from and commute with the previous (candidate) stabilizers: $K_z$, $h$ and $\EQ$.
$\EC$ and $\ES$ can therefore be stabilizers since they commute.
We can do DPT again with $H_0'' = - h\, (\cdots) - \EQ\, (\cdots) - \ES\, (\cdots) - \EC\, (\cdots)$
  and $H_1'' = H_{\h} + \cdots$.
$\EQyz$ is the highest energy operator generated by DPT and can therefore be a stabilizer.
Therefore, if $K_z \gg \max(K_x, K_y, t, h)$ and $h \gg \EX$ and $\EQ \gg t$, then 3+1D topological order will be stabilized.

In summary, a sufficient condition for 3+1D $Z_2$ topological order is
\begin{align}
  &\left[ \min(K_z, h) \gg \max(K_x, K_y, t) \text{ and } \EQ \gg t \right] \text{ or} \nonumber\\
  &\left[ K_z \gg \max(K_x, K_y, t, h) \text{ and } h \gg \EX \text{ and } \EQ \gg t \right] \label{eq:3+1D condition}
\end{align}
when generic (but arbitrarily small) perturbations (such as $H'$ (\eqnref{eq:H'})) are included in the Hamiltonian $H'$ (\eqnref{eq:H'}).

\subsection{YZ-plane 2+1D \ZTwo topological order}
\label{app:YZ}

This phase occurs when $K_z$, $h$, $t$, $\EQyz$, and $\EXyz$ are the stabilizers.
Let us determine sufficient conditions for this.
We'll need to do DPT in order to produce $\EQyz$ and $\EXyz$.

Let us first try DPT with $H_0 = - K_z\, (\cdots) - t\, (\cdots) + h\, (\cdots)$.
To ensure that DPT converges, we'll need to assume $\min(K_z, t, h) \gg \max(K_x, K_y)$.
$\EXyz$ and $\EC$ are the highest energy operators generated by DPT.
$\EXyz$ and $\EC$ can therefore be stabilizers since they commute.
The current stabilizers ($K_z$, $h$, $\EXyz$, and $\EC$) describe a phase that is unstable to perturbations
  and has unphysical degeneracy $\sim 2^{L^2}$ on a torus due to the $L^2$ missing $\EQyz$ operators on any single YZ plane.
However, with generic perturbations (for which we choose $H_{\h}$ (\eqnref{eq:H_h'}) as an example),
  we can do DPT again with $H_0' = - \EXyz\, (\cdots) - \EC\, (\cdots)$
  and $H_1' = H_{\h} + \cdots$
  where the last ``$\cdots$'' denotes unimportant subleading terms.
$\EQyz$ is the highest energy operator generated by DPT and can therefore be a stabilizer.
Therefore, if $\min(K_z, t, h) \gg \max(K_x, K_y)$, then the desired phase will be stabilized.

Let us now try DPT with $H_0 = - K_z\, (\cdots) - t\, (\cdots)$.
To ensure that DPT converges, we'll need to assume $\min(K_z, t) \gg \max(K_x, K_y, h)$.
$E_{+\Et}$ is the highest energy operator generated by DPT.
But $h$ anticommutes with $E_{+\Et}$, and can therefore only be a stabilizer if $h \gg E_{+\Et}$.
$\EX$ and $\EC$ are the next highest energy operators generated by DPT that are independent from and commute with the previous (candidate) stabilizers: $K_z$, $t$, and $h$.
$\EX$ and $\EC$ can therefore be stabilizers since they commute.
We can do DPT again with $H_0' = - h\, (\cdots) - \EX\, (\cdots) - \EC\, (\cdots)$
  and $H_1' = H_{\h} + \cdots$.
$\EQyz$ is the highest energy operator generated by DPT and can therefore be a stabilizer.
Therefore, if $\min(K_z, t) \gg \max(K_x, K_y, h)$ and $h \gg \EX$, then the desired phase will be stabilized.

Let us now try DPT with $H_0 = - K_z\, (\cdots) - h\, (\cdots)$.
To ensure that DPT converges, we'll need to assume $\min(K_z, h) \gg \max(K_x, K_y, t)$.
$\EQ$ is the highest energy operator generated by DPT.
But $t$ anticommutes with $\EQ$, and can therefore only be a stabilizer if $t \gg \EQ$.
$\EXyz$ and $\EC$ are the next highest energy operators generated by DPT that are independent from and commute with the previous (candidate) stabilizers: $K_z$, $t$, and $h$.
$\EXyz$ and $\EC$ can therefore be stabilizers since they commute.
We can do DPT again with $H_0' = - t\, (\cdots) - \EXyz\, (\cdots) - \EC\, (\cdots)$
  and $H_1' = H_{\h} + \cdots$.
$\EQyz$ is the highest energy operator generated by DPT and can therefore be a stabilizer.
Therefore, if $\min(K_z, h) \gg \max(K_x, K_y, t)$ and $t \gg \EQ$, then the desired phase will be stabilized.

Let us now try DPT with $H_0 = - K_z\, (\cdots)$.
To ensure that DPT converges, we'll need to assume $K_z \gg \max(K_x, K_y, t, h)$.
$\EQ$ and $\EX$ are the highest energy operators generated by DPT.
But $t$ anticommutes with $\EQ$, and can therefore only be a stabilizer if $t \gg \EQ$.
And $h$ anticommutes with $\EX$      , and can therefore only be a stabilizer if $h \gg \EX$.
We can do DPT again with $H_0' = - t\, (\cdots) - h\, (\cdots)$ and $H_1' = - \EQ\, (\cdots) - \EX\, (\cdots) + \cdots$
  where the last ``$\cdots$'' denotes unimportant subleading terms.
The previous assumptions are sufficient to ensure that DPT converges.
$\EC$ and $\EXyz$ are the highest energy operators generated by DPT that are independent from and commute with the previous (candidate) stabilizers: $K_z$, $t$, and $h$.
$\EC$ and $\EXyz$ can therefore be stabilizers since they commute.
We can do DPT again with $H_0'' = - t\, (\cdots) - h\, (\cdots) - \EXyz\, (\cdots) - \EC\, (\cdots)$
  and $H_1'' = H_{\h} + \cdots$.
$\EQyz$ is the highest energy operator generated by DPT and can therefore be a stabilizer.
Therefore, if $K_z \gg \max(K_x, K_y, t, h)$ and $t \gg \EQ$ and $h \gg \EX$, then the desired phase will be stabilized.

In summary, a sufficient condition for 2+1D $Z_2$ topological order on YZ planes is
\begin{align}
  &\left[ \min(K_z, t, h) \gg \max(K_x, K_y) \right] \text{ or} \nonumber\\
  &\left[ \min(K_z, t) \gg \max(K_x, K_y, h) \text{ and } h \gg \EX \right] \text{ or} \nonumber\\
  &\left[ \min(K_z, h) \gg \max(K_x, K_y, t) \text{ and } t \gg \EQ \right] \text{ or} \nonumber\\
  &\left[      K_z  \gg \max(K_x, K_y, t, h) \text{ and } t \gg \EQ \text{ and } h \gg \EX \right] \label{eq:YZ QSL condition}
\end{align}
when generic (but arbitrarily small) perturbations (such as $H'$ (\eqnref{eq:H'})) are included in the Hamiltonian $H'$ (\eqnref{eq:H'}).
Physically, when these perturbations are small, $\EC \gg \EQyz$ and this implies that in the excited states, flux excitations $\EQyz$ will want to be bound to a partner flux excitation on a neighboring layer.
However, this does not change the resulting phase of the ground state.

\subsection{\texorpdfstring{$\EQ$ and $\EX$}{Energy} Calculation}
\label{app:energies}

In this section, we'll calculate $\EQ$ and $\EX$ using degenerate perturbation theory (\appref{app:DPT}).
$\EQ$ and $\EX$ appear in the effective Hamiltonian $H^\text{eff}$ at fourth order.
  and the only term at this order that contributes is $\scP (H_1 \scD)^3 H_1 \scP$.
We will focus on the $\EQ$ or $\EX$ term that is generated for a single hexagon with indices labeled as shown below.
\begin{center}
  \includegraphics[width=.2\columnwidth]{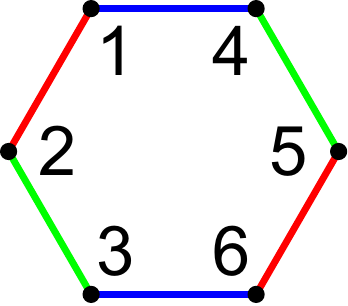}
\end{center}
Thus, we only need to consider the terms of $H_1$ that act only on these six sites.
And we can erase all Pauli operators from $H_0$ which aren't on one of these six sites.
$\EQ$ is centered on plaquettes (\figref{fig:toricCode}) and is calculated assuming $t \ll K_z$.
Thus, we can take $t=0$ when calculating $\EQ$,
  but we must consider the $h$ term acting on sites 1 and 4, and 3 and 6.
$\EX$       is centered on vertices   (\figref{fig:toricCode}) and is calculated assuming $h \ll K_z$.
Thus, we can take $h=0$ when calculating $\EX$,
  but we must consider the $t$ term acting on sites 2 and 5.
Therefore, we can calculate $\EQ$ and $\EX$ at the same time by using the $H_0$ and $H_1$ below and taking either $t$ or $h$ equal to zero.
\begin{align}
  H_0 &= - (K_z + \frac{t}{2})\, (\sigma^z_2 + \sigma^z_5) - K_z\, (\sigma^z_1 \sigma^z_4 + \sigma^z_3 \sigma^z_6) \nonumber\\
      &\,\quad - \frac{h}{2}\, (\sigma^x_1 \sigma^y_4 + \sigma^y_1 \sigma^x_4 + \sigma^x_3 \sigma^y_6 + \sigma^y_3 \sigma^x_6) \nonumber\\
  H_1 &= - K_x\, (\sigma^x_1 \sigma^x_2 + \sigma^x_5 \sigma^x_6) - K_y\, (\sigma^y_2 \sigma^y_3 + \sigma^y_4 \sigma^y_5) \nonumber
\end{align}
The effective Hamiltonian is:
\begin{widetext}
\begin{align}
  H^\text{eff}_4 &= \scP H_1 \scD H_1 \scD H_1 \scD H_1 \scP \nonumber\\
  &= 8 K_y \sigma^y_2 \sigma^y_3 \frac{-1}{4K_z+ t+ h}
       K_x \sigma^x_5 \sigma^x_6 \;\;\;\, \frac{-1}{4K_z+2t   }\;\;\;\,
       K_y \sigma^y_4 \sigma^y_5 \frac{-1}{4K_z+ t+ h}
       K_x \sigma^x_1 \sigma^x_2 \nonumber\\
  &+ 8 K_y \sigma^y_2 \sigma^y_3 \frac{-1}{4K_z+ t+ h}
       K_y \sigma^y_4 \sigma^y_5 \frac{-1}{8K_z+2t+2h}
       K_x \sigma^x_5 \sigma^x_6 \frac{-1}{4K_z+ t+ h}
       K_x \sigma^x_1 \sigma^x_2 \nonumber\\
  &= - E \sigma^x_1 \sigma^z_2 \sigma^y_3 \sigma^y_4 \sigma^z_5 \sigma^x_6 \nonumber\\
  &= - E \sigma^y_1 \sigma^z_2 \sigma^x_3 \sigma^x_4 \sigma^z_5 \sigma^y_6 \label{eq:op}\\
  E &= \frac{4 K_x^2 K_y^2}{(4 K_z + t + h)^2} \left( \frac{1}{2 K_z + t} - \frac{1}{4 K_z + t + h} \right) \label{eq:E}
\end{align}
\end{widetext}
$\EQ$ is given by $E$ in \eqnref{eq:E} with $t=0$,
  which matches \eqnref{eq:energies} asymptotically.
$\EX$ is given by $E$ in \eqnref{eq:E} with $h=0$,
  which also matches \eqnref{eq:energies} asymptotically.
Note that we used the fact that $\sigma^z_1 \sigma^z_4 = \sigma^z_3 \sigma^z_6 = 1$ in the low energy Hilbert space to get \eqnref{eq:op} from the previous line, which matches the form of $\sQ{}{p}$ and $\sX{}{\iota}$ in \figref{fig:toricCode}.

\section{Degenerate Perturbation Theory}
\label{app:DPT}

In this section, we review the Schrieffer-Wolff formulation of degenerate perturbation theory \cite{SchriefferWolff,BravyiDPT}, which we use in the rest of the paper.
We begin by splitting a Hamiltonian $H$ into an unperturbed Hamiltonian $H_0$ and a perturbation $\lambda H_1$ where $\lambda \ll 1$.
\begin{equation}
  H = H_0 + \lambda H_1
\end{equation}
We will focus on the case where $H_0$ has a degenerate ground state manifold with energy $E_0$.
The goal is to derive an effective Hamiltonian $H^\text{eff}$ which describes the physics of the ground state manifold in the presence of the perturbation $\lambda H_1$.

We define an operator $\scP$ which projects onto the ground state manifold:
\begin{equation}
  \scP H_0 = H_0 \scP = E_0 \scP
\end{equation}
and define the effective Hamiltonian $H^\text{eff}$ using a canonical transformation on $H$
\begin{align}
  H^\text{eff} &\equiv e^{+S} H e^{-S} \\
            &= H + [S,H] + \frac{1}{2!} [S,[S,H]] + \frac{1}{3!} [S,[S,[S,H]]] + \cdots \nonumber\\
            &= H_0 + \lambda   \left([S_1,H_0] + \underbrace{H_1}_{X_1} \right) \nonumber\\
          &\;\;\, + \lambda^2 \left([S_2,H_0] + \underbrace{[S_1,H_1] + \frac{1}{2!} [S_1,[S_1,H_0]]}_{X_2} \right) + \cdots \nonumber\\
  S &= \sum_{n=1}^\infty \lambda^n S_n \label{eq:S}\\
  X_n &\equiv \frac{1}{n!} (\partial_\lambda^n H^\text{eff})_{\lambda \rightarrow 0} - [S_n, H_0] \nonumber
\end{align}
The canonical transformation is generated by an antihermitian operator $S$.
We also define $X_n$ as shown above.
With $S$ defined as in \eqnref{eq:S}, in order for $H^\text{eff}$ to capture the desired physics,
  it is sufficient to require that the effective Hamiltonian commutes with the ground state projection operator $\scP$:
\begin{equation}
  [\scP, H^\text{eff}] = 0 \label{eq:H^eff}
\end{equation}
This condition implies that the unitary rotation $e^S$ has rotated $H$ into $H^\text{eff}$ with a block diagonal form such that the original ground state manifold is completely decoupled from the high energy Hilbert space.
A solution to \eqnref{eq:H^eff} exists in which $S_n$ is defined iteratively in terms of $S_1, S_2, \cdots, S_{n-1}$:
\begin{align}
  S_n  &= \scP X_n \scD - \scD X_n \scP \label{eq:Sn}\\
  \scD &= - \frac{1-\scP}{H_0 - E_0} \nonumber
\end{align}
The disadvantage of this solution is that $S$ is a nonlocal operator due to the presence of the nonlocal projection operator $\scP$.
Thus, when we calculate a series expansion for $H^\text{eff}$ (\eqnref{eq:H_eff}), it will not be explicitly local.
The locality of $H^\text{eff}$ will only result from a precise cancellation due to minus signs.

We now define some compact notation:
\begin{align}
  \langle \langle \nu_1, \nu_2, \cdots, \nu_n \rangle \rangle
    &\equiv \scD^{(\nu_1)} H_1 \scD^{(\nu_2)} H_1 \cdots \scD^{(\nu_n)} \\
  \langle \nu_1, \nu_2, \cdots, \nu_n \rangle\;
    &\equiv \langle \langle 0, \nu_1, \nu_2, \cdots, \nu_n, 0 \rangle \rangle \nonumber\\
    &=      \scP H_1 \scD^{(\nu_1)} H_1 \scD^{(\nu_2)} H_1 \cdots \scD^{(\nu_n)} H_1 \scP \nonumber\\
  \scD^{(\nu)} &\equiv
    \begin{cases}
      \scP     & \nu = 0 \\
      \scD^\nu & \nu > 0
    \end{cases} \nonumber
\end{align}
Using this notation and iterating \eqnref{eq:Sn} on a computer gives the following expansion for the canonical transformation generator $S$ and the effective Hamiltonian $H^\text{eff}$ projected into the low energy Hilbert space:
\begin{widetext}
\begin{align}
  S_1 &= \scP H_1 \scD - \scD H_1 \scP \\
      &= \langle\langle0,1\rangle\rangle - \langle\langle1,0\rangle\rangle \nonumber\\
  S_2 &= \scP H_1 \scD H_1 \scD - \scD H_1 \scD H_1 \scP - \scP H_1 \scP H_1 \scD^2 + \scD^2 H_1 \scP H_1 \scP \nonumber\\
      &= \langle \langle 0,1,1\rangle \rangle - \langle \langle 1,1,0\rangle \rangle - \langle \langle 0,0,2\rangle \rangle + \langle \langle 2,0,0\rangle \rangle \nonumber\\
  S_3 &= \langle \langle 0,1,1,1\rangle \rangle - \langle \langle 1,1,1,0\rangle \rangle \nonumber\\
      &- \langle \langle 0,0,1,2\rangle \rangle - \langle \langle 0,0,2,1\rangle \rangle - \langle \langle 0,1,0,2\rangle \rangle - \frac{1}{3} \langle \langle 0,2,0,1\rangle \rangle + \frac{1}{3} \langle \langle 1,0,2,0\rangle \rangle + \langle \langle 1,2,0,0\rangle \rangle + \langle \langle 2,0,1,0\rangle \rangle + \langle \langle 2,1,0,0\rangle \rangle \nonumber\\
      &+ \langle \langle 0,0,0,3\rangle \rangle - \langle \langle 3,0,0,0\rangle \rangle \nonumber\\
  S_4 &= \langle \langle 0,1,1,1,1\rangle \rangle - \langle \langle 1,1,1,1,0\rangle \rangle \nonumber\\
      &- \langle \langle 0,0,1,1,2\rangle \rangle - \langle \langle 0,0,1,2,1\rangle \rangle - \langle \langle 0,0,2,1,1\rangle \rangle - \langle \langle 0,1,0,1,2\rangle \rangle - \langle \langle 0,1,0,2,1\rangle \rangle - \langle \langle 0,1,1,0,2\rangle \rangle - \frac{1}{3} \langle \langle 0,1,2,0,1\rangle \rangle \nonumber\\
      &\quad\quad - \frac{1}{3} \langle \langle 0,2,0,1,1\rangle \rangle - \frac{1}{3} \langle \langle 0,2,1,0,1\rangle \rangle + \frac{1}{3} \langle \langle 1,0,1,2,0\rangle \rangle + \frac{1}{3} \langle \langle 1,0,2,1,0\rangle \rangle + \frac{1}{3} \langle \langle 1,1,0,2,0\rangle \rangle + \langle \langle 1,1,2,0,0\rangle \rangle \nonumber\\
      &\quad\quad + \langle \langle 1,2,0,1,0\rangle \rangle + \langle \langle 1,2,1,0,0\rangle \rangle + \langle \langle 2,0,1,1,0\rangle \rangle + \langle \langle 2,1,0,1,0\rangle \rangle + \langle \langle 2,1,1,0,0\rangle \rangle \nonumber\\
      &+ \langle \langle 0,0,0,2,2\rangle \rangle + \langle \langle 0,0,2,0,2\rangle \rangle + \frac{1}{3} \langle \langle 0,2,0,0,2\rangle \rangle - \frac{1}{3} \langle \langle 2,0,0,2,0\rangle \rangle - \langle \langle 2,0,2,0,0\rangle \rangle - \langle \langle 2,2,0,0,0\rangle \rangle \nonumber\\
      &+ \langle \langle 0,0,0,1,3\rangle \rangle + \langle \langle 0,0,0,3,1\rangle \rangle + \langle \langle 0,0,1,0,3\rangle \rangle + \frac{1}{3} \langle \langle 0,0,3,0,1\rangle \rangle + \langle \langle 0,1,0,0,3\rangle \rangle + \frac{1}{3} \langle \langle 0,3,0,0,1\rangle \rangle \nonumber\\
      &\quad\quad - \frac{1}{3} \langle \langle 1,0,0,3,0\rangle \rangle - \frac{1}{3} \langle \langle 1,0,3,0,0\rangle \rangle - \langle \langle 1,3,0,0,0\rangle \rangle - \langle \langle 3,0,0,1,0\rangle \rangle - \langle \langle 3,0,1,0,0\rangle \rangle - \langle \langle 3,1,0,0,0\rangle \rangle \nonumber\\
      &- \langle \langle 0,0,0,0,4\rangle \rangle + \langle \langle 4,0,0,0,0\rangle \rangle \nonumber
\end{align}

\begin{align}
  &\!\!\!\!\!\!\!\!\!\!\! \scP H^\text{eff} \scP = \sum_{n=0}^\infty \lambda^n H^\text{eff}_n \label{eq:H_eff}\\
  H^\text{eff}_0 &= E_0 \nonumber\\
  H^\text{eff}_1 &= \scP H_1 \scP
               = \langle\rangle \nonumber\\
  H^\text{eff}_2 &= \scP H_1 \scD H_1 \scP
               = \langle1\rangle \nonumber\\
  H^\text{eff}_3 &= \scP H_1 \scD H_1 \scD H_1 \scP - \frac{1}{2} \scP H_1 \scP H_1 \scD^2 H_1 \scP - \frac{1}{2} \scP H_1 \scD^2 H_1 \scP H_1 \scP \nonumber\\
              &= \langle1,1\rangle - \frac{1}{2} \langle0,2\rangle - \frac{1}{2} \langle2,0\rangle \nonumber\\
  H^\text{eff}_4 &= \langle1,1,1\rangle - \frac{1}{2} \langle0,1,2\rangle - \frac{1}{2} \langle0,2,1\rangle - \frac{1}{2} \langle1,0,2\rangle - \frac{1}{2} \langle1,2,0\rangle - \frac{1}{2} \langle2,0,1\rangle - \frac{1}{2} \langle2,1,0\rangle + \frac{1}{2} \langle0,0,3\rangle + \frac{1}{2} \langle3,0,0\rangle \nonumber\\
  H^\text{eff}_5 &= \langle 1,1,1,1\rangle - \frac{1}{2} \langle 0,1,1,2\rangle - \frac{1}{2} \langle 0,1,2,1\rangle - \frac{1}{2} \langle 0,2,1,1\rangle - \frac{1}{2} \langle 1,0,1,2\rangle - \frac{1}{2} \langle 1,0,2,1\rangle - \frac{1}{2} \langle 1,1,0,2\rangle - \frac{1}{2} \langle 1,1,2,0\rangle \nonumber\\
              &\quad\quad - \frac{1}{2} \langle 1,2,0,1\rangle - \frac{1}{2} \langle 1,2,1,0\rangle - \frac{1}{2} \langle 2,0,1,1\rangle - \frac{1}{2} \langle 2,1,0,1\rangle - \frac{1}{2} \langle 2,1,1,0\rangle \nonumber\\
              &+ \frac{1}{2}\langle 0,0,2,2\rangle + \frac{3}{8} \langle 0,2,0,2\rangle + \frac{1}{4} \langle 2,0,0,2\rangle + \frac{3}{8} \langle 2,0,2,0\rangle + \frac{1}{2} \langle 2,2,0,0\rangle \nonumber\\
              &+ \frac{1}{2} \langle 0,0,1,3\rangle + \frac{1}{2} \langle 0,0,3,1\rangle + \frac{1}{2} \langle 0,1,0,3\rangle + \frac{1}{2} \langle 1,0,0,3\rangle + \frac{1}{2} \langle 1,3,0,0\rangle + \frac{1}{2} \langle 3,0,0,1\rangle + \frac{1}{2} \langle 3,0,1,0\rangle + \frac{1}{2} \langle 3,1,0,0\rangle \nonumber\\
              &- \frac{1}{2} \langle 0,0,0,4\rangle - \frac{1}{2} \langle 4,0,0,0\rangle \nonumber\\
  H^\text{eff}_6 &= \langle 1,1,1,1,1\rangle - \frac{1}{2} \langle 0,1,1,1,2\rangle - \frac{1}{2} \langle 0,1,1,2,1\rangle - \frac{1}{2} \langle 0,1,2,1,1\rangle - \frac{1}{2} \langle 0,2,1,1,1\rangle - \frac{1}{2} \langle 1,0,1,1,2\rangle - \frac{1}{2} \langle 1,0,1,2,1\rangle \nonumber\\
              &\quad\quad - \frac{1}{2} \langle 1,0,2,1,1\rangle - \frac{1}{2} \langle 1,1,0,1,2\rangle - \frac{1}{2} \langle 1,1,0,2,1\rangle - \frac{1}{2} \langle 1,1,1,0,2\rangle - \frac{1}{2} \langle 1,1,1,2,0\rangle - \frac{1}{2} \langle 1,1,2,0,1\rangle \nonumber\\
              &\quad\quad - \frac{1}{2} \langle 1,1,2,1,0\rangle - \frac{1}{2} \langle 1,2,0,1,1\rangle - \frac{1}{2} \langle 1,2,1,0,1\rangle - \frac{1}{2} \langle 1,2,1,1,0\rangle - \frac{1}{2} \langle 2,0,1,1,1\rangle - \frac{1}{2} \langle 2,1,0,1,1\rangle \nonumber\\
              &\quad\quad - \frac{1}{2} \langle 2,1,1,0,1\rangle - \frac{1}{2} \langle 2,1,1,1,0\rangle \nonumber\\
              &+ \frac{1}{2} \langle 0,0,1,2,2\rangle + \frac{1}{2} \langle 0,0,2,1,2\rangle + \frac{1}{2} \langle 0,0,2,2,1\rangle + \frac{1}{2} \langle 0,1,0,2,2\rangle + \frac{3}{8} \langle 0,1,2,0,2\rangle + \frac{3}{8} \langle 0,2,0,1,2\rangle + \frac{3}{8} \langle 0,2,0,2,1\rangle \nonumber\\
              &\quad\quad + \frac{3}{8} \langle 0,2,1,0,2\rangle + \frac{1}{2} \langle 1,0,0,2,2\rangle + \frac{3}{8} \langle 1,0,2,0,2\rangle + \frac{1}{4} \langle 1,2,0,0,2\rangle + \frac{3}{8} \langle 1,2,0,2,0\rangle + \frac{1}{2} \langle 1,2,2,0,0\rangle \nonumber\\
              &\quad\quad + \frac{1}{4} \langle 2,0,0,1,2\rangle + \frac{1}{4} \langle 2,0,0,2,1\rangle + \frac{1}{4} \langle 2,0,1,0,2\rangle + \frac{3}{8} \langle 2,0,1,2,0\rangle + \frac{3}{8} \langle 2,0,2,0,1\rangle + \frac{3}{8} \langle 2,0,2,1,0\rangle \nonumber\\
              &\quad\quad + \frac{1}{4} \langle 2,1,0,0,2\rangle + \frac{3}{8} \langle 2,1,0,2,0\rangle + \frac{1}{2} \langle 2,1,2,0,0\rangle + \frac{1}{2} \langle 2,2,0,0,1\rangle + \frac{1}{2} \langle 2,2,0,1,0\rangle + \frac{1}{2} \langle 2,2,1,0,0\rangle \nonumber\\
              &+ \frac{1}{2} \langle 0,0,1,1,3\rangle + \frac{1}{2} \langle 0,0,1,3,1\rangle + \frac{1}{2} \langle 0,0,3,1,1\rangle + \frac{1}{2} \langle 0,1,0,1,3\rangle + \frac{1}{2} \langle 0,1,0,3,1\rangle + \frac{1}{2} \langle 0,1,1,0,3\rangle + \frac{1}{2} \langle 1,0,0,1,3\rangle \nonumber\\
              &\quad\quad + \frac{1}{2} \langle 1,0,0,3,1\rangle + \frac{1}{2} \langle 1,0,1,0,3\rangle + \frac{1}{2} \langle 1,1,0,0,3\rangle + \frac{1}{2} \langle 1,1,3,0,0\rangle + \frac{1}{2} \langle 1,3,0,0,1\rangle + \frac{1}{2} \langle 1,3,0,1,0\rangle \nonumber\\
              &\quad\quad + \frac{1}{2} \langle 1,3,1,0,0\rangle + \frac{1}{2} \langle 3,0,0,1,1\rangle + \frac{1}{2} \langle 3,0,1,0,1\rangle + \frac{1}{2} \langle 3,0,1,1,0\rangle + \frac{1}{2} \langle 3,1,0,0,1\rangle + \frac{1}{2} \langle 3,1,0,1,0\rangle \nonumber\\
              &\quad\quad + \frac{1}{2} \langle 3,1,1,0,0\rangle \nonumber\\
              &- \frac{1}{2} \langle 0,0,0,2,3\rangle - \frac{1}{2} \langle 0,0,0,3,2\rangle - \frac{1}{2} \langle 0,0,2,0,3\rangle - \frac{3}{8} \langle 0,0,3,0,2\rangle - \frac{3}{8} \langle 0,2,0,0,3\rangle + \frac{1}{8} \langle 0,2,0,3,0\rangle - \frac{1}{8} \langle 0,3,0,0,2\rangle \nonumber\\
              &\quad\quad+ \frac{1}{8} \langle 0,3,0,2,0\rangle - \frac{1}{4} \langle 2,0,0,0,3\rangle - \frac{1}{8} \langle 2,0,0,3,0\rangle - \frac{3}{8} \langle 2,0,3,0,0\rangle - \frac{1}{2} \langle 2,3,0,0,0\rangle - \frac{1}{4} \langle 3,0,0,0,2\rangle \nonumber\\
              &\quad\quad - \frac{3}{8} \langle 3,0,0,2,0\rangle - \frac{1}{2} \langle 3,0,2,0,0\rangle - \frac{1}{2} \langle 3,2,0,0,0\rangle \nonumber\\
              &- \frac{1}{2} \langle 0,0,0,1,4\rangle - \frac{1}{2} \langle 0,0,0,4,1\rangle - \frac{1}{2} \langle 0,0,1,0,4\rangle - \frac{1}{2} \langle 0,1,0,0,4\rangle - \frac{1}{2} \langle 1,0,0,0,4\rangle - \frac{1}{2} \langle 1,4,0,0,0\rangle - \frac{1}{2} \langle 4,0,0,0,1\rangle \nonumber\\
              &\quad\quad - \frac{1}{2} \langle 4,0,0,1,0\rangle - \frac{1}{2} \langle 4,0,1,0,0\rangle - \frac{1}{2} \langle 4,1,0,0,0\rangle \nonumber\\
              &+ \frac{1}{2} \langle 0,0,0,0,5\rangle + \frac{1}{2} \langle 5,0,0,0,0\rangle \nonumber
\end{align}
\end{widetext}

\section{Monte Carlo of Dual Ising Theories}
\label{app:MC}

In this appendix, we use classical Monte Carlo to provide evidence that the weakly coupled stacks of transverse-field Ising models that describe the dual phase transitions in \figref{fig:duality} are likely to be first order.
There are two cases to consider, 1) a weakly coupled 1D stack of 2+1D Ising models (as in \secref{sec:duality}), and 2) a weakly coupled 2D grid of 1+1D Ising models (as in \secref{sec:duality'}).
A classical Ising model in 3D and 2D will be used to simulate the 2+1D and 1+1D (respectively) zero temperature transverse field Ising model layers.

$I$ and $J$ will be used to denote the 3D or 2D lattice coordinate,
  while $L$ will denote the layer (of the 1D stack or 2D grid).
Thus, we will be interested in the following classical layered Ising-like model (with inverse temperature $\beta=1$):
\begin{align}
 H_\text{Ising}
   &= -J \sum_L     \sum_{\langle I J \rangle} \sigma_{L I} \sigma_{L J} \label{eq:Ising}\\
   &  -K \sum_{\langle LL' \rangle} \sum_{\langle I J \rangle} \sigma_{L I} \sigma_{L J} \sigma_{L' I} \sigma_{L' J} \nonumber\\
   &= - \sum_L \sum_{\langle I J \rangle} \underbrace{\left( J + K \sum_{\langle LL' \rangle}^L \sigma_{L' I} \sigma_{L' J} \right)}_{J_{LIJ}^\text{eff}} \sigma_{L I} \sigma_{L J} \label{eq:Ising'}
\end{align}
$\sum_L$ sums over the layers and $\sum_{\langle I J \rangle}$ sums over nearest-neighbor sites.
The $J$ term is a nearest-neighbor ferromagnetic Ising coupling that only couples spins on the same layer,
  where $J$ is roughly related by $J \sim \EQ / t$ in \eqnref{eq:H dual} or $J \sim h/\EX$ in \eqnref{eq:H dual'}.
The $K$ term is a 4-spin term which couples together neighbor-neighbor spins or neighboring layers,
  and roughly corresponds to a product of two $\sQ{\tilde}{p}$ or two $\tQ{\tilde}{p}$ operators in \secref{sec:duality},
  or $J \sim \EQyz/\EX$ in \eqnref{eq:H dual'}.
It is useful to rewrite $H_\text{Ising}$ as in \eqnref{eq:Ising'} to reveal that the $K$ term results in a new effective $J$ term ($J_{LIJ}^\text{eff}$) which is site and layer dependent.
$\sum_{\langle LL' \rangle}^L$ sums over all layers $L'$ which neighbor the layer $L$.

If $K=0$, then the layers are decoupled and every layer independently undergoes an order to disorder transition.
If $K$ is small and positive, then we can see from the form of $J_{LIJ}^\text{eff}$ that when a layer transitions into its ordered state,
  the average $J_{LIJ}^\text{eff}$ for neighboring layers will increase.
Thus, positive $K$ causes the effective $J$ to increase rapidly at the disorder to order transitions when $J$ is increased.
It is reasonable that this could lead to a first order transition.

If $K$ is small and negative, then when a layer transitions into its ordered state,
  the average $J_{LIJ}^\text{eff}$ for neighboring layers will decrease.
This makes it difficult for the system to order all at once.
What happens instead for intermediate $J$ is that only half the layers order such that ordered layers neighbor disordered layers, and visa versa.
Thus, $K$ makes the average $J_{LIJ}^\text{eff}$ smaller only for the disordered layers.
In \secref{app:interPhases}, we will briefly discuss what these phases correspond to in the original model.

From the form of \eqnref{eq:Ising'} we see that each layer is an Ising model with a site-dependent coupling $J_{LIJ}^\text{eff}$.
Therefore, we can apply the global Wolff update algorithm \cite{Wolff1989} to this model one layer at a time.
Thus, it is easy to generalize Wolff update code to efficiently study this model even when the correlation length along the layers is large.
However, since the update is only applied to one layer at a time, the update is effectively local across the layers.
Thus, if the correlation length across the layers is large, the simulation will be inefficient.
This can occur when the layer coupling $K$ is large;
  however, since we are interested in weak layer coupling $K$, this isn't a major issue.

\begin{figure*}
1D stack of 2+1D layers with $K = +1/32$: \\
\includegraphics[width=.28\textwidth]{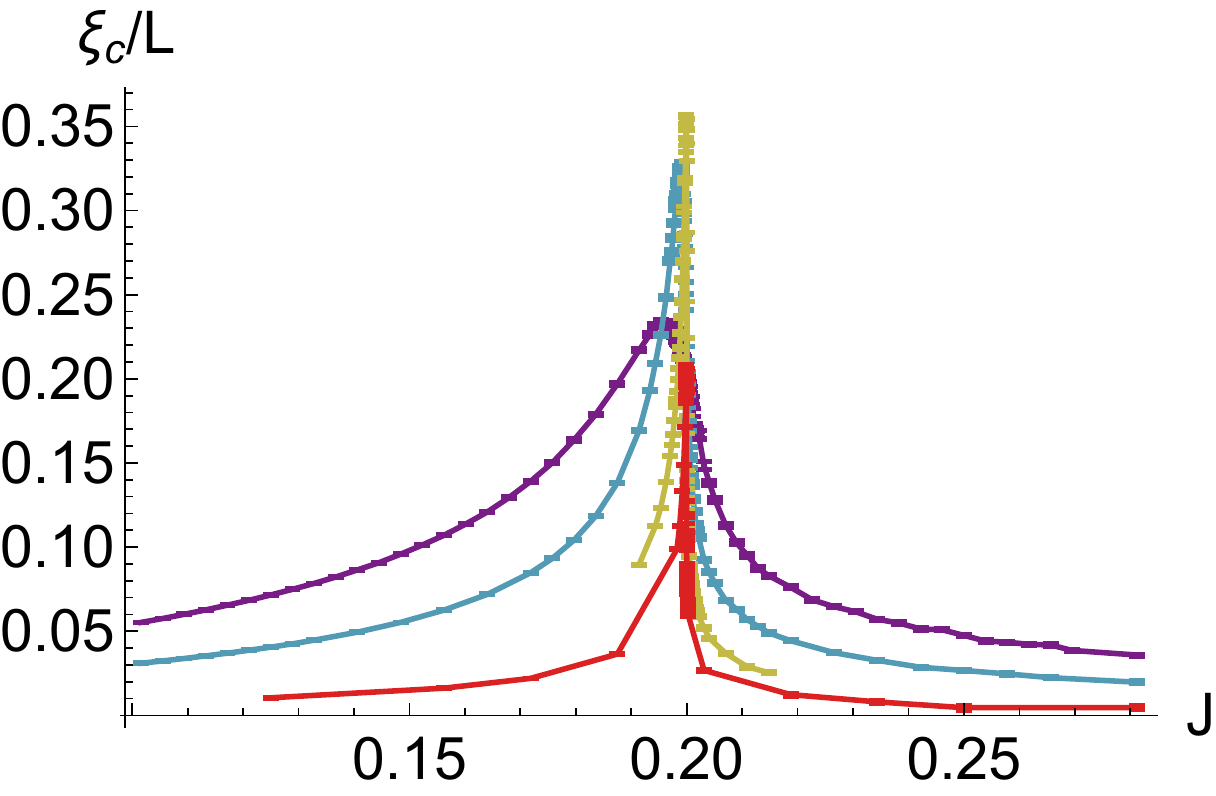}
\includegraphics[width=.28\textwidth]{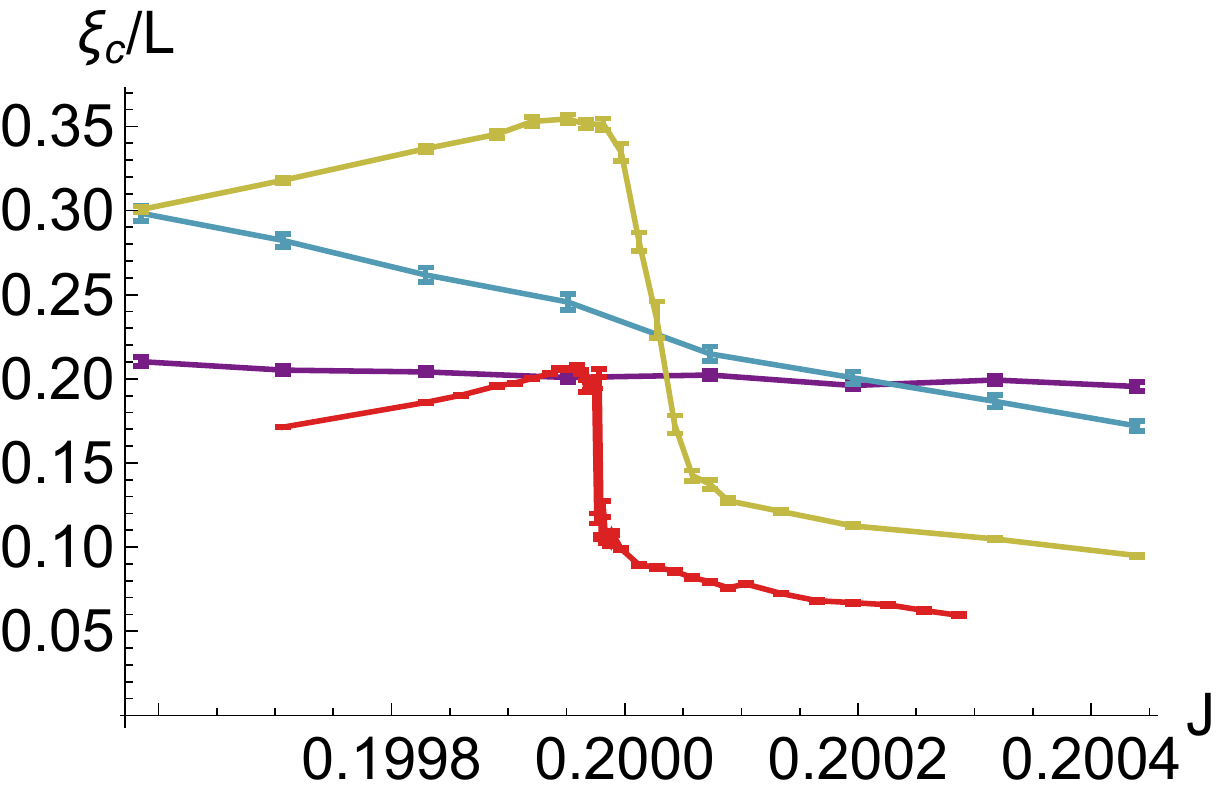}
\includegraphics[width=.28\textwidth]{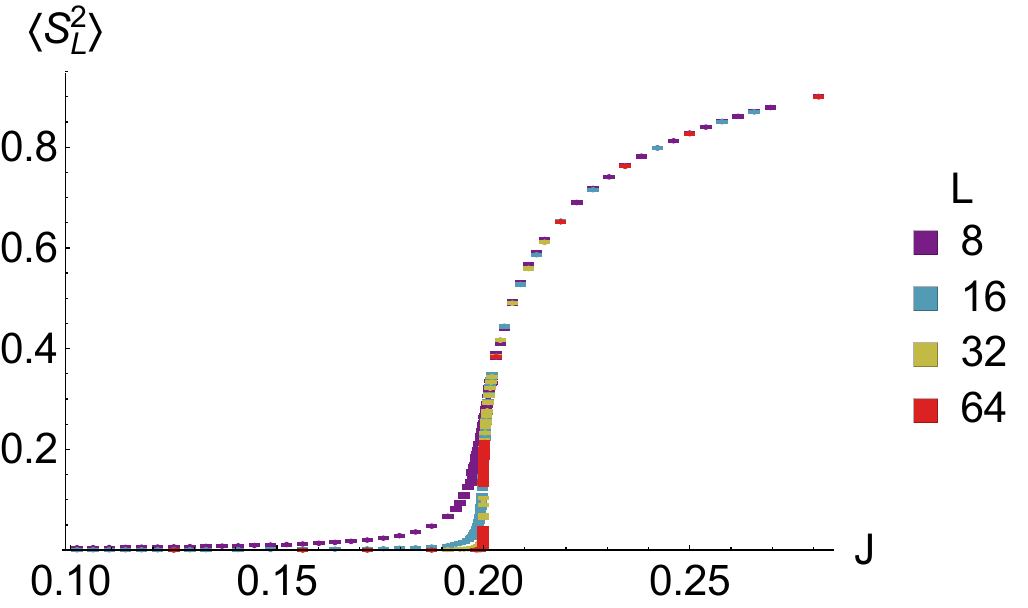} \\ \vspace{.3cm}
1D stack of 2+1D layers with $K = -1/4$: \\
\includegraphics[width=.28\textwidth]{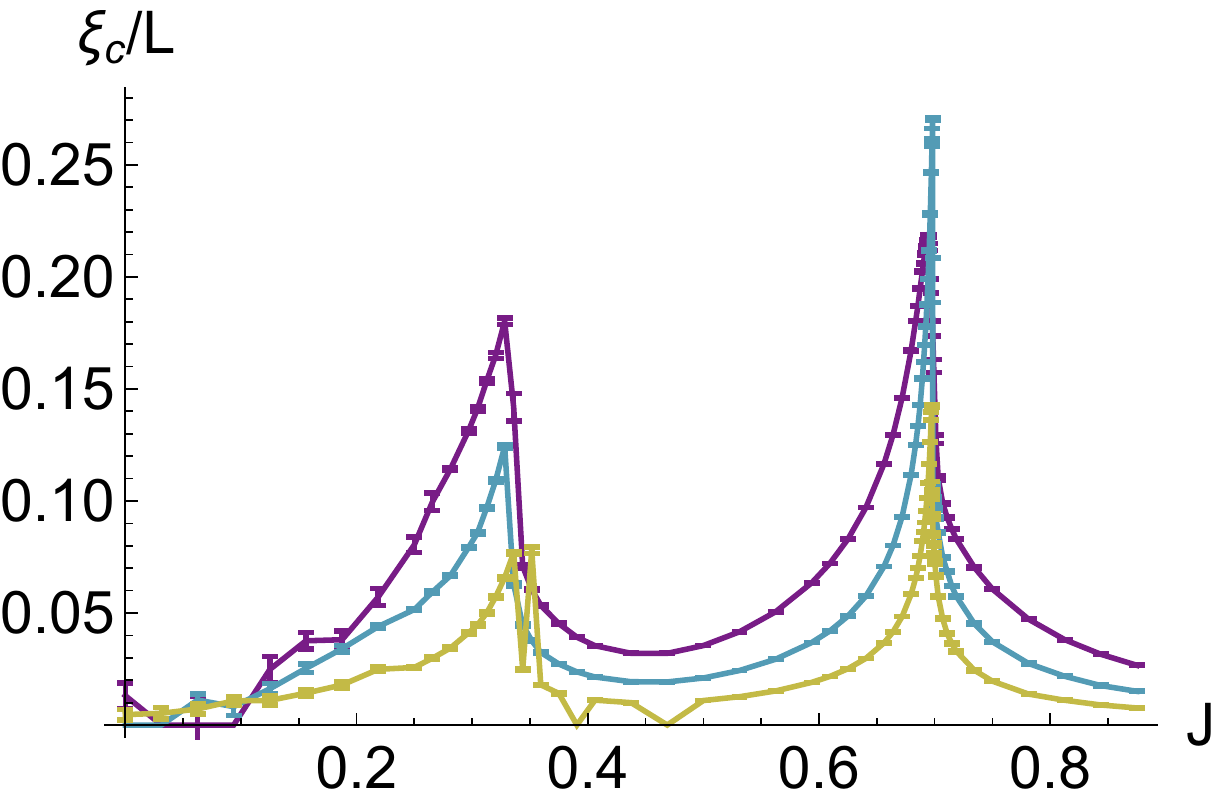}
\includegraphics[width=.28\textwidth]{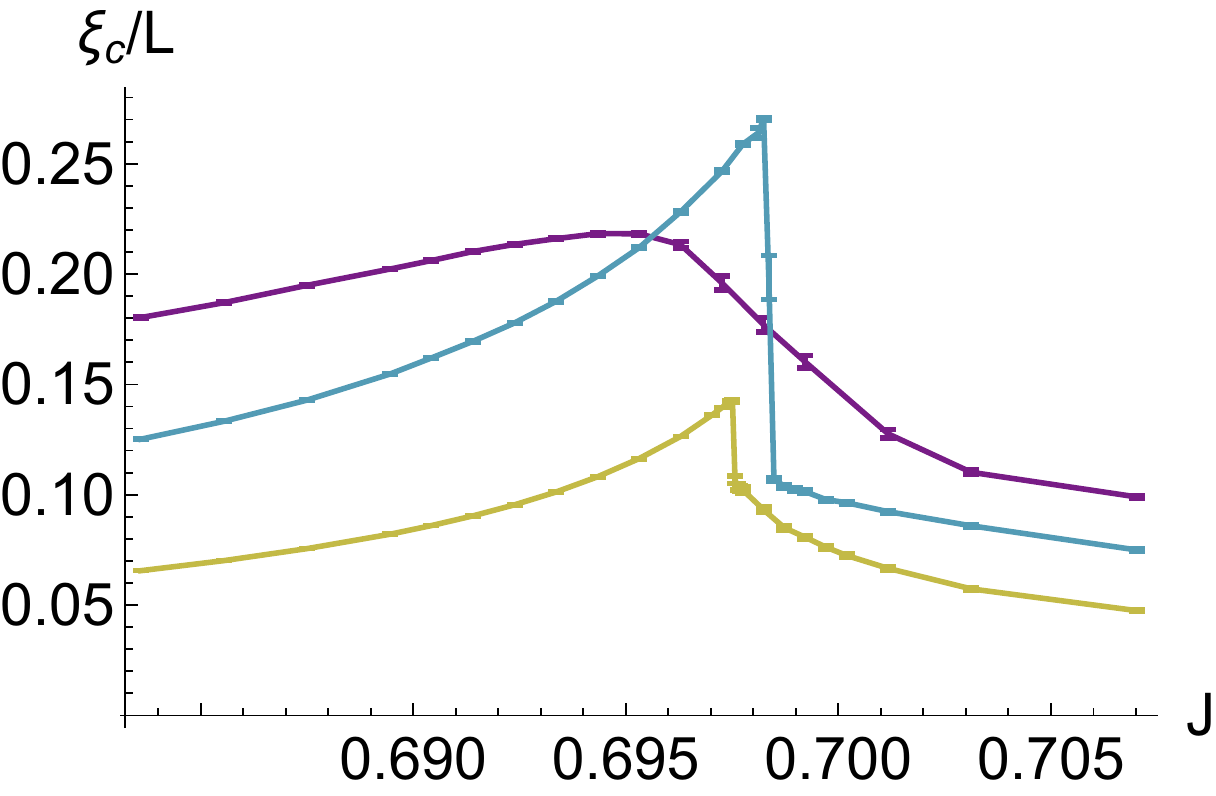}
\includegraphics[width=.28\textwidth]{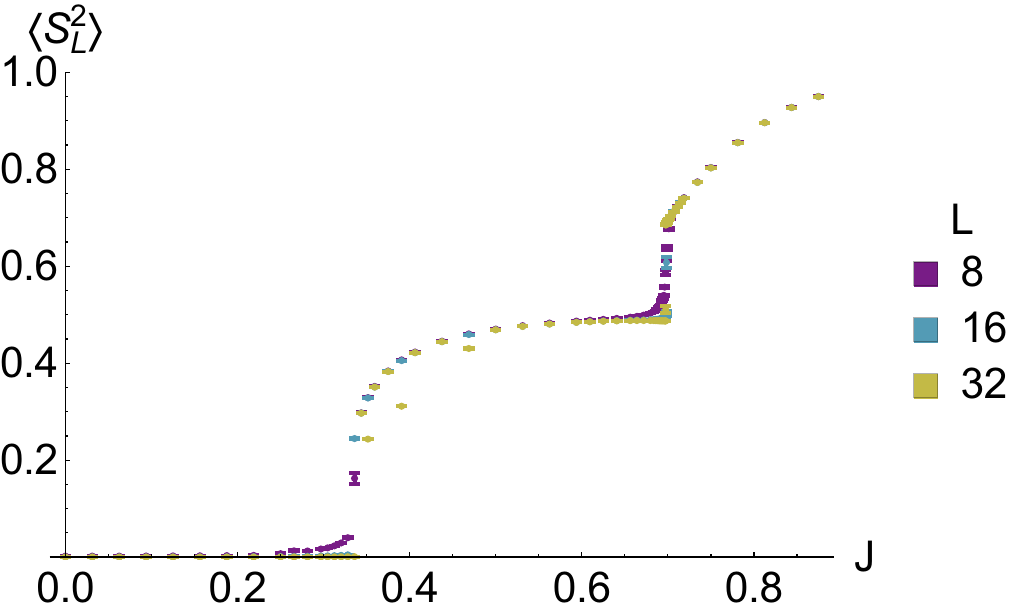} \\ \vspace{.3cm}
2D stack of 1+1D layers with $K = +1/16$: \\
\includegraphics[width=.28\textwidth]{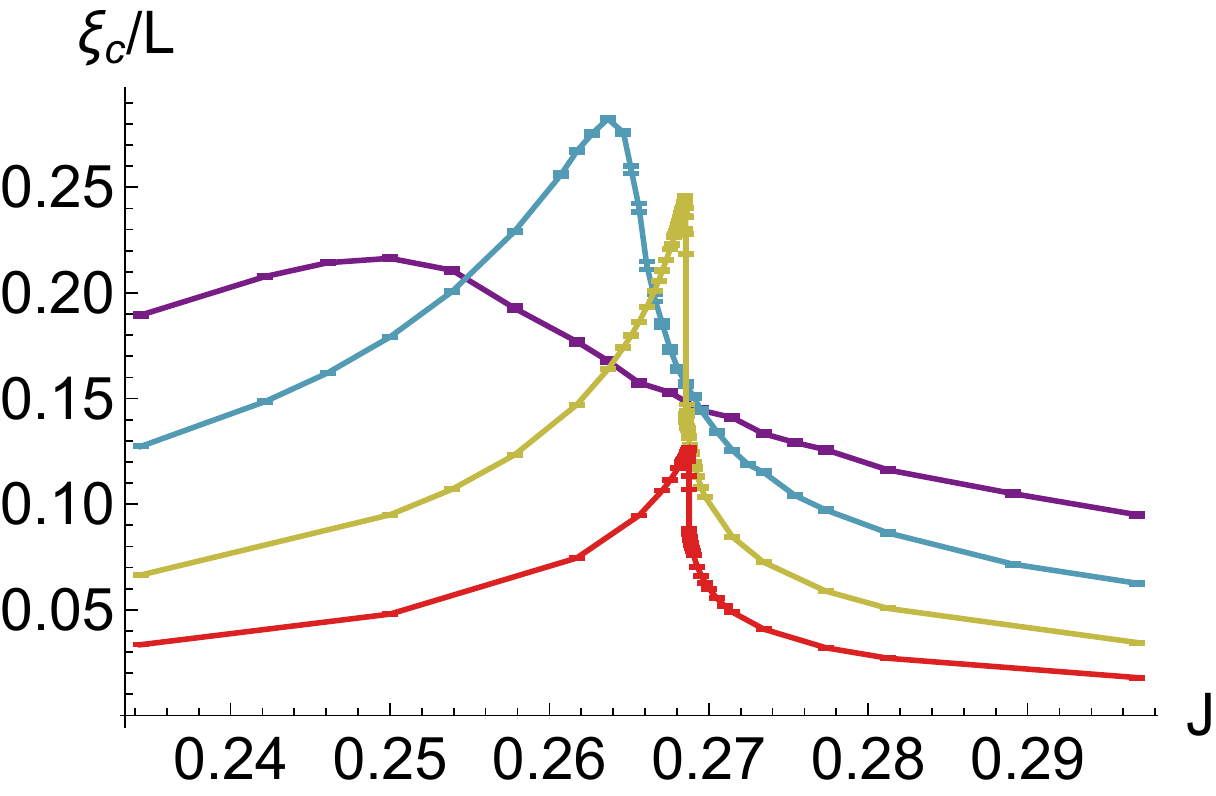}
\includegraphics[width=.28\textwidth]{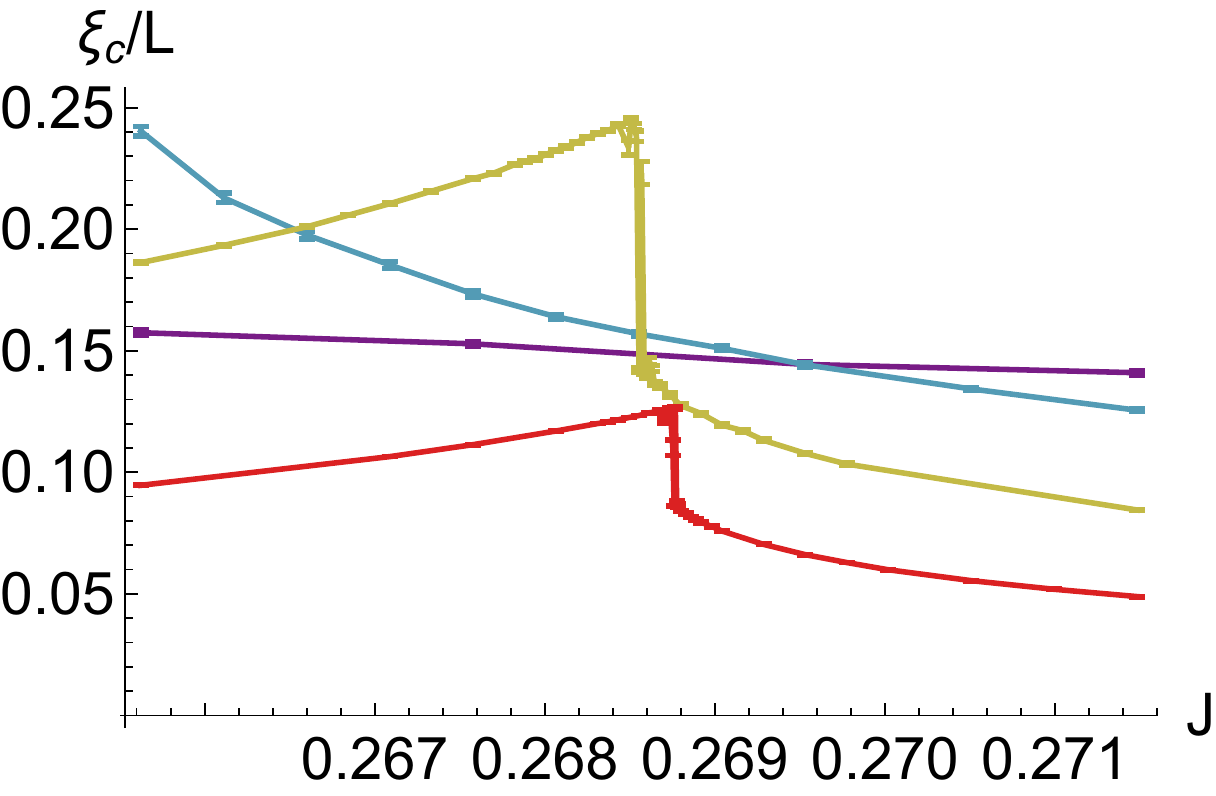}
\includegraphics[width=.28\textwidth]{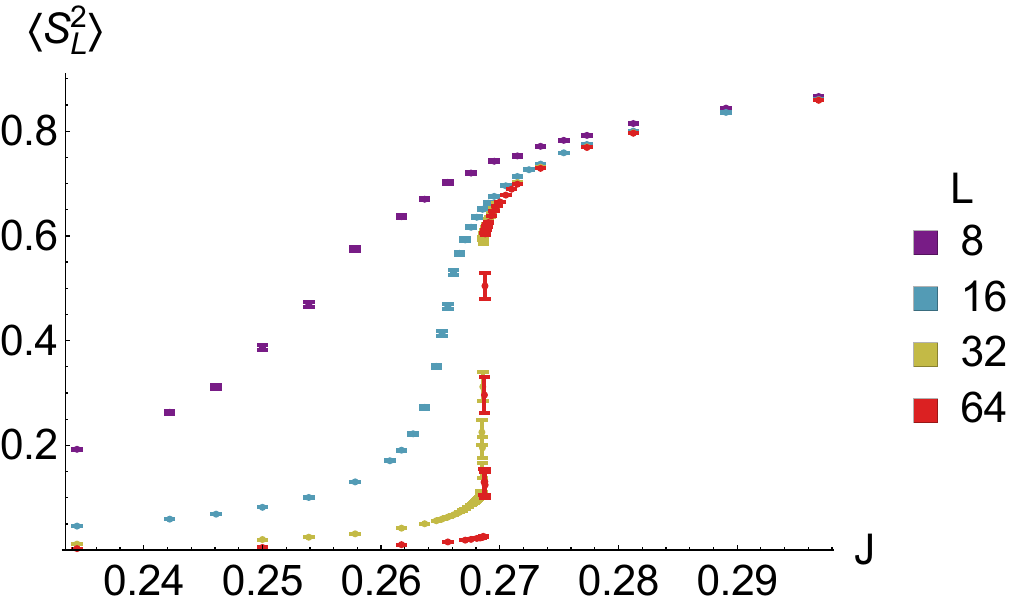} \\ \vspace{.3cm}
2D stack of 1+1D layers with $K = -1/8$: \\
\includegraphics[width=.28\textwidth]{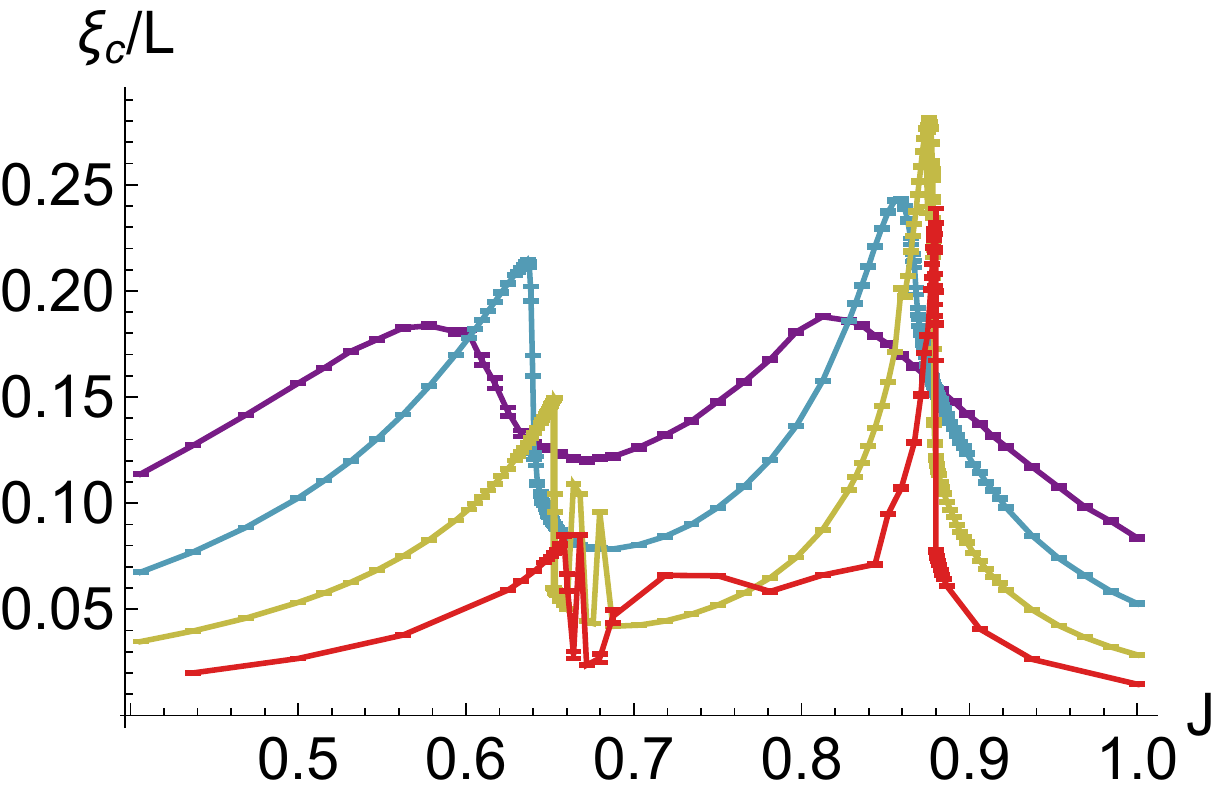}
\includegraphics[width=.28\textwidth]{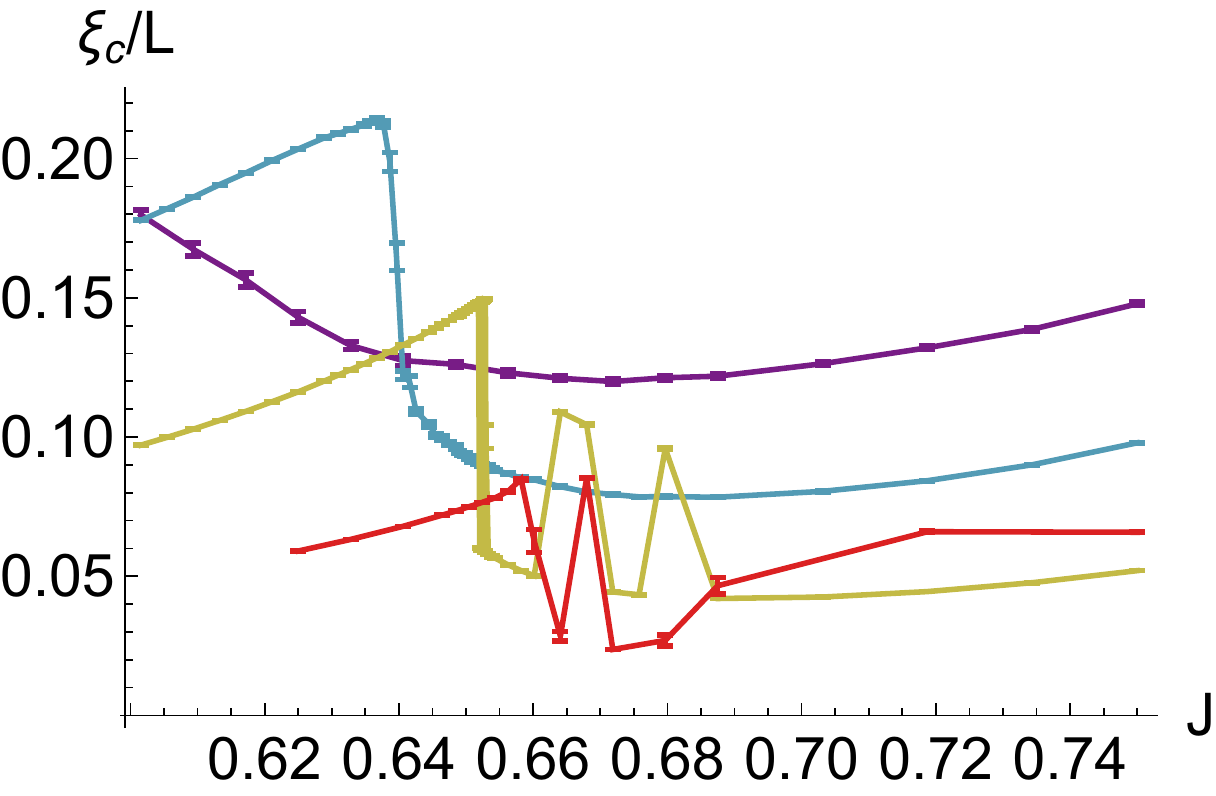}
\includegraphics[width=.28\textwidth]{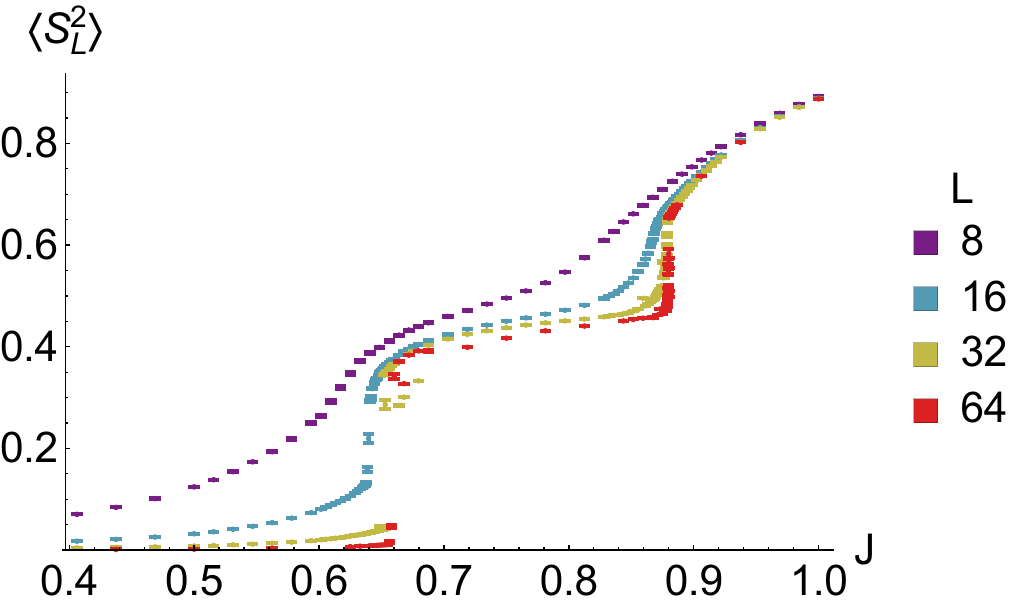} \\ \vspace{.3cm}
2D stack of 1+1D layers with $K = -1/4$: \\
\hspace{.3\textwidth}
\includegraphics[width=.28\textwidth]{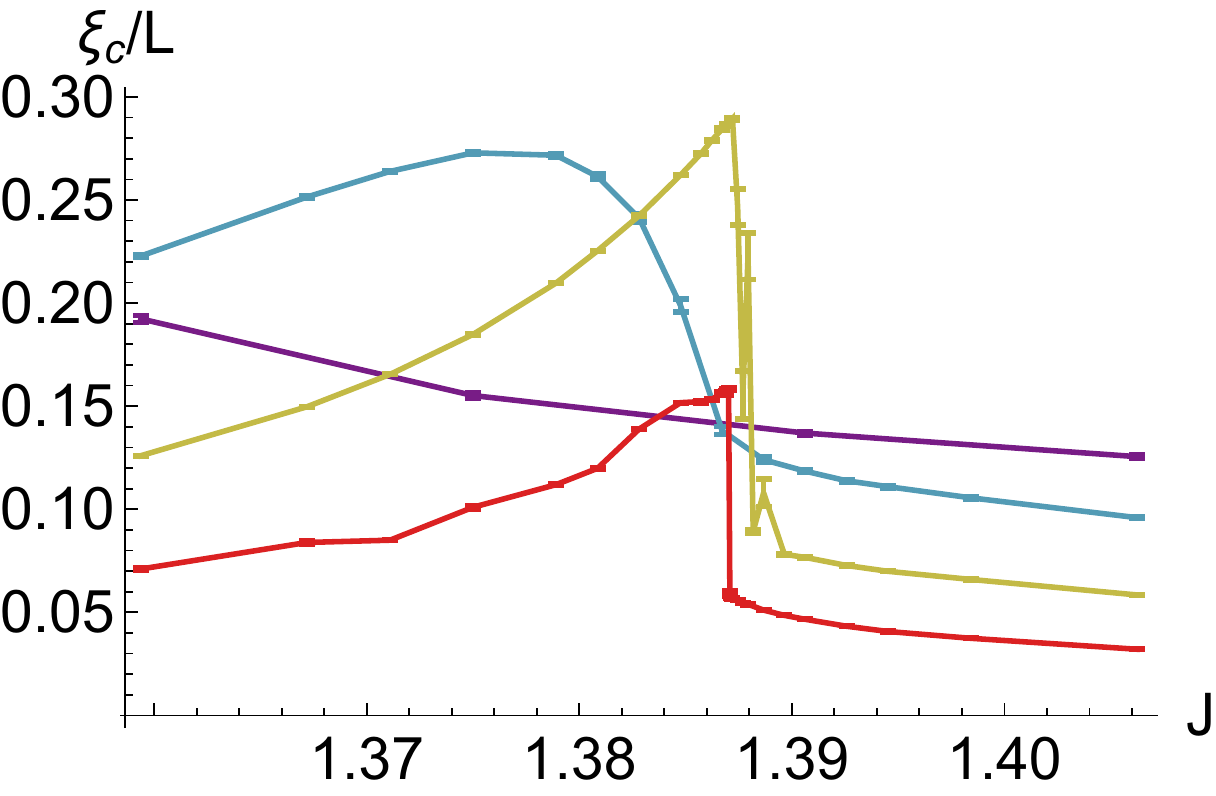}
\includegraphics[width=.28\textwidth]{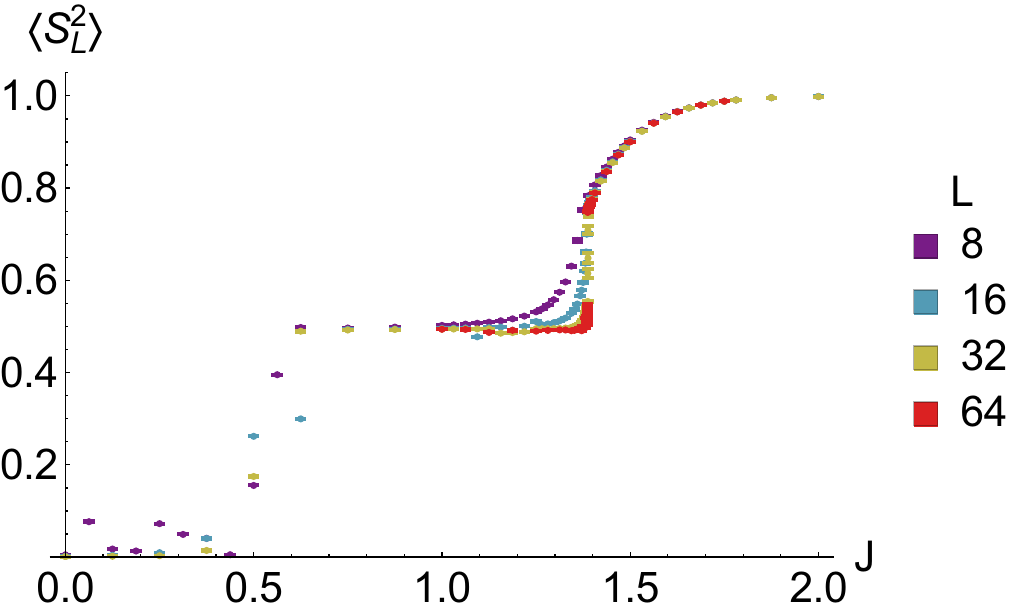}
\caption{
The first column of plots shows $\xi_c/L$ vs the ferromagnetic Ising coupling $J$,
  where $\xi_c$ is the connected correlation length between two spins on the same layer.
The second column is a closeup of a phase transition.
The third column is the squared layer magnetization $\langle S_L^2 \rangle$ vs $J$,
  where $S_L \equiv \frac{1}{N_I} \sum_I \sigma_{L I}$ is the average magnetization on a given layer.
The system size is $L \times L \times L \times L$.
Each row corresponds to a different dimension for each layer, or different sign of $K$.
All phase transitions appear to be first order since at each transition,
  $\xi_c/L$ tends to decrease as the system length $L$ increases.
An exception is the second transition for the 2D stack of 1+1D layers with $K = -1/8$ (4th row and 1st column of plots, near $J \sim 0.88$).
However, when $K$ is increased to $K = -1/4$, the transition looks first order.
(When $K = -1/4$, we cannot accurately measure the correlation length below $J \sim 1.2$ due to large autocorrelation times in this region.)
We expect that larger system sizes would show that the $K = -1/8$ transition is also second order.
Some of the data is a little messy (e.g. not smooth vs $J$) due to a loss of ergodicity (due to large autocorrelation times) near the first order transitions,
  which is a well known shortcoming of the Wolff MC update algorithm near a first order phase transition.
(The one standard deviation statistical error bars do not account for this error.)
All data points used at least $2^{12}$ MC sweeps.
The second row used $2^{16}$ sweeps.
}\label{fig:MC}
\end{figure*}

After applying this algorithm \cite{MC:github}, we find that all of the transitions appear to be first order.
The simulation results are summarized in \figref{fig:MC}.
The figure shows that at every phase transition, $\xi_c/L$ tends to decrease as the system size $L$ increases,
  where $\xi_c$ is the connected correlation length between two spins on the same layer.
\footnote{The connected correlation length $\xi_c$ was calculated using Eqn. 73 in \cite{Sandvik2011}}
(At a continuous phase transition, $\xi_c/L$ would instead asymptote to a constant.)
We also find other signs of discontinuous phase transitions, such as magnetization histograms with two bumps,
  which indicates the presence of metastable states and hysteresis effects which occur in discontinuous transitions.
However, we can not rule out the possibility that smaller $K$ could instead result in a continuous phase transition.
Indeed, our smaller $K$ looks second order at the currently simulated lattice sizes;
  however, we expect that larger lattices would then likely show signs of discontinuous phase transitions.

\subsection{Intermediate Phases}
\label{app:interPhases}

\begin{figure}
1D stack of 2+1D layers with $K = -1/4$: \\
\includegraphics[width=.3\textwidth]{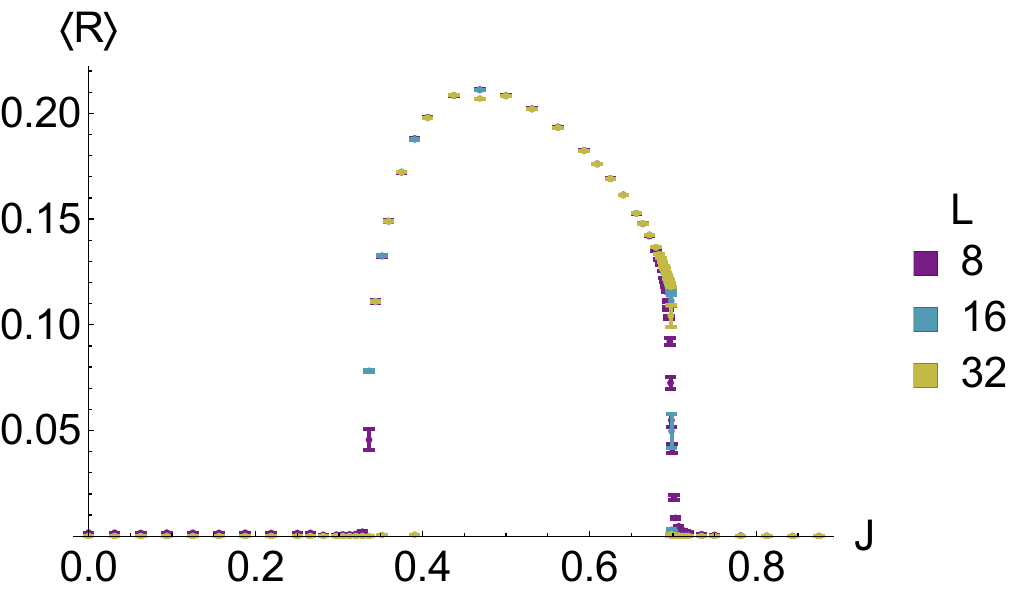} \\ \vspace{.5cm}
2D stack of 1+1D layers with $K = -1/8$: \\
\includegraphics[width=.3\textwidth]{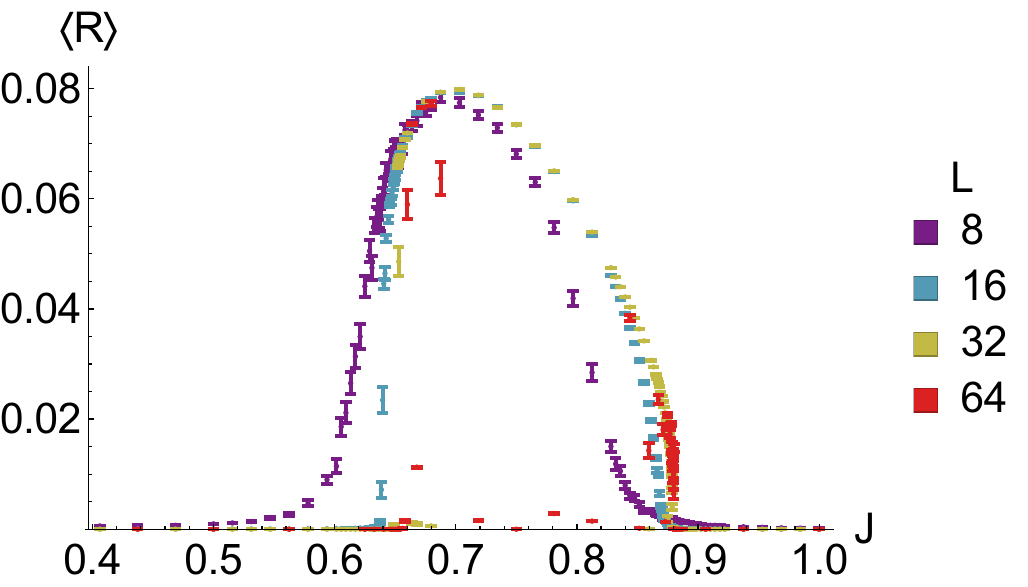}
\caption{
We show that the order parameter $R$ (\eqnref{eq:R}) describes the intermediate phase.
}\label{fig:intermediate}
\end{figure}

From \figref{fig:MC} we see that when the layer coupling $K$ is negative, there is an intermediate phase where the squared layer magnetization $\langle S_L^2 \rangle \sim 1/2$,
  where $S_L \equiv \frac{1}{N_I} \sum_I \sigma_{L I}$ is the average magnetization on a given layer $L$.
As previously mentioned, this phase spontaneously breaks translation symmetry and is characterized by a checkerboard pattern where half of the layers order, while the other half remain disordered.
This can be described by the following order parameter
\begin{align}
  R  &\equiv \frac{1}{N'} \sum_L \sum_{\langle I J \rangle} (-1)^L \sigma_{LI} \sigma_{LJ} \label{eq:R}\\
  N' &\equiv \sum_L \sum_{\langle I J \rangle} 1 \nonumber
\end{align}
In \figref{fig:intermediate} we show that this order parameter $R$ is nonzero only in the intermediate phase.

These intermediate phases of the layered Ising models are dual phases of phases in the original fracton model.
We can determine which phase results in the original model by adding a term to the dual Ising model which explicitly breaks the symmetry that is spontaneously broken in the intermediate phase,
  and then apply the appropriate duality mapping (\eqnref{eq:duality}, \ref{eq:duality'} or \ref{eq:duality global}) in reverse.
To explicitly break the symmetry, we add the following term to $H_\text{Ising}$ (\eqnref{eq:Ising}):
\begin{align}
  H_\text{inter}^\text{dual} = - F \sum_L^\text{odd} \sum_{\langle IJ \rangle} \sigma_{LI} \sigma_{LJ}
\end{align}
where $F$ is a large constant and $\sum_L^\text{odd}$ only sums over every other layer (in a checkerboard pattern).
If we are interested in the 1+1D duality (\secref{sec:duality}), then we apply \eqnref{eq:duality} and find that $H_\text{inter}^\text{dual}$ is dual to
\begin{align}
  H_\text{inter}^\text{1+1D} = -F \sum_{\text{x-axis } \ell}^{y+z \text{ odd}} \bhat\rho^z_\ell
\end{align}
where $\sum_{\text{x-axis } \ell}^{y+z \text{ odd}}$ only sums over x-axis links $\ell$ with odd $y+z$.
To describe the intermediate phase between X-cube order and YZ-plane 2+1D $Z_2$ QSL (intermediate \#4 in \tabref{tab:degen}), we can add $H_\text{inter}^\text{1+1D}$ to $\bhat{H}_\text{X-cube}$ (\eqnref{eq:X-cube}) and apply degenerate perturbation theory (DPT) using $H_\text{inter}^\text{1+1D}$ as the unperturbed Hamiltonian.
This will generate a Hamiltonian which describes the intermediate phase.
This process can be repeated to generate Hamiltonians for the possible intermediate phases for all four phase transitions shown in \figref{fig:duality}.
We do not write down these Hamiltonians here, nor have we analyzed what kind of topological order is present in these models.
However, we have calculated their ground state degeneracies, which we show in \tabref{tab:degen}.
This was done via a computer program \cite{degeneracy:github} on finite lattices; we then extrapolated to arbitrary lattice sizes.

\begin{table*}
\begin{center}
\begin{tabular}{c||c|c|c|c|c}
            & XY\&XZ plane      &                                &                        &                       & YZ-plane \\
 phase      & 2+1D $Z_2$ QSL    & intermediate \#1               & 3+1D $Z_2$ QSL         & intermediate \#2      & 2+1D $Z_2$ QSL \\\hline
 degeneracy & $2^{2L_y + 2L_z}$ & $2 \times 2^6$                 & $2^3$                  & $2 \times 2^{L_x}$    & $2^{2L_x}$ \\ \hline\hline
 phase      &                   & intermediate \#3               & X-cube                 & intermediate \#4      & \\\hline
 degeneracy &                   & $2 \times 2^{L_x+2L_y+2L_z-3}$ & $2^{2L_x+2L_y+2L_z-3}$ & $2 \times 2^{2L_x+3}$ & 
\end{tabular}
\end{center}
\caption{
Ground state degeneracies for the four phases in \figref{fig:duality}, along with the four possible intermediate phases that can occur.
For example, intermediate \#3 can occur between the 2+1D $Z_2$ QSL and X-cube phases.
The degeneracies of all of the intermediate phases take the form $2 \times \cdots$.
The first factor of two in the degeneracies is not topological,
  and occurs due to the spontaneous translation symmetry breaking of the intermediate phase.
The rest of the degeneracy is topological.
}\label{tab:degen}
\end{table*}


\end{document}